\documentclass[12pt,english]{article}
\usepackage{times}
\usepackage[T1]{fontenc}
\usepackage{geometry}
\usepackage{array}
\usepackage{graphicx}
\usepackage{setspace}
\usepackage{amssymb}

\makeatletter

\providecommand{\tabularnewline}{\\}

\usepackage{bm}

\usepackage{babel}
\makeatother
\begin{document}

\section*{\noindent Planning of \emph{EM} Skins for Improved Quality-of-Service
in Urban Areas}

\noindent ~

\noindent \vfill

\noindent A. Benoni,$^{(1)}$ M. Salucci,$^{(1)}$ \emph{Member, IEEE},
G. Oliveri,$^{(1)}$ \emph{Senior Member, IEEE}, P. Rocca,$^{(1)(2)}$
\emph{Senior Member, IEEE}, B. Li,$^{(3)(4)}$ and A. Massa,$^{(1)(3)(5)}$
\emph{Fellow, IEEE}

\noindent \vfill

\noindent ~

\noindent {\footnotesize $^{(1)}$} \emph{\footnotesize ELEDIA Research
Center} {\footnotesize (}\emph{\footnotesize ELEDIA}{\footnotesize @}\emph{\footnotesize UniTN}
{\footnotesize - University of Trento)}{\footnotesize \par}

\noindent {\footnotesize DICAM - Department of Civil, Environmental,
and Mechanical Engineering}{\footnotesize \par}

\noindent {\footnotesize Via Mesiano 77, 38123 Trento - Italy}{\footnotesize \par}

\noindent \textit{\emph{\footnotesize E-mail:}} {\footnotesize \{}\emph{\footnotesize arianna}{\footnotesize .}\emph{\footnotesize benoni}{\footnotesize ,}
\emph{\footnotesize marco.salucci}{\footnotesize ,} \emph{\footnotesize giacomo.oliveri}{\footnotesize ,}
\emph{\footnotesize paolo.rocca}{\footnotesize ,} \emph{\footnotesize andrea.massa}{\footnotesize \}@}\emph{\footnotesize unitn.it}{\footnotesize \par}

\noindent {\footnotesize Website:} \emph{\footnotesize www.eledia.org/eledia-unitn}{\footnotesize \par}

\noindent {\footnotesize ~}{\footnotesize \par}

\noindent {\footnotesize $^{(2)}$} \emph{\footnotesize ELEDIA Research
Center} {\footnotesize (}\emph{\footnotesize ELEDIA@XIDIAN} {\footnotesize -
Xidian University)}{\footnotesize \par}

\noindent {\footnotesize P.O. Box 191, No.2 South Tabai Road, 710071
Xi'an, Shaanxi Province - China}{\footnotesize \par}

\noindent {\footnotesize E-mail:} \emph{\footnotesize paolo.rocca@xidian.edu.cn}{\footnotesize \par}

\noindent {\footnotesize Website:} \emph{\footnotesize www.eledia.org/eledia-xidian}{\footnotesize \par}

\noindent {\footnotesize ~}{\footnotesize \par}

\noindent {\footnotesize $^{(3)}$} \emph{\footnotesize ELEDIA Research
Center} {\footnotesize (}\emph{\footnotesize ELEDIA@TSINGHUA} {\footnotesize -
Tsinghua University)}{\footnotesize \par}

\noindent {\footnotesize 30 Shuangqing Rd, 100084 Haidian, Beijing
- China}{\footnotesize \par}

\noindent {\footnotesize E-mail: \{}\emph{\footnotesize andrea.massa}{\footnotesize ,}
\emph{\footnotesize libaozhu}{\footnotesize \}}\emph{\footnotesize @tsinghua.edu.cn}{\footnotesize \par}

\noindent {\footnotesize Website:} \emph{\footnotesize www.eledia.org/eledia-tsinghua}{\footnotesize \par}

\noindent {\footnotesize ~}{\footnotesize \par}

\noindent {\footnotesize $^{(4)}$} \emph{\footnotesize Beijing National
Research Center for Information Science and Technology} {\footnotesize (}\emph{\footnotesize BNRist}{\footnotesize )
- Tsinghua University}{\footnotesize \par}

\noindent \emph{\footnotesize 30 Shuangqing Road, 100084, Haidian,
Beijing - China}{\footnotesize \par}

\noindent {\footnotesize E-mail:} \emph{\footnotesize libaozhu@tsinghua.edu.cn}{\footnotesize \par}

\noindent {\footnotesize ~}{\footnotesize \par}

\noindent {\footnotesize $^{(5)}$} \emph{\footnotesize ELEDIA Research
Center} {\footnotesize (}\emph{\footnotesize ELEDIA}{\footnotesize @}\emph{\footnotesize UESTC}
{\footnotesize - UESTC)}{\footnotesize \par}

\noindent {\footnotesize School of Electronic Engineering, Chengdu
611731 - China}{\footnotesize \par}

\noindent \textit{\emph{\footnotesize E-mail:}} \emph{\footnotesize andrea.massa@uestc.edu.cn}{\footnotesize \par}

\noindent {\footnotesize Website:} \emph{\footnotesize www.eledia.org/eledia}{\footnotesize -}\emph{\footnotesize uestc}{\footnotesize \par}

\noindent \vfill

\noindent \emph{This work has been submitted to the IEEE for possible
publication. Copyright may be transferred without notice, after which
this version may no longer be accessible.}

\noindent \vfill

\newpage
\section*{Planning of \emph{EM} Skins for Improved Quality-of-Service in Urban
Areas}

~

~

~

\begin{flushleft}A. Benoni, M. Salucci, G. Oliveri, P. Rocca, B. Li,
and A. Massa\end{flushleft}

\vfill

\begin{abstract}
\noindent The optimal planning of electromagnetic skins (\emph{EMS}s)
installed on the building facades to enhance the received signal strength,
thus the wireless coverage and/or the quality-of-service (\emph{QoS})
in large-scale urban areas, is addressed. More specifically, a novel
instance of the System-by-Design (\emph{SbD}) paradigm is proposed
towards the implementation of a smart electromagnetic environment
(\emph{SEME}) where low-cost passive static reflective skins are deployed
to enhance the level of the power received within selected regions-of-interest
(\emph{RoI}s). Thanks to the \emph{ad-hoc} customization of the \emph{SbD}
functional blocks, which includes the exploitation of a digital twin
(\emph{DT}) for the accurate yet fast assessment of the wireless coverage
condition, effective solutions are yielded. Numerical results, dealing
with real-world test-beds, are shown to assess the capabilities, the
potentialities, and the current limitations of the proposed \emph{EMS}s
planning strategy.

\vfill
\end{abstract}
\noindent \textbf{Key words}: Smart \emph{EM} Environment (\emph{SEME}),
\emph{EM} Skins (\emph{EMS}s), System-by-Design (\emph{SbD}), Genetic
Algorithms (\emph{GA}s), Global Optimization, Wireless Network Planning.

\newpage
\section{Introduction}

\noindent The smart electromagnetic environment (\emph{SEME}) is without
any doubt a promising and revolutionizing concept for the design of
future wireless communications systems \cite{Di Renzo 2019}-\cite{Oliveri 2021}.
It is based on the idea that the environment should be no more regarded
as an uncontrollable impairment to the overall quality-of-service
(\emph{QoS}). Conversely, it should be exploited as a powerful {}``tool''
to enable unprecedented manipulations of the complex electromagnetic
(\emph{EM}) phenomena for enhancing the overall coverage, the data
throughput, and the \emph{QoS} \cite{Massa 2021}\cite{Oliveri 2021}.
As a matter of fact, fitting ever-growing needs for ubiquitous connectivity
and low latency/resiliency of forthcoming communication standards,
also beyond the fifth-generation (\emph{5G}) \cite{Huang 2020}-\cite{Hong 2021},
will be possible only if the propagation scenario will play a fundamental
role in counteracting the distortions, the delays, the losses, and
the fading of the \emph{EM} waves radiated by the base-stations (\emph{BTS}s).

\noindent A first step towards this path is to address the synthesis
of the \emph{BTS} in an unconventional way by fitting user-defined
requirements on the \emph{QoS}, while bypassing the optimization of
standard free-space line-of-sight (\emph{LOS}) key performance indicators
(\emph{KPI}s) (e.g., gain, sidelobe level, and half-power beamwidth).
Indeed, these latter do not take into account the presence of the
environment as a stakeholder of the overall system performance \cite{Massa 2021}\cite{Oliveri.2019}.
Within this framework, the approach in \cite{Massa 2021} optimizes
the \emph{BTS} excitations by opportunistically exploiting the \emph{EM}
interactions with the surrounding obstacles to fulfil user-defined
radiation masks.

\noindent Otherwise, many studies have been recently carried out on
the possibility to improve the performance of a wireless communication
system by using reconfigurable intelligent surfaces (\emph{RIS}s)
\cite{Basar 2019}-\cite{Huang 2019}. Such a technology consists
of engineered tunable reflecting/refracting metasurfaces \cite{Pitilakis 2021}-\cite{Fang 2021}
that adaptively generate anomalous reflection/transmission of the
impinging \emph{EM} waves coming from the \emph{BTS}s. Therefore,
\emph{RIS}s are exploited to redirect the scattered \emph{EM} wave
towards arbitrary directions, not compliant with the classical Snell's
laws, where the received power would be otherwise weak/insufficient
to support a desired throughput and \emph{QoS}.

\noindent The development of effective \emph{RIS}-based solutions
has \emph{}benefited from the many similarities with the well-established
theory of both reflectarrays (\emph{RA}s) \cite{Yang 2017}\cite{Nayeri 2015}
and transmittarrays (\emph{TA}s) \cite{Tang 2021}. As a matter of
fact, the design of advanced metasurfaces with tunable magnitude/phase
modulation has been performed by properly extending the \emph{RA}s/\emph{TA}s
synthesis concepts in order to take into account the presence of finite-size
arrangements of sub-wavelengths metallic elements mounted on wall
surfaces \cite{Diaz-Rubio 2021}. However, some unsolved challenges
need to be still faced to make \emph{RIS}s an attractive technology
for large-scale urban deployments \cite{Trichopoulos 2021}\cite{Bjorson 2020}.
Indeed, new technological advancements are expected to enable the
installation of \emph{RIS}s in wide regions as well as conformal to
irregular surfaces with cost-efficient manufacturing, installation,
and maintenance, while consuming a low power. Moreover, new switching
topologies and materials (e.g., graphene and liquid crystals) are
under investigation to improve the sub-optimal performance of \emph{PIN}
diodes and varactors, which are currently employed to implement the
\emph{RIS} reconfigurability in sub-6GHz and millimeter wave/terahertz
systems \cite{Trichopoulos 2021}\cite{Fang 2021}\cite{Tang 2021}. 

\noindent Static passive \emph{EM} skins (\emph{EMS}s) \cite{Oliveri 2021}
are a promising simpler, lighter, and cheaper alternative to \emph{RIS}s
for increasing the wireless coverage and/or reducing the occurrence
of {}``blind spots'' in urban scenarios. \emph{EMS}s leverage on
the capabilities of passive modulated metasurfaces to control the
\emph{EM} interactions through a proper synthesis of their micro-scale
physical structure \cite{Oliveri 2021}. The absence of diodes, varactors,
phase shifters, amplifiers, and other components makes them particularly
attractive for a low-cost deployment/maintenance in large-scale environments.

\noindent However, while the facades of the buildings are strategic
(e.g., no other costs for realizing customized supporting infrastructures)
for the installation of \emph{EMS}s, a suitable selection of the minimum
number of buildings where the \emph{EMS}s should be installed is mandatory
to yield reliable as well as feasible solutions for recovering/yielding
the desired \emph{QoS} within specific regions-of-interest (\emph{RoI}s).

\noindent Within this context, this paper addresses, for the first
time to the best of the authors' knowledge, the planning of \emph{EMS}s
in real-world urban scenarios. The proposed strategy is not customized
to a specific technological implementation of the \emph{EMS}s and
it gives the wireless operator a full control of which {}``candidate''
facades/buildings can be used to mount the \emph{EMS}s. More specifically,
the problem at hand is formulated as a global optimization one, which
is efficiently solved within the System-by-Design (\emph{SbD}) framework
\cite{Massa 2021b}\cite{Salucci 2021} to yield an optimal (i.e.,
max-coverage-improvement and lowest-cost) \emph{EMS}s configuration
that provides the desired level of received power within the \emph{RoI}s.
Towards this end, a proper selection, customization, and interconnection
of the functional blocks of the \emph{SbD} scheme is carried out starting
from the definition of a suitable binary representation of the solution
space, which is then effectively explored with a customized implementation
of the binary genetic algorithm (\emph{BGA}) \cite{Rocca 2009}-\cite{Goldberg 1989}.
Moreover, a fast surrogate of the accurate, but time-consuming, ray-tracing
(\emph{RT})-based \emph{EM} coverage simulator is built according
to the learning-by-examples (\emph{LBE}) paradigm \cite{Massa 2018b}. 

\noindent The paper is organized as follows. Section \ref{sec:Mathematical-Formulation}
describes the mathematical formulation of the problem at hand, while
the \emph{SbD}-based planning strategy is detailed in Sect. \ref{sec:SbD-Solution-Approach}.
Numerical results are then shown (Sect. \ref{sec:Numerical-Validation})
to assess the effectiveness and the potentialities as well as the
current limitations of the proposed approach for the deployment of
\emph{EMS}s in real-world urban scenarios. Finally, some conclusions
and final remarks are drawn (Sect. \ref{sec:Conclusions}).

\section{\noindent Mathematical Formulation \label{sec:Mathematical-Formulation}}

\noindent Let us consider a large-scale urban propagation scenario
$\Xi$ served by a \emph{BTS} antenna located at the position $\mathbf{r}_{\Psi}$
{[}$\mathbf{r}_{\Psi}=\left(x_{\Psi},y_{\Psi},z_{\Psi}\right)${]}
(Fig. 1) and working at the operating frequency $f$. Due to obstructions
(caused by buildings/vegetation and other shadowing obstacles), reflections
(due to reflective surfaces), refractions (owing to the presence of
media characterized by different propagation velocities), and diffractions
(generated by edges), the \emph{EM} waves radiated by the \emph{BTS}
towards the mobile terminals propagate in non-line-of-sight (\emph{NLOS})
conditions and multi-path phenomena arise. As a consequence, there
is a set of $S$ ($S\geq1$) \emph{RoI}s, $\underline{\Omega}$ $=$\{$\Omega^{\left(s\right)}$;
$s=1,...,S$\}, within the urban scenario $\Xi$ (Fig. 1) where the
received power \cite{Liao 1977}\cite{Balanis 2016} turns out to
be lower than the minimum coverage threshold, $\mathcal{P}_{th}$,
which guarantees a target throughput \emph{}and a suitable \emph{QoS}
to the end-users\begin{equation}
\mathcal{P}_{0}\left(\mathbf{r}\right)<\mathcal{P}_{th}\,\,\,\,\,\mathbf{r}\in\Omega^{\left(s\right)}\,(s=1,...,S).\label{eq:P-lower-than-threshold}\end{equation}
In order to restore the wireless coverage condition ($\mathcal{P}\left(\mathbf{r}\right)\ge\mathcal{P}_{th}$)
within the area $\Xi$ served by the \emph{BTS}, a set of \emph{EMS}s
is deployed to reflect an adequate level of power towards each $s$-th
($s=1,...,S$) \emph{RoI} $\Omega^{\left(s\right)}$. It is worth
pointing out that the introduction of such field manipulation devices
to implement a \emph{SEME} cannot be arbitrary since \emph{EMS}s can
be mounted only on the facades of the buildings by also taking into
account the architectural constraints. Moreover, the number of \emph{EMS}s
must be kept as low as possible to reduce the overall cost as well
as to minimize the environmental impact. 

\noindent Owing to the {}``feasibility'' constraint, a set of $W^{\left(s\right)}$
{}``candidate'' building walls, $\underline{\tau}^{\left(s\right)}$
$=$ \{$\tau_{w}^{\left(s\right)}$; $w=1,...,W^{\left(s\right)}$\},
in the neighborhood of each $s$-th ($s=1,...,S$) \emph{RoI} $\Omega^{\left(s\right)}$,
$\Pi^{\left(s\right)}$ (Fig. 2), is selected for the installation
of \emph{EMS}s by the network operator. Thus, there are $K$ ($K=\sum_{s=1}^{S}W^{\left(s\right)}$)
admissible locations for deploying the \emph{EMS}s in the urban scenario
at hand. Accordingly, the \emph{EMS}s planning problem can be stated
as follows

\begin{quotation}
\noindent \textbf{Optimal} \textbf{\emph{EMSs}} \textbf{Planning Problem}
\textbf{(}\textbf{\emph{OPP}}\textbf{)} - Given $K$ admissible sites,
determine the locations and the layouts of the minimum number $Q$
($Q\ll K$) of \emph{EMS}s so that the power $\mathcal{P}\left(\mathbf{r}\right)$
received within the $s$-th ($s=1,...,S$) \emph{RoI} ($\mathbf{r}\in\Omega^{\left(s\right)}$)
fulfils the coverage/\emph{QoS} condition $\mathcal{P}\left(\mathbf{r}\right)\geq\mathcal{P}_{th}$.
\end{quotation}
\noindent To solve such an \emph{OPP}, an innovative instance of the
\emph{SbD} strategy is applied (Sect. \ref{sec:SbD-Solution-Approach}).

\section{\emph{SbD} Solution Approach \label{sec:SbD-Solution-Approach}}

Within the \emph{SbD} framework \cite{Massa 2021}, the \emph{EMSs}
planning is carried out by implementing the following \emph{SbD} blocks
(Fig. 3):

\begin{enumerate}
\item \emph{EMSs Design} (\emph{EMSD} - Sect. \ref{sub:SRSD-Block}) - The
purpose of this block is the synthesis of the complete set $\underline{\Gamma}$
of $K$ {}``admissible'' \emph{EMSs}, $\underline{\Gamma}$ $=$
\{$\Gamma_{w}^{\left(s\right)}$; $w=1,...,W^{\left(s\right)}$; $s=1,...,S$\},
starting from the knowledge of the locations of the \emph{BTS}, $\mathbf{r}_{\Psi}$,
of the selected \emph{RoI}s, \{$\mathbf{r}_{\Omega}^{\left(s\right)}$;
$s=1,...,S$\}, and of the \emph{EMSs} barycenters, \{$\mathbf{r}_{w}^{\left(s\right)}$;
$w=1,...,W^{\left(s\right)}$; $s=1,...,S$\}, on the {}``candidate''
building walls ($\mathbf{r}_{w}^{\left(s\right)}\in\tau_{w}^{\left(s\right)}$);
\item \emph{Problem Formulation} (\emph{PF} - Sect. \ref{sub:PF-Block})
- This block implements two different tasks. On the one hand, it defines
the set of $U$ degrees-of-freedom (\emph{DoF}s) to yield the most
suitable encoding of the \emph{OPP} unknowns, $\underline{\chi}$
$=$ \{$\chi_{u}$; $u=1,...,U$\}. On the other hand, it mathematically
formulates the \emph{OPP} as a global optimization problem by defining
a fitness function, $\varphi\left\{ \underline{\chi}\right\} $, that
measures the mismatch between the \emph{OPP} objectives and the \emph{EMS}s
configuration coded by the \emph{DoF} vector $\underline{\chi}$;
\item \emph{Fitness Function Evaluation} (\emph{FFE} - Sect. \ref{sub:FFE-Block})
- This block is aimed at efficiently evaluating the fitness associated
to each trial solution, $\underline{\chi}$, of the \emph{OPP}. To
reduce the heavy computational load of a full-wave \emph{EM} prediction
of the wireless coverage within the $S$ \emph{RoI}s for any trial
deployment of the \emph{EMSs}, a fast yet reliable digital twin (\emph{DT})
is exploited to assess the coverage condition in quasi real-time;
\item \emph{Solution Space Exploration} (\emph{SSE} - \ref{sub:SSE-Block})
- This block performs an effective sampling of the $U$-dimensional
solution space to find the global optimum solution, $\underline{\chi}^{\left(opt\right)}$,
that fulfils the project requirements by maximizing the fitness function
$\varphi\left\{ \underline{\chi}\right\} $ (i.e., $\underline{\chi}^{\left(opt\right)}=\arg\left[\max_{\underline{\chi}}\varphi\left\{ \underline{\chi}\right\} \right]$).
On the one hand, the implementation of the \emph{SSE} block is based
on the identification of the most effective optimization {}``engine''
to deal with the \emph{OPP-DoF}s defined by the \emph{PF} block. On
the other hand, it leverages on the fast predictions of the received
power level generated by the \emph{FFE} block to determine $\underline{\chi}^{\left(opt\right)}$
with a non-negligible time saving with respect to a standard integration
of a full-wave \emph{EM} solver within an optimization tool.
\end{enumerate}
In the following, a detailed description of each \emph{SbD} functional
block is provided.

\subsection{\emph{EMSD} Block \label{sub:SRSD-Block}}

Let us consider the design of the $\left(w,s\right)$-th ($w=1,...,W^{\left(s\right)}$,
$s=1,...,S$) \emph{EMS}, $\Gamma_{w}^{\left(s\right)}$, to be mounted
on the building facade $\tau_{w}^{\left(s\right)}$ at the position
$\mathbf{r}_{w}^{\left(s\right)}$ {[}$\mathbf{r}_{w}^{\left(s\right)}=\left(x_{w}^{\left(s\right)},y_{w}^{\left(s\right)},z_{w}^{\left(s\right)}\right)$
- Fig. 4{]} for enhancing the strength of the signal received at the
$s$-th \emph{RoI}, $\Omega^{\left(s\right)}$, centered at $\mathbf{r}_{\Omega}^{\left(s\right)}$
{[}$\mathbf{r}_{\Omega}^{\left(s\right)}=\left(x_{\Omega}^{\left(s\right)},y_{\Omega}^{\left(s\right)},z_{\Omega}^{\left(s\right)}\right)$
- Fig. 4{]}. Without loss of generality and by assuming a local coordinate
system $\left(x',y',z'\right)$ with origin in $\mathbf{r}_{w}^{\left(s\right)}$
(Fig. 4), the \emph{EM} wave radiated by the \emph{BTS} towards \emph{}$\Gamma_{w}^{\left(s\right)}$
is modeled as a monochromatic plane wave at frequency $f$ with incident
wave vector equal to\begin{equation}
\mathbf{k}_{\Psi}^{\left(w,s\right)}=-\frac{2\pi}{\lambda}\left[\sin\left(\theta_{\Psi}^{\left(w,s\right)}\right)\cos\left(\varphi_{\Psi}^{\left(w,s\right)}\right)+\sin\left(\theta_{\Psi}^{\left(w,s\right)}\right)\sin\left(\varphi_{\Psi}^{\left(w,s\right)}\right)+\cos\left(\theta_{\Psi}^{\left(w,s\right)}\right)\right]\label{eq:incident-wave-vector}\end{equation}
where $\theta_{\Psi}^{\left(w,s\right)}$and $\varphi_{\Psi}^{\left(w,s\right)}$
are the elevation and the azimuth coordinates of the angle of incidence,
respectively, whose expressions are

\begin{equation}
\left\{ \begin{array}{l}
\theta_{\Psi}^{\left(w,s\right)}=\mathcal{F}_{\theta}\left(\mathbf{r}_{\Psi}^{'}\right)\\
\varphi_{\Psi}^{\left(w,s\right)}=\mathcal{F}_{\varphi}\left(\mathbf{r}_{\Psi}^{'}\right),\end{array}\right.\label{eq:theta-phi-incident}\end{equation}
$\mathcal{F}_{\theta}\left(\mathbf{r}\right)$ and $\mathcal{F}_{\varphi}\left(\mathbf{r}\right)$
being the Cartesian-to-Polar operators equal to\begin{equation}
\left\{ \begin{array}{l}
\mathcal{F}_{\theta}\left(\mathbf{r}\right)=\arccos\left(\frac{z}{\sqrt{x^{2}+y^{2}+z^{2}}}\right)\\
\mathcal{F}_{\varphi}\left(\mathbf{r}\right)=\arctan\left(\frac{y}{x}\right).\end{array}\right.\label{eq:}\end{equation}
Moreover, $\mathbf{r}_{\Psi}^{'}$ {[}$\mathbf{r}_{\Psi}^{'}=\left(x_{\Psi}^{'},y_{\Psi}^{'},z_{\Psi}^{'}\right)${]}
denotes the position of the \emph{BTS} as seen from the \emph{EMS}
$\Gamma_{w}^{\left(s\right)}$ (Fig. 4) and its Cartesian coordinates
are\begin{equation}
\left\{ \begin{array}{l}
x_{\Psi}^{'}=\mathcal{G}_{x}\left(\mathbf{r}_{\Psi},\mathbf{r}_{w}^{\left(s\right)}\right)\\
y_{\Psi}^{'}=\mathcal{G}_{y}\left(\mathbf{r}_{\Psi},\mathbf{r}_{w}^{\left(s\right)}\right)\\
z_{\Psi}^{'}=\mathcal{G}_{z}\left(\mathbf{r}_{\Psi},\mathbf{r}_{w}^{\left(s\right)}\right)\end{array}\right.,\label{eq:}\end{equation}
$\mathcal{G}_{c}$ being the $c$-th ($c=x$, $y$, $z$) transformation
operator defined as\begin{equation}
\left\{ \begin{array}{l}
\mathcal{G}_{x}\left(\mathbf{r},\mathbf{r}_{w}^{\left(s\right)}\right)=\left(x-x_{w}^{\left(s\right)}\right)\cos\left(\alpha_{w}^{\left(s\right)}\right)+\left(y-y_{w}^{\left(s\right)}\right)\sin\left(\alpha_{w}^{\left(s\right)}\right)\\
\mathcal{G}_{y}\left(\mathbf{r},\mathbf{r}_{w}^{\left(s\right)}\right)=\left(z-z_{w}^{\left(s\right)}\right)\\
\mathcal{G}_{z}\left(\mathbf{r},\mathbf{r}_{w}^{\left(s\right)}\right)=\left(x-x_{w}^{\left(s\right)}\right)\sin\left(\alpha_{w}^{\left(s\right)}\right)-\left(y-y_{w}^{\left(s\right)}\right)\cos\left(\alpha_{w}^{\left(s\right)}\right)\end{array}\right.\label{eq:}\end{equation}
where $\alpha_{w}^{\left(s\right)}$ is the orientation angle of the
building wall $\tau_{w}^{\left(s\right)}$ with respect to the global
$x$-axis (Fig. 4).

\noindent Analogously, the angular direction of the \emph{EM} wave
reflected from the \emph{EMS} $\Gamma_{w}^{\left(s\right)}$ towards
the barycenter of $\Omega^{\left(s\right)}$, $\left(\theta_{\Omega}^{\left(w,s\right)},\,\varphi_{\Omega}^{\left(w,s\right)}\right)$,
turns out to be\begin{equation}
\left\{ \begin{array}{l}
\theta_{\Omega}^{\left(w,s\right)}=\mathcal{F}_{\theta}\left(\mathbf{r}_{\Omega}^{\left(s\right)'}\right)\\
\varphi_{\Omega}^{\left(w,s\right)}=\mathcal{F}_{\varphi}\left(\mathbf{r}_{\Omega}^{\left(s\right)'}\right)\end{array}\right.\label{eq:theta-phi-reflected}\end{equation}
$\mathbf{r}_{\Omega}^{\left(s\right)'}$ $=$ \{$\mathcal{G}_{x}\left(\mathbf{r}_{\Omega}^{\left(s\right)},\mathbf{r}_{w}^{\left(s\right)}\right)$,
$\mathcal{G}_{y}\left(\mathbf{r}_{\Omega}^{\left(s\right)},\mathbf{r}_{w}^{\left(s\right)}\right)$,
$\mathcal{G}_{z}\left(\mathbf{r}_{\Omega}^{\left(s\right)},\mathbf{r}_{w}^{\left(s\right)}\right)$\}
being the position of $\Omega^{\left(s\right)}$ in the \emph{EMS}
local system of coordinates (Fig 4).

\noindent Once the directions of incidence, $\left(\theta_{\Psi}^{\left(w,s\right)},\,\varphi_{\Psi}^{\left(w,s\right)}\right)$,
and of reflection, $\left(\theta_{\Omega}^{\left(w,s\right)},\,\varphi_{\Omega}^{\left(w,s\right)}\right)$,
of the impinging wave from the \emph{BTS} are defined, the \emph{EMS}
$\Gamma_{w}^{\left(s\right)}$ ($w=1,...,W^{\left(s\right)}$, $s=1,...,S$)
is designed according to the two-step synthesis procedure described
in \cite{Oliveri 2021}. Shortly, the {}``reference'' electric/magnetic
current distributions on the surface of $\Gamma_{w}^{\left(s\right)}$,
which radiate in far-field a pencil beam pointed towards the \emph{RoI},
$\left(\theta_{\Omega}^{\left(w,s\right)},\,\varphi_{\Omega}^{\left(w,s\right)}\right)$,
are computed. Then, the pattern of the metallizations \cite{Oliveri 2021}
that compose the $\left(w,s\right)$-th ($w=1,...,W^{\left(s\right)}$;
$s=1,...,S$) \emph{EMS}, $\Gamma_{w}^{\left(s\right)}$, which is
realized in low-cost \emph{PCB} technology, is derived by optimizing
the $O$ geometric descriptors of the $L$ unit cells of the \emph{EMS},
$\underline{d}_{w}^{\left(s\right)}=\left\{ d_{w,ol}^{\left(s\right)};\, o=1,...,O;\, l=1,...,L\right\} $,
so that the electric/magnetic current distributions induced on the
surface $\Gamma_{w}^{\left(s\right)}$ by the impinging wave from
the \emph{BTS} match the {}``reference'' ones.

\subsection{\emph{PF} Block \label{sub:PF-Block}}

According to the \emph{SbD} paradigm \cite{Massa 2021}, the \emph{PF}
block is aimed at formulating the \emph{OPP} into a proper mathematical
framework to enable its reliable and cost-effective solution. Towards
this end and owing to the problem at hand, a \emph{binary encoding}
is adopted to define the set of $U$ \emph{DoF}s. More in detail,
a deployment of \emph{EMS}s in the urban scenario is coded with the
$K$-size (i.e., $U=K$) binary chromosome $\underline{\chi}$ $=$
\{$\chi_{w}^{\left(s\right)}$; $w=1,...,W^{\left(s\right)}$; $s=1,...,S$\}
whose $\left(w,s\right)$-th ($w=1,...,W^{\left(s\right)}$, $s=1,...,S$)
entry is equal to $0$ / $1$ ($\chi_{w}^{\left(s\right)}=1$/$\chi_{w}^{\left(s\right)}=0$)
when the $\left(w,s\right)$-th \emph{EMS}, $\Gamma_{w}^{\left(s\right)}$,
designed in the \emph{EMSD} block (Sect. \ref{sub:SRSD-Block}), \emph{}is
installed/not-installed on the corresponding building facade $\tau_{w}^{\left(s\right)}$.%
\footnote{\noindent For the sake of notation conciseness, the following notation
will be also used in the description of the optimization strategy:
$\underline{\chi}$ $=$ \{$\chi_{k}$; $k=1,...,K$\}, $\chi_{k}$
$=$ \{$0$, $1$\} when the $k$-th \emph{EMS} , $\Gamma_{k}$, is
installed/not-installed on the $k$-th building facade, $\tau_{k}$.%
}

\noindent The arising binary-coded planning problem is then formulated
by the \emph{PF} block as a global optimization task by properly defining
the fitness function $\varphi$. Such a performance index mathematically
models the underlying physics by quantifying the fulfilment of the
\emph{QoS} requirement by a trial solution $\underline{\chi}$ (i.e.,
a trial \emph{EMS}s deployment). More specifically, the fitness of
a guess $\underline{\chi}$, $\varphi\left\{ \underline{\chi}\right\} $,
is given by the inverse of a two-term single-objective cost function
$\Phi\left\{ \underline{\chi}\right\} $ ($\varphi\left\{ \underline{\chi}\right\} \triangleq\frac{1}{\Phi\left\{ \underline{\chi}\right\} }$)\begin{equation}
\Phi\left\{ \underline{\chi}\right\} =\Phi_{cov}\left\{ \underline{\chi}\right\} +\Phi_{cost}\left\{ \underline{\chi}\right\} .\label{eq:total-fitness}\end{equation}
The \emph{coverage term} $\Phi_{cov}\left\{ \underline{\chi}\right\} $
measures the mismatch between the power received within the $S$ \emph{RoI}s
and the threshold value $\mathcal{P}_{th}$ through the following
expression\begin{equation}
\Phi_{cov}\left\{ \underline{\chi}\right\} \triangleq\frac{1}{M}\sum_{s=1}^{S}\sum_{m=1}^{M^{\left(s\right)}}\frac{\left|\mathcal{P}_{th}-\mathcal{P}\left(\left.\mathbf{r}_{m}^{\left(s\right)}\right|\underline{\chi}\right)\right|}{\left|\mathcal{P}_{th}\right|}\mathcal{H}\left\{ \mathcal{P}_{th}-\mathcal{P}\left(\left.\mathbf{r}_{m}^{\left(s\right)}\right|\underline{\chi}\right)\right\} \label{eq:coverage-term}\end{equation}
where $M^{\left(s\right)}$ is the number of receivers that lay in
the positions, \{$\mathbf{r}_{m}^{\left(s\right)}$; $m=1,...,M^{\left(s\right)}$\}
{[}$\mathbf{r}_{m}^{\left(s\right)}=\left(x_{m}^{\left(s\right)},\, y_{m}^{\left(s\right)},\, h\right)${]},
of the $s$-th ($s=1,...,S$) \emph{RoI} at a height $h$ above the
ground level (i.e., $z=0$ - Fig. 1), $M=\sum_{s=1}^{S}M^{\left(s\right)}$,
and $\mathcal{H}$ is the Heaviside function (i.e., $\mathcal{H}\left\{ a\right\} =1$
if $a>0$ and $\mathcal{H}\left\{ a\right\} =0$, otherwise). 

\noindent The second term in (\ref{eq:total-fitness}) is proportional
to the number of deployed \emph{EMS}s and it is defined as follows\begin{equation}
\Phi_{cost}\left\{ \underline{\chi}\right\} \triangleq\frac{Q}{K},\label{eq:cost-term}\end{equation}
$Q$ being equal to the $\ell_{0}$-norm of $\underline{\chi}$ ,
$Q=\left\Vert \underline{\chi}\right\Vert _{0}$.

\noindent The fittest solution of the \emph{OPP}, $\underline{\chi}^{\left(opt\right)}$,
is the global minimum of (\ref{eq:total-fitness})\begin{equation}
\underline{\chi}^{\left(opt\right)}=\arg\left[\min_{\underline{\chi}}\Phi\left\{ \underline{\chi}\right\} \right],\label{eq:}\end{equation}
that is the best trade-off between the maximum power received in the
\emph{RoI}s and the minimum number of installed \emph{EMS}s. Indeed,
the presence of a larger number of \emph{EMSs} generally would imply
a greater level of power in the \emph{RoI}s ($\Phi_{cov}\left\{ \underline{\chi}\right\} $
$\downarrow$), but at the cost of a higher implementation cost and
a heavier wireless network complexity as well as a bigger architectural/environmental
impact ($\Phi_{cost}\left\{ \underline{\chi}\right\} $ $\uparrow$).

\subsection{\emph{FFE} Block \label{sub:FFE-Block}}

\noindent The computation of the fitness of each trial solution $\underline{\chi}$,
$\varphi\left\{ \underline{\chi}\right\} $, could represent the main
bottleneck of the overall \emph{OPP} solution strategy, especially
if repeated many times as in the \emph{SSE} block for the exploration
of the $K$-dimensional solution space (Sect. \ref{sub:SSE-Block}),
unless suitable countermeasures are undertaken. As a matter of fact,
even though the evaluation of the term \emph{}(\ref{eq:cost-term})
is immediate since it only depends on the number of non-null entries
of the binary vector $\underline{\chi}$, on the contrary, the prediction
of the level of power within the $S$ \emph{RoI}s, to assess the fulfilment
of the \emph{QoS} requirements, would be computationally heavy whether
using full-wave \emph{EM} simulation tools based on \emph{RT} techniques
due to the large scale of the urban scenario at hand (Fig. 1). 

\noindent In order to minimize the computational load, the \emph{FFE}
block is responsible for the off-line generation of a \emph{DT} to
estimate the coverage term (\ref{eq:coverage-term}) (i.e., $\widetilde{\Phi}_{cov}\left\{ \underline{\chi}\right\} \approx\Phi_{cov}\left\{ \underline{\chi}\right\} $)
so to efficiently predict the cost function value, $\Phi\left\{ \underline{\chi}\right\} $,
as follows (Fig. 3)\begin{equation}
\widetilde{\Phi}\left\{ \underline{\chi}\right\} =\widetilde{\Phi}_{cov}\left\{ \underline{\chi}\right\} +\Phi_{cost}\left\{ \underline{\chi}\right\} .\label{eq:total-predicted-fitness}\end{equation}
Towards this end, a Gaussian Process (\emph{GP}) \cite{Massa 2018b}\cite{Forrester 2008}\cite{Jones 1998}
is used to build the \emph{DT} from the information embedded within
an (off-line generated) training set of $T$ input/output (\emph{I/O})
pairs\begin{equation}
\mathbb{T}=\left\{ \left(\underline{\chi}^{\left(t\right)};\,\Phi_{cov}\left\{ \underline{\chi}^{\left(t\right)}\right\} \right);\, t=1,...,T\right\} \label{eq:training set}\end{equation}
randomly-chosen among the full set of $B=2^{K}$ \emph{EMSs} configurations
($T\ll B$). More specifically, the computationally-fast guess of
(\ref{eq:coverage-term}) is given by\begin{equation}
\widetilde{\Phi}_{cov}\left\{ \underline{\chi}\right\} =\nu+\left(\underline{\rho}\right)^{'}\underline{\underline{C}}^{-1}\left(\underline{\Phi}-\underline{1}\nu\right)\label{eq:coverage-term-prediction}\end{equation}
where $\underline{\Phi}=\left\{ \Phi_{cov}\left\{ \underline{\chi}^{\left(t\right)}\right\} ;\, t=1,...,T\right\} ^{'}$
and $\nu=\left(\underline{1}^{'}\underline{\underline{C}}^{-1}\underline{1}\right)\underline{1}^{'}\underline{\underline{C}}^{-1}\underline{\Phi}$,
$.^{'}$ being the transpose operator and $\underline{1}$ is a $\left(T\times1\right)$-size
vector of unitary entries. Moreover, $\underline{\underline{C}}$
$=$ \{$\Lambda\left(\underline{\chi}^{\left(p\right)},\underline{\chi}^{\left(q\right)}\right)$;
$p,\, q=1,...,T$\} and $\underline{\rho}=\left\{ \Lambda\left(\underline{\chi},\,\underline{\chi}^{\left(t\right)}\right);\, t=1,...,T\right\} $
are the training correlation matrix and the correlation vector of
$\underline{\chi}$, respectively, the correlation between two input
samples $\underline{\chi}^{\left(a/b\right)}\in\mathbb{R}^{K}$ being
computed as\begin{equation}
\Lambda\left(\underline{\chi}^{\left(a\right)},\underline{\chi}^{\left(b\right)}\right)=\exp\left(-\sum_{k=1}^{K}\beta_{k}\left|\chi_{k}^{\left(a\right)}-\chi_{k}^{\left(b\right)}\right|^{\gamma_{k}}\right)\label{eq:}\end{equation}
where $\underline{\beta}=\left\{ \beta_{k}>0;\, k=1,...,K\right\} $
and $\underline{\gamma}=\left\{ 1\leq\gamma_{k}\leq2;\, k=1,...,K\right\} $
are \emph{GP} hyper-parameters determined during the off-line training
phase by maximizing the likelihood function of $\mathbb{T}$ \cite{Massa 2018b}\cite{Forrester 2008}\cite{Jones 1998}.

\subsection{\emph{SSE} Block \label{sub:SSE-Block}}

By following the \emph{SbD} guidelines \cite{Massa 2021} and according
to the \emph{no-free-lunch theorem} for optimization \cite{Wolpert 1997},
the implementation of the \emph{SSE} block is driven by the nature
of the fitness function and of the solution space defined by the \emph{PF}
block (Sect. \ref{sub:PF-Block}). As a matter of fact, it is profitable
to select the most suitable optimization engine that assures a proper
balance between exploration and local sampling \emph{}of the solution
space, while hill-climbing \emph{}local minima of the cost function,
to converge towards the global optimum of (\ref{eq:total-fitness}),
$\underline{\chi}^{\left(opt\right)}$. Moreover, the optimization
tool is required to properly handle (i.e., without considering time-expensive
coding/decoding operations) the binary nature of the \emph{DoF}s of
the \emph{OPP}.

\noindent Owing to these considerations, a binary genetic algorithm
(\emph{BGA})-based strategy is adopted to find $\underline{\chi}^{\left(opt\right)}$
by evolving a population of $P$ binary individuals, $\mathbb{P}_{i}$
$=$ \{$\underline{\chi}_{i}^{\left(p\right)}$; $p=1,...,P$\} ($i=1,...,I$;
$i$ being the iteration index) during $I$ iterations and according
to the concepts of natural selection and genetic pressure \cite{Rocca 2009}\cite{Goudos 2021}.
Moreover, the \emph{BGA} is here customized to take full advantage
of the \emph{SbD} framework for obtaining a considerable time saving
by avoiding the time-costly assessment of each ($i$, $p$)-th ($i=1,...,I$;
$p=1,...,P$) trial solution, $\underline{\chi}_{i}^{\left(p\right)}$,
with a full-wave \emph{EM} simulation. Towards this end, the iterative
minimization of (\ref{eq:total-fitness}) leverages on the profitable
interconnection of the \emph{SSE} block with the \emph{DT} derived
in \emph{}the \emph{FFE} block (Sect. \ref{sub:FFE-Block}). Furthermore,
the proposed implementation of the \emph{BGA} benefits from the knowledge
of the training set $\mathbb{T}$ to enhance the convergence rate
of the optimization process according to the {}``schemata theorem''
of \emph{GA}s \cite{Goldberg 1989}, which states that ''\emph{above
average schemata receive exponentially increasing trials in subsequent
generations}''. Accordingly, since the probability to yield {}``good''
schemata \cite{Goldberg 1989} by randomly-selecting $P$ binary chromosomes
from the whole set of $B$ ($B\triangleq2^{K}$) admissible binary
sequences ($P\ll B$) is generally low, the individuals of the initial
population $\mathbb{P}_{0}$ are chosen among the fittest ones of
the $T$ solutions of the training set, \{$\underline{\chi}^{\left(t\right)}$;
$t=1,...,T$\}.

\noindent The sequence of procedural steps carried out in the \emph{SEE}
block is summarized in the following.

\begin{enumerate}
\item \emph{Initialization} ($i=0$) - Sort the training solutions of $\mathbb{T}$,
\{$\underline{\chi}^{\left(t\right)}$; $t=1,...,T$\}, according
to their fitness values, $\varphi\left\{ \underline{\chi}^{\left(t\right)}\right\} $
($t=1,...,T$), and build the sorted set $\mathbb{R}_{0}$ ($\mathbb{R}_{0}=\left\{ \underline{\chi}_{0}^{\left(r\right)};\, r=1,...,T\right\} $
being $\underline{\chi}_{0}^{\left(1\right)}$ $=$ $\arg$($\min_{t=1,...,T}${[}$\Phi\left\{ \underline{\chi}^{\left(t\right)}\right\} ${]})
and $\underline{\chi}_{0}^{\left(T\right)}$ $=$ $\arg$ ($\max_{t=1,...,T}$
{[}$\Phi\left\{ \underline{\chi}^{\left(t\right)}\right\} ${]}).
Create the initial population $\mathbb{P}_{0}$ ($\mathbb{P}_{0}=\left\{ \underline{\chi}_{0}^{\left(p\right)};\, p=1,...,P\right\} $)
by randomly selecting $P$ individuals out of the $\frac{T}{2}$ elements
of $\mathbb{R}_{0}$ with fitness values above the median;
\item \emph{Optimization Loop} ($i=1,...,I$)

\begin{enumerate}
\item Generate a new population of offsprings, $\mathbb{P}_{i}$, by applying
the genetic operators to the population of the parents, $\mathbb{P}_{\left(i-1\right)}$,
as follows\begin{equation}
\mathbb{P}_{i}=\mathcal{M}\left\{ \mathcal{C}\left\{ \mathcal{S}\left\{ \mathbb{P}_{\left(i-1\right)}\right\} \right\} \right\} \label{eq:}\end{equation}
where the {}``mating pool'' $\mathbb{P}_{\left(i-1\right)}^{\mathcal{S}}$
($\mathbb{P}_{\left(i-1\right)}^{\mathcal{S}}\triangleq\mathcal{S}\left\{ \mathbb{P}_{\left(i-1\right)}\right\} $)
is derived by undergoing $\mathbb{P}_{\left(i-1\right)}$ to the roulette
wheel selection, $\mathcal{S}$ \cite{Rocca 2009}, while the single-point
cross-over, $\mathcal{C}$, is applied with probability $\delta^{\mathcal{C}}$
to generate the intermediate set of individuals, $\mathbb{P}_{\left(i-1\right)}^{\mathcal{C}}$
($\mathbb{P}_{\left(i-1\right)}^{\mathcal{C}}\triangleq\mathcal{C}\left\{ \mathbb{P}_{\left(i-1\right)}^{\mathcal{S}}\right\} $),
which is successively processed by the mutation operator, $\mathcal{M}$,
with mutation and bit-mutation probabilities equal to $\delta_{1}^{\mathcal{M}}$
and $\delta_{2}^{\mathcal{M}}$, respectively \cite{Rocca 2009};
\item Exploit the \emph{FFE} block (Sect. \ref{sub:FFE-Block}) to predict,
according to (\ref{eq:coverage-term-prediction}) and (\ref{eq:total-predicted-fitness}),
the fitness function of each individual of $\mathbb{P}_{i}$, \{$\varphi\left\{ \underline{\chi}_{i}^{\left(p\right)}\right\} $;
$p=1,...,P$\}. Select through elitism \cite{Rocca 2009} the fittest
individual generated until the current $i$-th iteration, $\underline{\chi}^{\left(i\right)}=\arg\left[\min_{p=1,...,P}\min_{j=1,...,i}\widetilde{\Phi}\left\{ \underline{\chi}_{j}^{\left(p\right)}\right\} \right]$;
\item Check the convergence condition (i.e., $i=I$ or $\Phi\left\{ \underline{\chi}^{\left(i\right)}\right\} \le\xi$,
$\xi$ being a user-defined threshold). If the convergence has been
reached, then goto to the {}``\emph{Output Phase}'' (3.), otherwise
update the iteration index ($i\to i+1$) and repeat the {}``\emph{Optimization
Loop}'' (2.);
\end{enumerate}
\item \emph{Output Phase} - Output the \emph{OPP} solution given by\begin{equation}
\underline{\chi}^{\left(opt\right)}=\arg\left[\min_{p=1,...,P}\widetilde{\Phi}\left\{ \underline{\chi}_{i}^{\left(p\right)}\right\} \right].\label{eq:}\end{equation}

\end{enumerate}

\section{\noindent Numerical Validation \label{sec:Numerical-Validation}}

\noindent The goal of this section is to assess the capabilities and
the potentialities of the approach to enhance the wireless coverage
and the overall \emph{QoS} in large urban scenarios thanks to the
optimal deployment of \emph{EMS}s on the building facades.

\noindent Such an assessment has been carried out in a real-world
scenario by considering the Gardolo district of the city of Trento
(Italy) as the benchmark test-bed (Fig. 5). As it can be inferred
from the satellite image {[}Fig. 5(\emph{a}){]} taken from the Google
Maps website \cite{Google Maps}, the selected area $\Xi$ is a square
region $1$ {[}Km{]}-sided that comprises several streets, a railway,
and a quite dense distribution of civil, commercial, and industrial
buildings.

\noindent The actual position of the \emph{BTS}, which serves the
users in $\Xi$, is $\mathbf{r}_{\Psi}=\left(3.95\times10^{2},\,5.79\times10^{2},\,30\right)$
{[}m{]} and it has been extracted from the official \emph{BTS} cartography
publicly accessible on the website of the city of Trento \cite{Cartografia-BTS-Trento}
{[}Fig. 5(\emph{b}){]}. The site consists of $V=3$ sectors having
an angular extension equal to $\Delta\phi_{\Psi}=120$ {[}deg{]} in
azimuth, pointed towards the directions $\phi_{\Psi}^{\left(v\right)}=\left(v-1\right)\times120$
{[}deg{]} ($v=1,...,V$), with a mechanical down-tilt of $\Delta\theta_{\Psi}=2$
{[}deg{]} in elevation \cite{Yang 2019}\cite{3G Americas 2010}\cite{4G Americas 2013}.
Each $v$-th ($v=1,...,V$) sector has been assumed to be illuminated
by a panel of the \emph{BTS}, which is composed by a planar array
(Fig. 6) of $N=\left(13\times2\right)$ $\frac{\lambda}{2}$-spaced
slot-coupled dual-polarized (slant-45) square patch radiators working
at $f=3.5$ {[}GHz{]} with a ground plane of size $\left(L_{y}\times L_{z}\right)=\left(1.75\times7\right)$
{[}$\lambda${]} \cite{3G Americas 2010}\cite{4G Americas 2013}
(see the inset in Fig. 6). Such an antenna array has been accurately
modeled in the Ansys \emph{HFSS} simulation suite \cite{HFSS 2021}
to take into account all mutual coupling effects. The co-polar gain
pattern for the $+45$-polarization operation, $G_{+45}\left(\theta,\,\phi\right)$
\cite{Quan 2014}, is shown in Fig. 6, the maximum gain being $G_{+45}^{\max}=\max_{\left(\theta,\,\phi\right)}G_{+45}\left(\theta,\,\phi\right)=16.3$
{[}dBi{]} %
\footnote{\noindent For symmetry reasons, the co-polar pattern for the $-45$
operation, $G_{-45}\left(\theta,\,\phi\right)$, coincides with $G_{+45}\left(\theta,\,\phi\right)$.
Accordingly, all the results reported in the following refers to the
$+45$ operation of the \emph{BTS}.%
}. 

\noindent As for the wireless coverage, the power distribution in
$\Xi$ (Fig. 5) has been computed with the \emph{RT}-based Altair
WinProp simulator \cite{WINPROP}. Towards this end, the exact position,
orientation, and dimensions of each building has been first extracted
from the OpenStreetMap (\emph{OSM}) Geographic Information System
(\emph{GIS}) database \cite{OpenStreetMap} {[}Fig. 5(\emph{b}){]},
then it has been imported into WinProp to generate the \emph{EM} simulation
scenario in Fig. 5(\emph{c}) where the buildings have been assumed
to be made of concrete with relative permittivity $\varepsilon_{r}=4$
and conductivity $\sigma=10^{-2}$ {[}S/m{]} \cite{ITU-R P.2040-1}\cite{Daniels 2004}.
Figure 7(\emph{a}) shows the power distribution (i.e., $\mathcal{P}_{0}\left(\mathbf{r}\right)$,
$\mathbf{r}\in\Xi$) computed at a standard user height of $h=1.5$
{[}m{]} \cite{3GPP 36.873} on a grid of points uniformly-spaced ($\Delta_{x}=\Delta_{y}=5$
{[}m{]} being the spacing along $x$- and $y$-axes), when feeding
the \emph{BTS} panels with an input power of $\mathcal{P}_{in}^{\left(v\right)}=20$
{[}W{]}, $v=1,...,V$ \cite{3G Americas 2010}\cite{4G Americas 2013}.
In addition to the standard attenuation due to the path loss, the
distribution of the power radiated by the \emph{BTS} turns out to
be strongly affected by the presence of the buildings, which cause
shadowing effects as well as wave-guiding phenomena (e.g., canyoning
along the main streets).

\noindent The \emph{RoI}s in $\Xi$ have been identified by thresholding
($\mathcal{P}_{th}=-65$ {[}dBm{]}%
\footnote{\noindent The value $\mathcal{P}_{th}=-65$ {[}dBm{]} is often assumed
as the reference signal received power (\emph{RSRP}) to support advanced
high-throughput wireless services (e.g., high-definition video streaming
\cite{GSMA 2020}).%
} being the value of the coverage threshold) the distribution of $\mathcal{P}_{0}$.
The binary map in Fig. 7(\emph{b}) shows $S=2$ \emph{RoI}s. The first
one has an area of $A\left(\Omega^{\left(1\right)}\right)=1225$ {[}$m^{2}${]}
($M^{\left(1\right)}=49$) and it is centered at $\mathbf{r}_{\Omega}^{\left(1\right)}=\left(411,\,698,\,1.5\right)$
{[}m{]} {[}Fig. 7(\emph{b}) and Figs. 8(\emph{a}){]}, while the other
is located at $\mathbf{r}_{\Omega}^{\left(2\right)}=\left(123,\,280,\,1.5\right)$
{[}m{]} and extends on a support of $A\left(\Omega^{\left(2\right)}\right)=1075$
{[}$m^{2}${]} ($M^{\left(2\right)}=43$) {[}Fig. 7(\emph{b}) and
Figs. 8(\emph{b}){]}.

\noindent According to the proposed planning method (Sect. \ref{sec:Mathematical-Formulation}),
a set of $W^{\left(s\right)}$ ($W^{\left(1\right)}=8$ {[}Fig. 8(\emph{c}){]};
$W^{\left(2\right)}=12$ {[}Fig. 8(\emph{d}){]}) building facades
has been chosen into each neighboring square region, $\Pi^{\left(s\right)}$,
of a \emph{RoI} $\Omega^{\left(s\right)}$($s=1,...,S$). Then, the
complete set of $K=20$ \emph{EMSs} has been off-line synthesized
in the \emph{EMSD} block (Fig. 3 - Sect. \ref{sub:SRSD-Block}) by
assuming an installation quota of $z_{w}^{\left(s\right)}=\left(H_{w}^{\left(s\right)}-2\right)$
{[}m{]} from the ground level (Fig. 4), $H_{w}^{\left(s\right)}$
being the height of the $\left(w,\, s\right)$-th ($w=1,...,W^{\left(s\right)}$,
$s=1,...,S$) wall, $\tau_{w}^{\left(s\right)}$, in the \emph{OSM}
database \cite{OpenStreetMap}. More specifically, each \emph{EMS}
has been implemented \cite{Oliveri 2021} with a properly-tailored
pattern of $L=\left(50\times50\right)$ square-shaped unit cells (i.e.,
$O=1$) printed on a support of $A\left(\Gamma_{w}^{\left(s\right)}\right)=\left(2.14\times2.14\right)$
{[}$m^{2}${]} ($w=1,...,W^{\left(s\right)}$, $s=1,...,S$) over
a Rogers RT/duroid 5870 substrate ($\varepsilon_{r}=2.33$, $\tan\delta=1.2\times10^{-3}$)
of thickness $3.7\times10^{-2}$ {[}$\lambda${]}.

\noindent The \emph{OPP} at hand is then solved by sampling the $B$-size
($B\approx1.05\times10^{6}$ - Tab. I) solution space of the admissible
\emph{EM}Ss deployments with the \emph{SbD}-based approach (Sect.
\ref{sec:SbD-Solution-Approach}). Towards this end, the \emph{SbD}
control parameters have been set according to the state-of-the-art
guidelines \cite{Massa 2021}\cite{Rocca 2009}: $T=4\times10^{3}$,
$P=40$, $I=10^{3}$, $\delta^{\mathcal{C}}=8\times10^{-1}$, $\delta_{1}^{\mathcal{M}}=10^{-1}$,
and $\delta_{2}^{\mathcal{M}}=10^{-2}$. The final ($i=I$) outcome
of the planning process is summarized in Fig. 9 where the thresholded
maps of the distribution of power received in $\Pi^{\left(1\right)}$
{[}Fig. 9(\emph{a}){]} and $\Pi^{\left(2\right)}$ {[}Fig. 9(\emph{b}){]}
are reported along with the positions of the selected \emph{EMSs}.
There are $Q^{\left(opt\right)}=7$ \emph{EMSs} (i.e., $\underline{\Gamma}^{\left(opt\right)}=\left\{ \Gamma_{1}^{\left(1\right)},\,\Gamma_{2}^{\left(1\right)},\,\Gamma_{7}^{\left(1\right)},\,\Gamma_{2}^{\left(2\right)},\,\Gamma_{4}^{\left(2\right)},\,\Gamma_{5}^{\left(2\right)},\,\Gamma_{6}^{\left(2\right)}\right\} $
- Tab. I), $\left.Q^{\left(opt\right)}\right|_{s=1}=3$ {[}i.e., $\left.\underline{\Gamma}^{\left(opt\right)}\right|_{s=1}=\left\{ \Gamma_{1}^{\left(1\right)},\,\Gamma_{2}^{\left(1\right)},\,\Gamma_{7}^{\left(1\right)}\right\} $
- Fig. 9(\emph{a}){]} and $\left.Q^{\left(opt\right)}\right|_{s=2}=4$
{[}$\left.\underline{\Gamma}^{\left(opt\right)}\right|_{s=2}=\left\{ \Gamma_{2}^{\left(2\right)},\,\Gamma_{4}^{\left(2\right)},\,\Gamma_{5}^{\left(2\right)},\,\Gamma_{6}^{\left(2\right)}\right\} $
- Fig. 9(\emph{b}){]} for $\Omega^{\left(1\right)}$ and $\Omega^{\left(2\right)}$,
respectively. Thanks to the reduction of the coverage term (\ref{eq:coverage-term})
of about two orders of magnitude with respect to the {}``nominal''
scenario without \emph{EMSs} (i.e., $\frac{\Phi_{cov}\left\{ \underline{\chi}=\underline{0}\right\} }{\Phi_{cov}\left\{ \underline{\chi}^{\left(opt\right)}\right\} }=1.19\times10^{-2}$
- Tab. I), there is a remarkable enhancement of the level of the power
received in the \emph{RoI}s. Indeed, the coverage condition always
holds true in $\Omega^{\left(1\right)}$ {[}i.e., $\mathcal{P}\left(\left.\mathbf{r}\right|\underline{\chi}^{\left(opt\right)}\right)>\mathcal{P}_{th}$,
$\mathbf{r}\in\Omega^{\left(1\right)}$ - Fig. 9(\emph{a}) vs. Fig.
8(\emph{a}){]}, while only few locations of $\Omega^{\left(2\right)}$
are under the power threshold $\mathcal{P}_{th}$, $\Delta A\left(\Omega^{\left(2\right)}\right)=86$
\% {[}$\Delta A\left(\Omega^{\left(s\right)}\right)\triangleq\frac{A\left(\left.\Omega^{\left(s\right)}\right|\underline{\chi}=\underline{0}\right)-A\left(\left.\Omega^{\left(s\right)}\right|\underline{\chi}^{\left(opt\right)}\right)}{A\left(\left.\Omega^{\left(s\right)}\right|\underline{\chi}=\underline{0}\right)}$;
($s=1,...,S$){]} being the widening of the coverage surface {[}Fig.
9(\emph{b}) vs. Fig. 8(\emph{b}){]} %
\footnote{\noindent It is worth pointing out that the coverage improvement in
$\Omega^{\left(2\right)}$ is intrinsically a harder task than that
for $\Omega^{\left(1\right)}$ because of the larger distance, $R$,
of the \emph{BTS} from \emph{}$\Omega^{\left(2\right)}$ (i.e., $R\left(\mathbf{r}_{\Psi},\,\mathbf{r}_{\Omega}^{\left(1\right)}\right)\approx123$
{[}m{]} vs. $R\left(\mathbf{r}_{\Psi},\,\mathbf{r}_{\Omega}^{\left(2\right)}\right)\approx405$
{[}m{]} - Fig. 7).%
}.

\noindent To point out the coverage improvement enabled by the \emph{EMSs},
the thresholded maps of the power gap $\Delta\mathcal{P}\left(\mathbf{r}\right)$
{[}$\Delta\mathcal{P}\left(\mathbf{r}\right)\triangleq\mathcal{P}\left(\left.\mathbf{r}\right|\underline{\chi}^{\left(opt\right)}\right)-\mathcal{P}_{0}\left(\mathbf{r}\right)${]}
are shown, as well {[}Figs. 9(\emph{c})-9(\emph{d}){]}. One can observe
that the received power has been increased (i.e., $\Delta\mathcal{P}\left(\mathbf{r}\right)>0$)
over a wide region around both the \emph{RoI}s centers, while the
red pixels always correspond to limited/negligible reductions of power
level (i.e., $\Delta\mathcal{P}\left(\mathbf{r}\right)\geq-1.5$ {[}dBm{]}). 

\noindent Let us now focus on $\Omega^{\left(1\right)}$ to investigate
on the {}``effect/impact'' of each $q$-th ($q=1,...,\left.Q^{\left(opt\right)}\right|_{s=1}=3$)
\emph{EMS} on the wireless coverage. Towards this purpose, Figure
10 gives the behavior of the cumulative density function (\emph{CDF})
of the received power, $\Theta$, which is defined as\begin{equation}
\Theta\left\{ \left.\mathcal{P}\left(\mathbf{r}\right)\right|\widehat{\mathcal{P}}\right\} =\textnormal{Pr}\left\{ \mathcal{P}\left(\mathbf{r}\right)\leq\widehat{\mathcal{P}}\right\} \label{eq:}\end{equation}
where $\textnormal{Pr}\left\{ \,.\,\right\} $ denotes the probability
function and $\widehat{\mathcal{P}}\in\left[-70,\,-50\right]$ {[}dBm{]},
computed over a circular region of radius $\zeta=40$ {[}m{]} and
centered in $\mathbf{r}_{\Omega}^{\left(1\right)}$ {[}Fig. 9(\emph{a}){]}.
It turns out that there is a progressive improvement of the wireless
coverage (i.e., $\left.\Theta\left\{ \left.\mathcal{P}\left(\mathbf{r}\right)\right|\mathcal{P}_{th}\right\} \right|_{Q^{\left(1\right)}-1}$
$>$ $\left.\Theta\left\{ \left.\mathcal{P}\left(\mathbf{r}\right)\right|\mathcal{P}_{th}\right\} \right|_{Q^{\left(1\right)}}$
$>$ $\left.\Theta\left\{ \left.\mathcal{P}\left(\mathbf{r}\right)\right|\mathcal{P}_{th}\right\} \right|_{Q^{\left(1\right)}+1}$)
starting from the {}``nominal'' case {[}i.e., $\Theta$\{$\left.\mathcal{P}_{0}\left(\mathbf{r}\right)\right|\mathcal{P}_{th}$\}
$=$ $30.2$ \% - Fig. 8(\emph{a}){]} up to the \emph{EMSs} planning
at convergence {[}i.e., $\Theta\left\{ \left.\mathcal{P}\left(\left.\mathbf{r}\right|\underline{\chi}^{\left(opt\right)}\right)\right|\mathcal{P}_{th}\right\} $
$=$ $0$ \% - Fig. 9(\emph{a}){]}. For completeness, Figure 11 shows
the thresholded maps for the two intermediate sub-optimal configurations
comprising $Q^{\left(1\right)}=1$ {[}$\left.\Theta\left\{ \left.\mathcal{P}\left(\mathbf{r}\right)\right|\mathcal{P}_{th}\right\} \right|_{Q^{\left(1\right)}=1}=14.2$
\% - Fig. 11(\emph{a}) and Fig. 10{]} and $Q^{\left(1\right)}=2$
\emph{EMSs} {[}$\left.\Theta\left\{ \left.\mathcal{P}\left(\mathbf{r}\right)\right|\mathcal{P}_{th}\right\} \right|_{Q^{\left(1\right)}=2}=2.5$
\% - Fig. 11(\emph{b}) and Fig. 10{]}. 

\noindent Similar results have been yielded for the \emph{RoI} $\Omega^{\left(2\right)}$
(Fig. 12), as well. More in detail, the \emph{SbD}-derived \emph{EMS}s
distribution reduces the probability of being below the coverage threshold
of $\mathcal{P}_{th}=-65$ {[}dBm{]} from $\left.\Theta\left\{ \left.\mathcal{P}\left(\mathbf{r}\right)\right|\mathcal{P}_{th}\right\} \right|_{Q^{\left(2\right)}=0}$
$=$ $26.0$ \% {[}Fig. 8(\emph{b}){]} down to $\left.\Theta\left\{ \left.\mathcal{P}\left(\mathbf{r}\right)\right|\mathcal{P}_{th}\right\} \right|_{Q^{\left(2\right)}=4}$
$=$ $4.7$ \% {[}Fig. 9(\emph{b}){]}, being $\left.\Theta\left\{ \left.\mathcal{P}\left(\mathbf{r}\right)\right|\mathcal{P}_{th}\right\} \right|_{Q^{\left(2\right)}=1}$
$=$ $25.2$ \% {[}Fig. 13(\emph{a}){]}, $\left.\Theta\left\{ \left.\mathcal{P}\left(\mathbf{r}\right)\right|\mathcal{P}_{th}\right\} \right|_{Q^{\left(2\right)}=2}$
$=$ $17.2$ \% {[}Fig. 13(\emph{b}){]}, and $\left.\Theta\left\{ \left.\mathcal{P}\left(\mathbf{r}\right)\right|\mathcal{P}_{th}\right\} \right|_{Q^{\left(2\right)}=3}$
$=$ $9.3$ \% {[}Fig. 13(\emph{c}){]}.

\noindent As for the computational issues, the \emph{SbD} method assures
a time saving of about $\Delta t_{sav}\approx90\%$%
\footnote{\noindent Considering that the average simulation time for evaluating
the received power distribution associated to one trial guess of the
\emph{SbD-DoF}s vector $\underline{\chi}$ is equal to $\Delta t_{sim}=75$
{[}sec{]} on a standard laptop equipped with an Intel(R) Core(TM)
i5-8250U CPU @ 1.60GHz and 16 {[}GB{]} of RAM memory, it turns out
that the time for a serial assessment of all $B$ configurations would
be equal to $\Delta t_{sim}^{enum}=\left(B\times\Delta t_{sim}^{avg}\right)\approx910$
{[}days{]}.%
} ($\Delta t_{sav}\triangleq\frac{\left(P\times I\right)-T}{\left(P\times I\right)}$)
\cite{Massa 2021} with respect to a standard optimization, mainly
thanks to the exploitation of the \emph{DT} for the coverage assessment
(Sect. \ref{sub:FFE-Block}) during the iterative process.

\noindent The second test case of the numerical validation is concerned
with a more challenging \emph{OPP}, the power threshold being set
to $\mathcal{P}_{th}=-60$ {[}dBm{]}. Owing to the harder requirement,
two additional \emph{RoI}s appear on the same scenario $\Xi$ of the
previous example ($S=4$ - Fig. 14), namely the \emph{RoI} $\Omega^{\left(3\right)}$
($\mathbf{r}_{\Omega}^{\left(3\right)}=\left(313,\,914,\,1.5\right)$
{[}m{]}, $M^{\left(3\right)}=74$, and $A\left(\Omega^{\left(3\right)}\right)=1850$
{[}$m^{2}${]}) and the \emph{RoI} $\Omega^{\left(4\right)}$ ($\mathbf{r}_{\Omega}^{\left(4\right)}=\left(363,\,396,\,1.5\right)$
{[}m{]}, $M^{\left(4\right)}=59$, and $A\left(\Omega^{\left(4\right)}\right)=1475$
{[}$m^{2}${]}). Therefore, $W^{\left(3\right)}=10$ {[}Fig. 15(\emph{a}){]}
and $W^{\left(4\right)}=8$ {[}Fig. 15(\emph{b}){]} new \emph{EMSs}
have been designed in the \emph{EMSD} block (Sect. \ref{sub:SRSD-Block})
for a potential deployment on the {}``candidate'' building facades
in $\Pi^{\left(3\right)}$ {[}Fig. 15(\emph{a}){]} and $\Pi^{\left(4\right)}$
{[}Fig. 15(\emph{b}){]}. Due to the higher cardinality of the solution
space (i.e., $K=38$ $\rightarrow$ $B=2.75\times10^{11}$ - Tab.
I), a bigger training set $\mathbb{T}$ of $T=7.6\times10^{3}$ \emph{I/O}
pairs has been generated to build the \emph{DT} (Sect. \ref{sub:FFE-Block}),
while the population size of the \emph{BGA} has been increased to
$P=76$ \cite{Massa 2021}\cite{Rocca 2009}.

\noindent At the convergence, the coverage maps in Fig. 16 have been
synthesized by installing $Q^{\left(opt\right)}=24$ \emph{EMS}s ($\Phi_{cost}\left\{ \underline{\chi}^{\left(opt\right)}\right\} =6.32\times10^{-1}$
- Tab. I). As expected, the number of \emph{EMSs} deployed for $\Omega^{\left(1\right)}$
{[}i.e., $\left.Q^{\left(opt\right)}\right|_{s=1}=6$ - Figs. 16(\emph{a})-16(\emph{c}){]}
and for $\Omega^{\left(2\right)}$ {[}i.e., $\left.Q^{\left(opt\right)}\right|_{s=2}=10$
- Figs. 16(\emph{d})-16(\emph{f}){]} is larger with respect to the
previous benchmark because of the more demanding requirement on $\mathcal{P}_{th}$.
Moreover, there has been an improvement of the \emph{QoS} within the
$S=4$ \emph{RoI}s (see the left column vs. the middle column in Fig.
16) as quantitatively assessed by the reduction of the value of coverage
term (\ref{eq:coverage-term}) with respect to the scenario without
\emph{EMS}s (i.e., $\frac{\Phi_{cov}\left\{ \underline{\chi}=\underline{0}\right\} }{\Phi_{cov}\left\{ \underline{\chi}^{\left(opt\right)}\right\} }=2.59\times10^{-1}$
- Tab. I). The effectiveness of the proposed planning method can be
also clearly inferred from the analysis of the $\Delta\mathcal{P}\left(\mathbf{r}\right)$
maps (right column of Fig. 16) where the power level increases within
large areas centered on the RoIs centers, \{$\mathbf{r}_{\Omega}^{\left(s\right)}$;
$s=1,...,S$\}.

\noindent The positive outcome on the \emph{EMS}s planning is also
confirmed, from a statistical viewpoint, by the \emph{CDF}s in Fig.
17. For instance, let us analyze the case of the \emph{RoI} $\Omega^{\left(4\right)}$.
It turns out that the deployment of $Q^{\left(4\right)}=\left.Q^{\left(opt\right)}\right|_{s=4}=3$
\emph{EMSs} reduces the probability of being below the \emph{QoS}
threshold of $\mathcal{P}_{th}=-60$ {[}dBm{]} from $\left.\Theta\left\{ \left.\mathcal{P}\left(\mathbf{r}\right)\right|\mathcal{P}_{th}\right\} \right|_{Q^{\left(4\right)}=0}$
$=$ $49.2$ \% {[}Fig. 16(\emph{l}){]} down to $\left.\Theta\left\{ \left.\mathcal{P}\left(\mathbf{r}\right)\right|\mathcal{P}_{th}\right\} \right|_{Q^{\left(4\right)}=3}$
$=$ $8.3$ \% {[}Fig. 16(\emph{n}){]} with a reduction of the {}``blind''
area of about $\Delta A\left\{ \Omega^{\left(4\right)}\right\} \approx86.4$
\%.

\section{\noindent Conclusions \label{sec:Conclusions}}

\noindent In the framework of the emerging \emph{SEME} paradigm, the
planning of passive/low-cost \emph{EMSs} to enhance the \emph{QoS}
in large-scale urban propagation scenarios has been addressed. An
innovative \emph{SbD}-based strategy has been proposed to solve the
arising \emph{OPP} by determining optimal trade-off solutions, which
jointly maximize the level of power received within {}``no-coverage/low-\emph{QoS}''
\emph{RoI}s and minimize the overall cost and environmental impact.

\noindent The numerical assessment, on a real-world test-bed, has
proved the feasibility of the proposed strategy for the \emph{SEME}
implementation as well as the effectiveness of the proposed planning
method. By considering both different \emph{RoI}s and distances from
the \emph{BTS} as well as various values of the coverage threshold
$\mathcal{P}_{th}$, effective \emph{EMS}s deployments have been obtained
with a significant computational efficiency, as well.

\noindent Future works, beyond the scope of this paper, will be aimed
at extending the proposed approach to deal with the planning of mixed
scenarios involving both \emph{RIS}s and \emph{IAB} nodes.

\section*{\noindent Acknowledgements}

\noindent This work benefited from the networking activities carried
out within the Project {}``CYBER-PHYSICAL ELECTROMAGNETIC VISION:
Context-Aware Electromagnetic Sensing and Smart Reaction (EMvisioning)''
(Grant no. 2017HZJXSZ){}`` funded by the Italian Ministry of Education,
University, and Research under the PRIN2017 Program (CUP: E64I19002530001).
Moreover, it benefited from the networking activities carried out
within the Project {}``SPEED'' (Grant No. 61721001) funded by National
Science Foundation of China under the Chang-Jiang Visiting Professorship
Program, the Project 'Inversion Design Method of Structural Factors
of Conformal Load-bearing Antenna Structure based on Desired EM Performance
Interval' (Grant no. 2017HZJXSZ) funded by the National Natural Science
Foundation of China, and the Project 'Research on Uncertainty Factors
and Propagation Mechanism of Conformal Loab-bearing Antenna Structure'
(Grant No. 2021JZD-003) funded by the Department of Science and Technology
of Shaanxi Province within the Program Natural Science Basic Research
Plan in Shaanxi Province. A. Massa wishes to thank E. Vico for her
never-ending inspiration, support, guidance, and help.

\newpage
\section*{FIGURE CAPTIONS}

\begin{itemize}
\item \textbf{Figure 1.} Pictorial sketch of the \emph{OPP} geometry.
\item \textbf{Figure 2.} Pictorial sketch of the {}``candidate'' building
walls. \{$\tau_{w}^{\left(s\right)}$; $w=1,...,W^{\left(s\right)}$\},
for the installation of \emph{EMS}s to improve the \emph{QoS} within
the $s$-th ($s=1,...,S$) \emph{RoI}, $\Omega^{\left(s\right)}$.
\item \textbf{Figure 3.} Block diagram of the \emph{SbD}-based approach
to the \emph{EMS}s planning.
\item \textbf{Figure 4.} Pictorial sketch of the local coordinate system
for the $\left(w,s\right)$-th ($w=1,...,W^{\left(s\right)}$; $s=1,...,S$)
\emph{EMS}, $\Gamma_{w}^{\left(s\right)}$.
\item \textbf{Figure 5.} \emph{Numerical Assessment} - Picture of (\emph{a})
the Google Map of the test-bed region (Gardolo district - Trento,
Italy), (\emph{b}) the corresponding cartography from the \emph{OSM}
database, and (\emph{c}) the WinProp simulation scenario.
\item \textbf{Figure 6.} \emph{Numerical Assessment} ($f=3.5$ {[}GHz{]},
$L_{y}=1.75$ {[}$\lambda${]}, $L_{z}=7$ {[}$\lambda${]}) - Model
of the planar array of dual-polarization (slant-45) slot-coupled patch
antennas along with the \emph{HFSS} full-wave simulated co-polar gain
pattern $G_{+45}\left(\theta,\phi\right)$.
\item \textbf{Figure 7.} \emph{Numerical Assessment} ($f=3.5$ {[}GHz{]},
$\mathcal{P}_{th}=-65$ {[}dBm{]}, $S=2$, $K=20$) - Picture of (\emph{a})
the reference/nominal received power distribution, $\mathcal{P}_{0}\left(\mathbf{r}\right)$,
and of (\emph{b}) the corresponding thresholded power map.
\item \textbf{Figure 8.} \emph{Numerical Assessment} ($f=3.5$ {[}GHz{]},
$\mathcal{P}_{th}=-65$ {[}dBm{]}, $S=2$, $K=20$) - Picture of (\emph{a})(\emph{b})
the thresholded power map for the \emph{Nominal Scenario} (i.e., w/o
\emph{EMS}s) and of (\emph{c})(\emph{d}) the admissible locations
for the \emph{EMS}s deployment in the neighborhood of (\emph{a})(\emph{c})
the \emph{RoI} $\Omega^{\left(1\right)}$, $\Pi^{\left(1\right)}$,
and of (\emph{b})(\emph{d}) the \emph{RoI} $\Omega^{\left(2\right)}$,
$\Pi^{\left(2\right)}$.
\item \textbf{Figure 9.} \emph{Numerical Assessment} ($f=3.5$ {[}GHz{]},
$\mathcal{P}_{th}=-65$ {[}dBm{]}, $S=2$, $K=20$) - Picture of the
map of (\emph{a})(\emph{b}) the thresholded power distribution for
the \emph{SbD}-optimized \emph{EMS}s deployment and of (\emph{c})(\emph{d})
the power gap distribution in the neighborhood of (\emph{a})(\emph{c})
the \emph{RoI} $\Omega^{\left(1\right)}$, $\Pi^{\left(1\right)}$,
and of (\emph{b})(\emph{d}) the \emph{RoI} $\Omega^{\left(2\right)}$,
$\Pi^{\left(2\right)}$.
\item \textbf{Figure 10.} \emph{Numerical Assessment} ($f=3.5$ {[}GHz{]},
$\mathcal{P}_{th}=-65$ {[}dBm{]}, $S=2$, $K=20$) - Plot of the
\emph{CDF} of the received power, $\Theta\left\{ \left.\mathcal{P}\left(\mathbf{r}\right)\right|\widehat{\mathcal{P}}\right\} $,
within a circular region centered on the \emph{RoI} $\Omega^{\left(1\right)}$,
$\mathbf{r}_{\Omega}^{\left(1\right)}$, of radius $\zeta=40$ {[}m{]}.
\item \textbf{Figure 11.} \emph{Numerical Assessment} ($f=3.5$ {[}GHz{]},
$\mathcal{P}_{th}=-65$ {[}dBm{]}, $S=2$, $K=20$) - Picture of the
thresholded power map in the neighborhood of the \emph{RoI} $\Omega^{\left(1\right)}$,
$\Pi^{\left(1\right)}$, when deploying (\emph{a}) $Q^{\left(1\right)}=1$
and (\emph{b}) $Q^{\left(1\right)}=2$ \emph{EMS}s according to the
\emph{SbD}-based planning method.
\item \textbf{Figure 12.} \emph{Numerical Assessment} ($f=3.5$ {[}GHz{]},
$\mathcal{P}_{th}=-65$ {[}dBm{]}, $S=2$, $K=20$) - Plot of the
\emph{CDF} of the received power, $\Theta\left\{ \left.\mathcal{P}\left(\mathbf{r}\right)\right|\widehat{\mathcal{P}}\right\} $,
within a circular region centered on the \emph{RoI} $\Omega^{\left(2\right)}$,
$\mathbf{r}_{\Omega}^{\left(2\right)}$, of radius $\zeta=40$ {[}m{]}.
\item \textbf{Figure 13.} \emph{Numerical Assessment} ($f=3.5$ {[}GHz{]},
$\mathcal{P}_{th}=-65$ {[}dBm{]}, $S=2$, $K=20$) - Picture of the
thresholded power map in the neighborhood of the \emph{RoI} $\Omega^{\left(2\right)}$,
$\Pi^{\left(2\right)}$, when deploying (\emph{a}) $Q^{\left(2\right)}=1$,
(\emph{b}) $Q^{\left(2\right)}=2$, and (\emph{c}) $Q^{\left(2\right)}=3$
\emph{EMS}s according to the \emph{SbD}-based planning method.
\item \textbf{Figure 14.} \emph{Numerical Assessment} ($f=3.5$ {[}GHz{]},
$\mathcal{P}_{th}=-60$ {[}dBm{]}, $S=4$, $K=38$) - Picture of the
thresholded power map for the \emph{Nominal Scenario} (i.e., w/o \emph{EMS}s).
\item \textbf{Figure 15.} \emph{Numerical Assessment} ($f=3.5$ {[}GHz{]},
$\mathcal{P}_{th}=-60$ {[}dBm{]}, $S=4$, $K=38$) - Sketch of the
admissible locations for the \emph{EMS}s deployment in the neighborhood
of (\emph{a})(\emph{c}) the \emph{RoI} $\Omega^{\left(3\right)}$,
$\Pi^{\left(3\right)}$, and of (\emph{b})(\emph{d}) the \emph{RoI}
$\Omega^{\left(4\right)}$, $\Pi^{\left(4\right)}$.
\item \textbf{Figure 16.} \emph{Numerical Assessment} ($f=3.5$ {[}GHz{]},
$\mathcal{P}_{th}=-60$ {[}dBm{]}, $S=4$, $K=38$) - Picture of (\emph{a})(\emph{b})(\emph{d})(\emph{e})(\emph{g})(\emph{h})(\emph{l})(\emph{m})
the thresholded power map for (\emph{a})(\emph{d})(\emph{g})(\emph{l})
the \emph{Nominal Scenario} (i.e., w/o \emph{EMS}s) and (\emph{b})(\emph{e})(\emph{h})(\emph{m})
the scenario with the \emph{SbD}-optimize \emph{EMS}s along with (\emph{c})(\emph{f})(\emph{i})(\emph{n})
the power gap distributions in the neighborhood of (\emph{a})-(\emph{c})
the \emph{RoI} $\Omega^{\left(1\right)}$, $\Pi^{\left(1\right)}$,
(\emph{d})-(\emph{f}) the \emph{RoI} $\Omega^{\left(2\right)}$, $\Pi^{\left(2\right)}$,
(\emph{g})-(\emph{i}) the \emph{RoI} $\Omega^{\left(3\right)}$, $\Pi^{\left(3\right)}$,
and (\emph{l})-(\emph{n}) the \emph{RoI} $\Omega^{\left(4\right)}$,
$\Pi^{\left(4\right)}$.
\item \textbf{Figure 17.} \emph{Numerical Assessment} ($f=3.5$ {[}GHz{]},
$\mathcal{P}_{th}=-60$ {[}dBm{]}, $S=4$, $K=38$) - Plot of the
\emph{CDF} of the received power, $\Theta\left\{ \left.\mathcal{P}\left(\mathbf{r}\right)\right|\widehat{\mathcal{P}}\right\} $,
within a circular region of radius $\zeta=40$ {[}m{]} centered on
(\emph{a}) the \emph{RoI} $\Omega^{\left(1\right)}$, $\mathbf{r}_{\Omega}^{\left(1\right)}$,
(\emph{b}) the \emph{RoI} $\Omega^{\left(2\right)}$, $\mathbf{r}_{\Omega}^{\left(2\right)}$,
(\emph{c}) the \emph{RoI} $\Omega^{\left(3\right)}$, $\mathbf{r}_{\Omega}^{\left(3\right)}$,
and (\emph{d}) the \emph{RoI} $\Omega^{\left(4\right)}$, $\mathbf{r}_{\Omega}^{\left(4\right)}$.
\end{itemize}

\section*{TABLE CAPTIONS}

\begin{itemize}
\item \textbf{Table I.} Descriptors of the urban scenario and of the \emph{OPP}.
\end{itemize}
\newpage
\begin{center}~\vfill\end{center}

\begin{center}\includegraphics[%
  width=0.90\columnwidth]{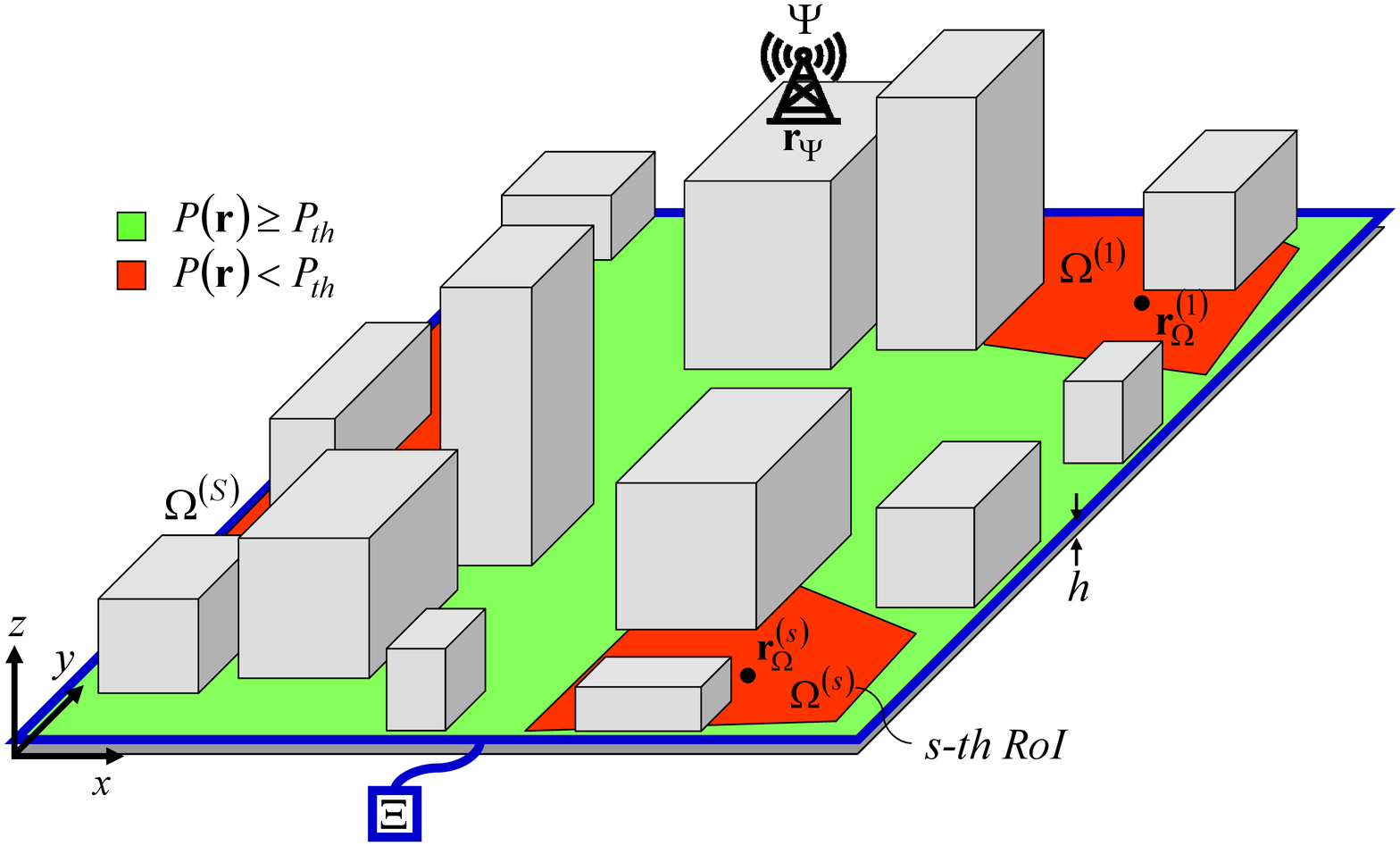}\end{center}

\begin{center}~\vfill\end{center}

\begin{center}\textbf{Fig. 1 - A. Benoni} \textbf{\emph{et al.}}\textbf{,}
\textbf{\emph{{}``}}Planning of \emph{EM} Skins for ...''\end{center}

\newpage
\begin{center}~\vfill\end{center}

\begin{center}\includegraphics[%
  width=0.90\columnwidth]{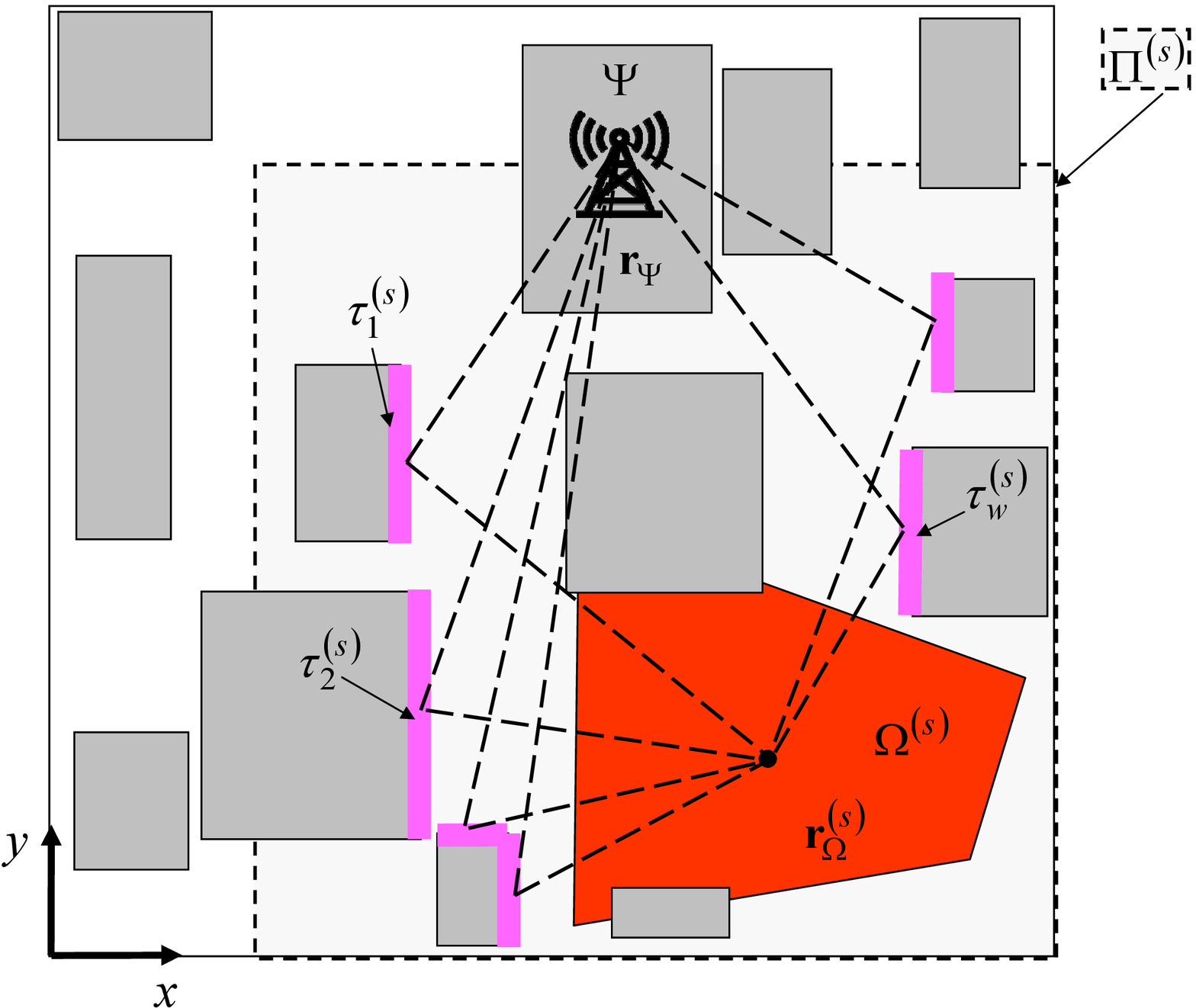}\end{center}

\begin{center}~\vfill\end{center}

\begin{center}\textbf{Fig. 2 - A. Benoni} \textbf{\emph{et al.}}\textbf{,}
\textbf{\emph{{}``}}Planning of \emph{EM} Skins for ...''\end{center}

\newpage
\begin{center}~\vfill\end{center}

\begin{center}\includegraphics[%
  width=0.90\columnwidth]{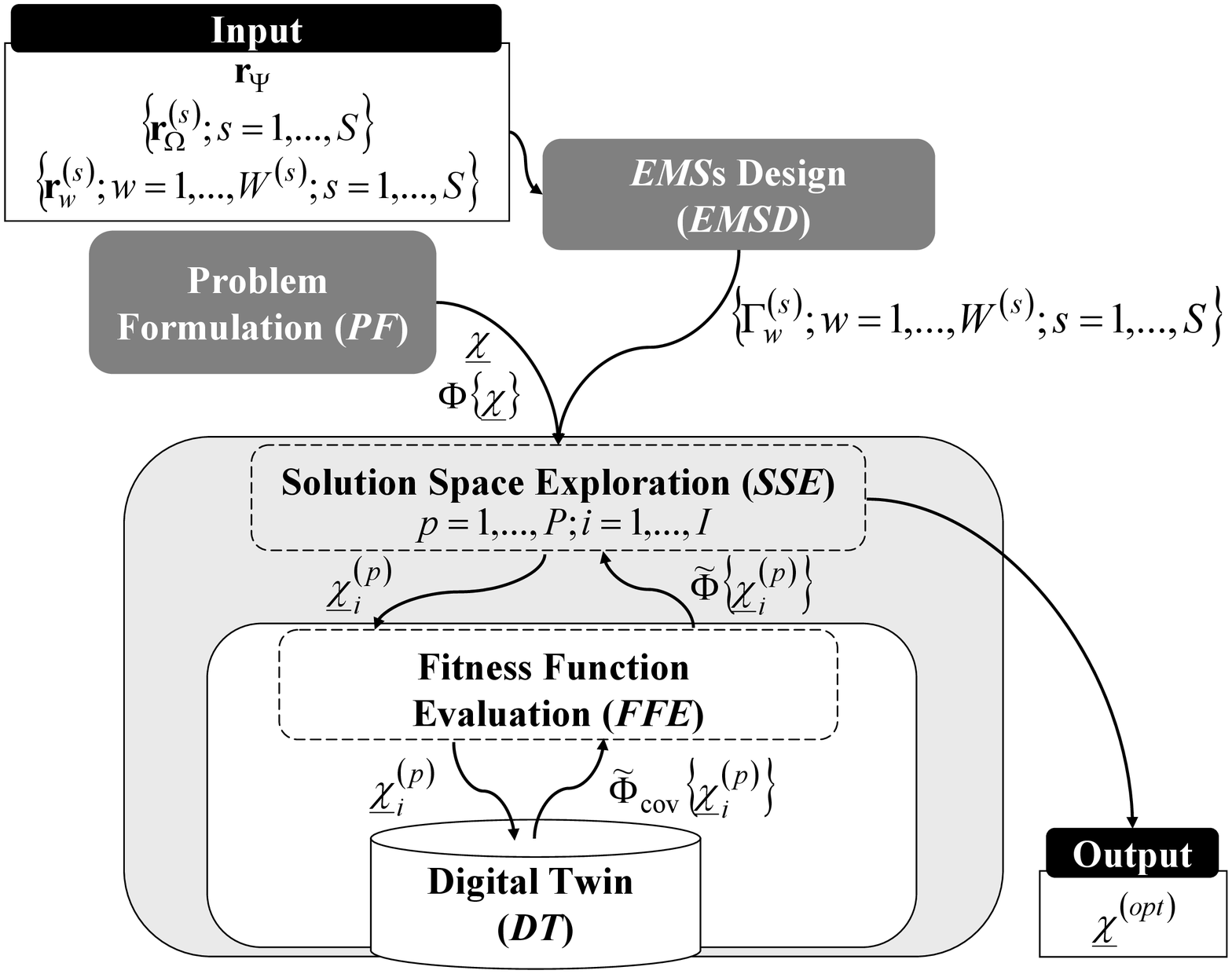}\end{center}

\begin{center}~\vfill\end{center}

\begin{center}\textbf{Fig. 3 - A. Benoni} \textbf{\emph{et al.}}\textbf{,}
\textbf{\emph{{}``}}Planning of \emph{EM} Skins for ...''\end{center}

\newpage
\begin{center}~\vfill\end{center}

\begin{center}\includegraphics[%
  width=0.90\columnwidth]{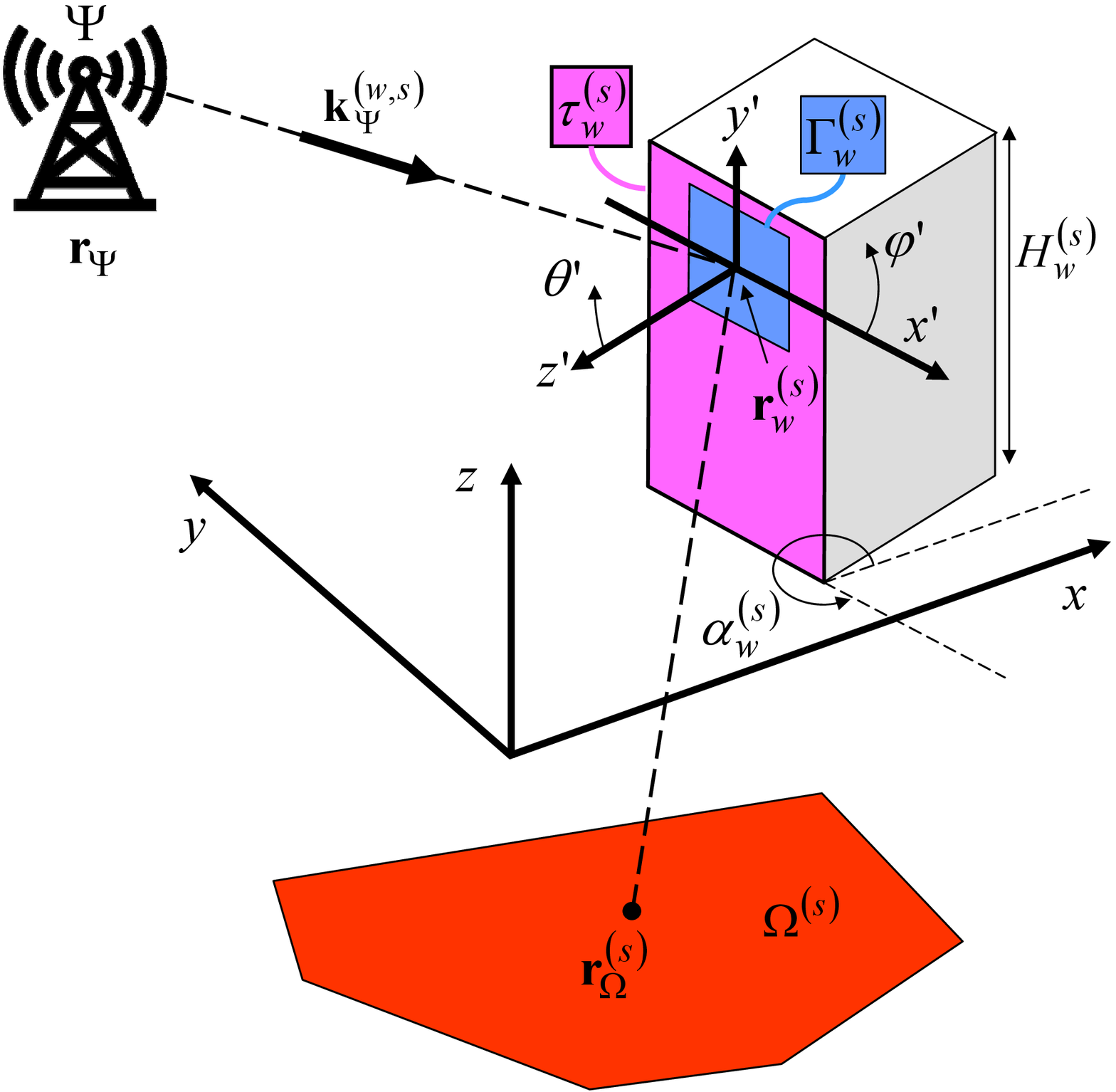}\end{center}

\begin{center}~\vfill\end{center}

\begin{center}\textbf{Fig. 4 - A. Benoni} \textbf{\emph{et al.}}\textbf{,}
\textbf{\emph{{}``}}Planning of \emph{EM} Skins for ...''\end{center}

\newpage
\begin{center}~\vfill\end{center}

\begin{center}\begin{tabular}{>{\centering}m{0.45\columnwidth}>{\centering}m{0.45\columnwidth}}
\multicolumn{2}{c}{\includegraphics[%
  width=0.65\columnwidth,
  keepaspectratio]{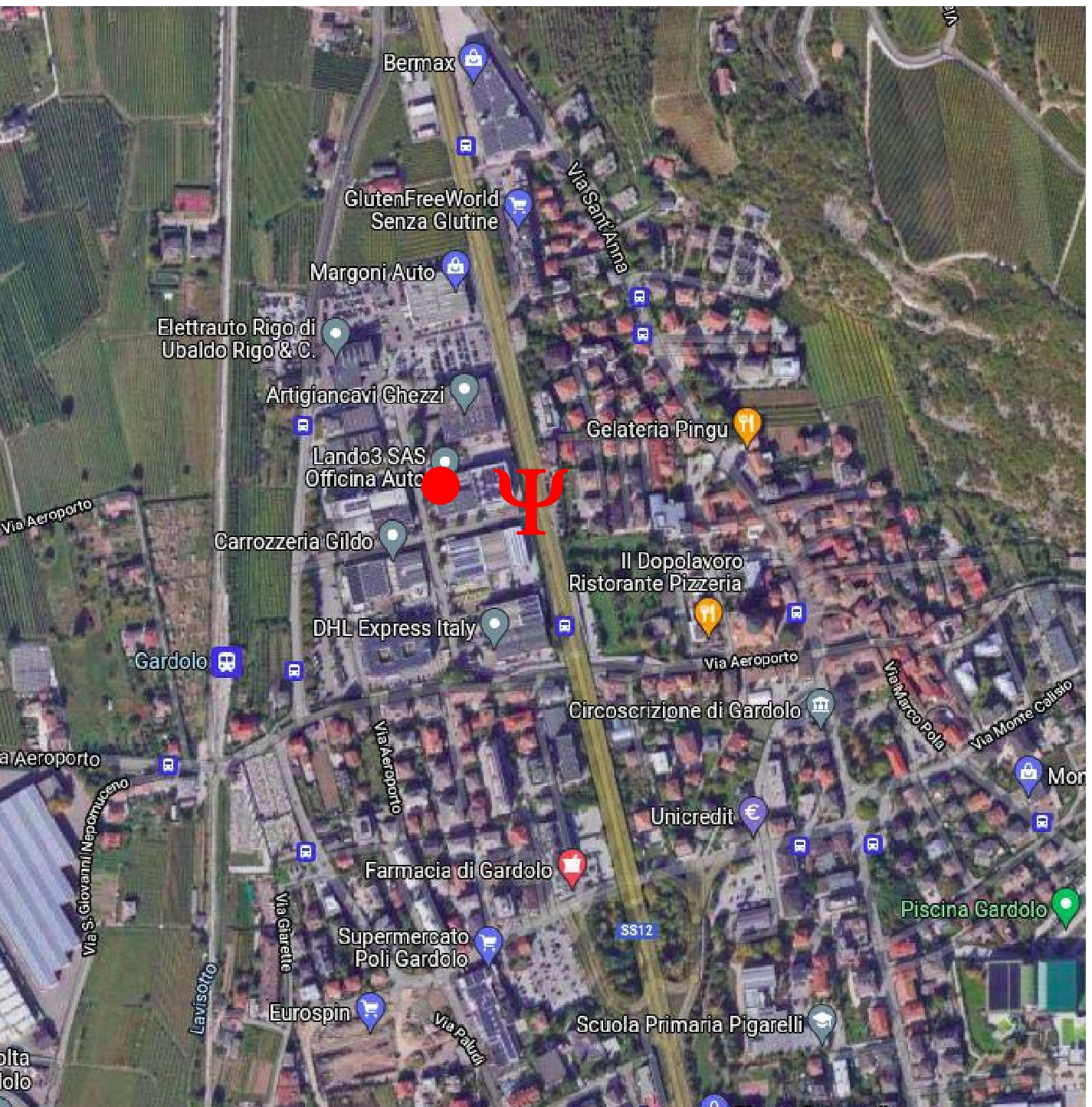}}\tabularnewline
\multicolumn{2}{c}{(\emph{a})}\tabularnewline
\multicolumn{2}{c}{}\tabularnewline
\includegraphics[%
  width=0.40\columnwidth]{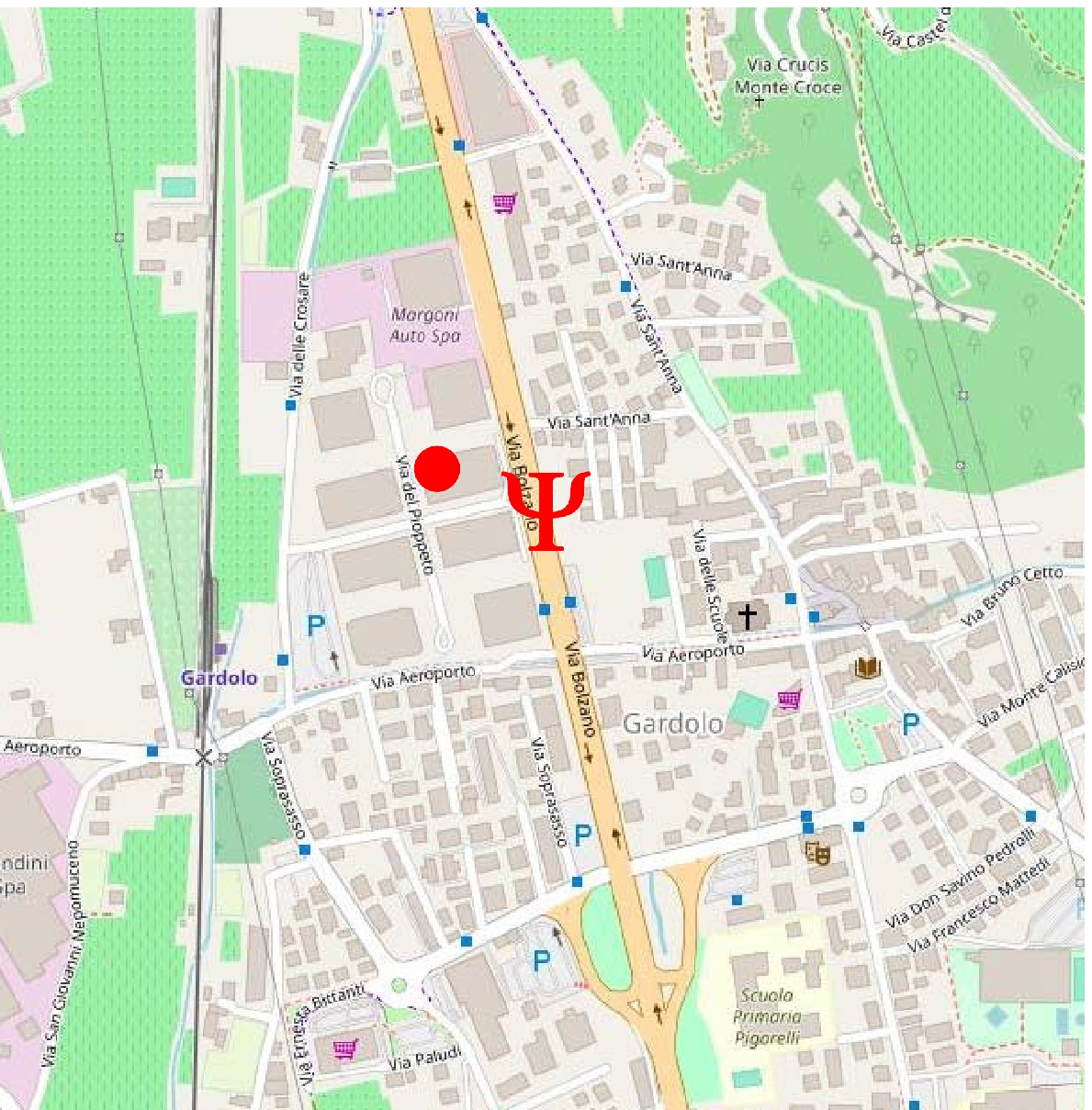}&
\includegraphics[%
  width=0.50\columnwidth]{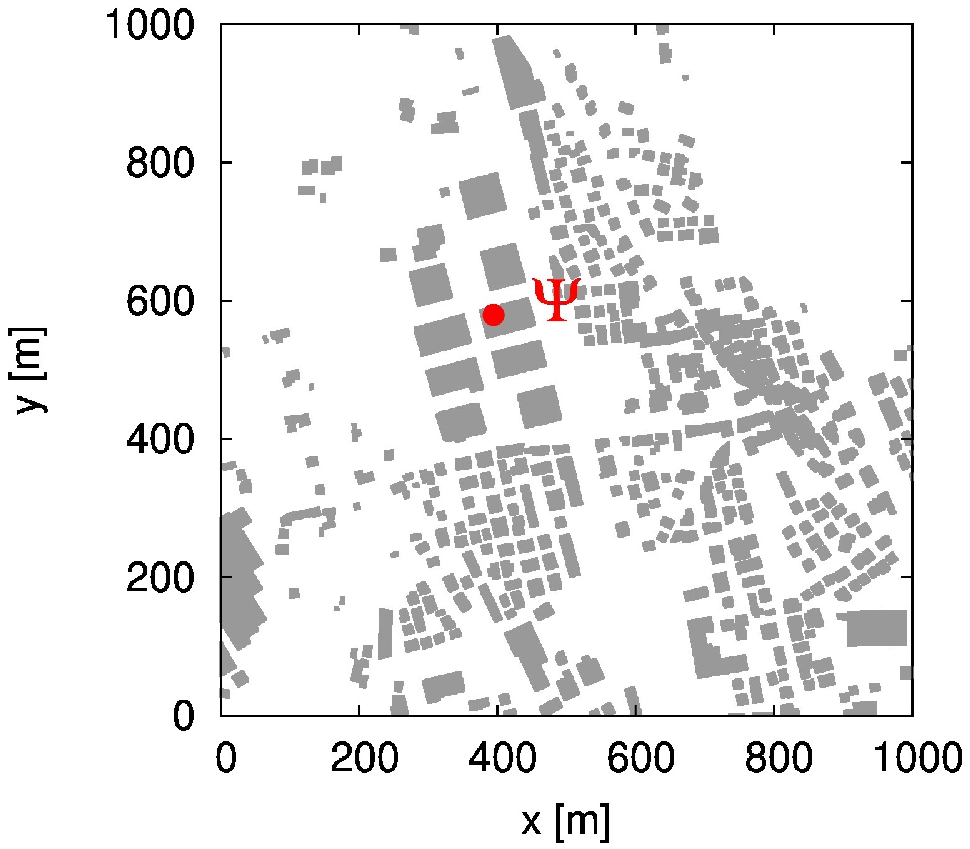}\tabularnewline
(\emph{b})&
(\emph{c})\tabularnewline
\end{tabular}\end{center}

\begin{center}~\vfill\end{center}

\begin{center}\textbf{Fig. 5 - A. Benoni} \textbf{\emph{et al.}}\textbf{,}
\textbf{\emph{{}``}}Planning of \emph{EM} Skins for ...''\end{center}

\newpage
\begin{center}~\vfill\end{center}

\begin{center}\includegraphics[%
  width=0.70\columnwidth]{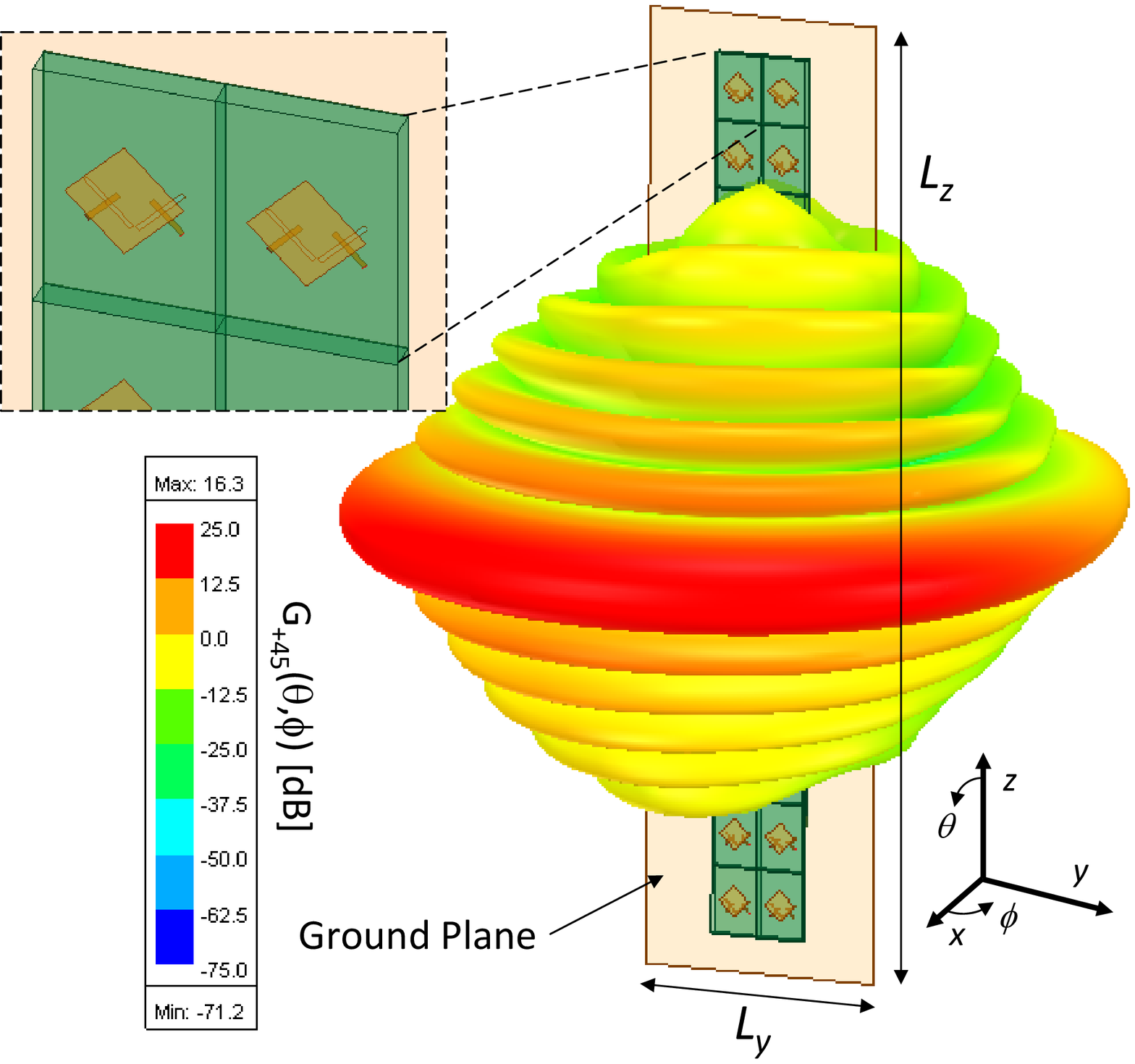}\end{center}

\begin{center}~\vfill\end{center}

\begin{center}\textbf{Fig. 6 - A. Benoni} \textbf{\emph{et al.}}\textbf{,}
\textbf{\emph{{}``}}Planning of \emph{EM} Skins for ...''\end{center}

\newpage
\begin{center}~\vfill\end{center}

\begin{center}\begin{tabular}{c}
\includegraphics[%
  width=0.70\columnwidth]{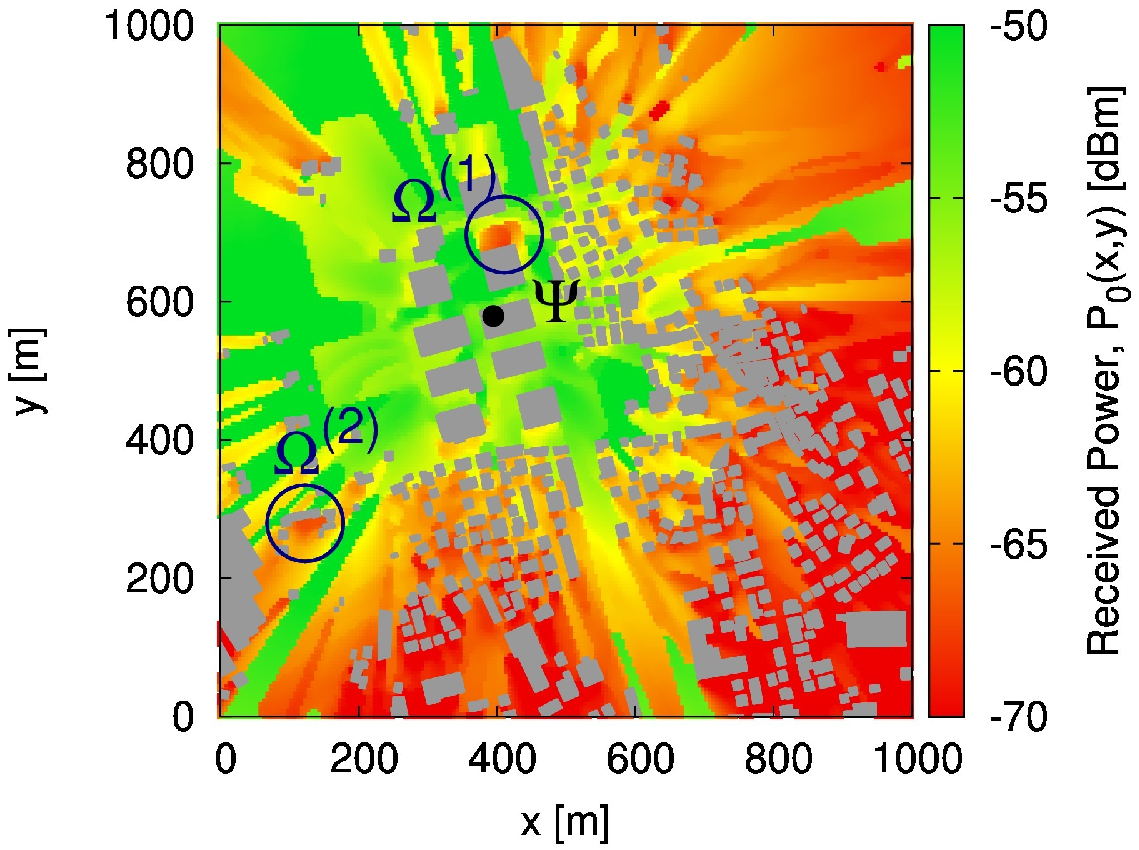}\tabularnewline
(\emph{a})\tabularnewline
\tabularnewline
\includegraphics[%
  width=0.65\columnwidth]{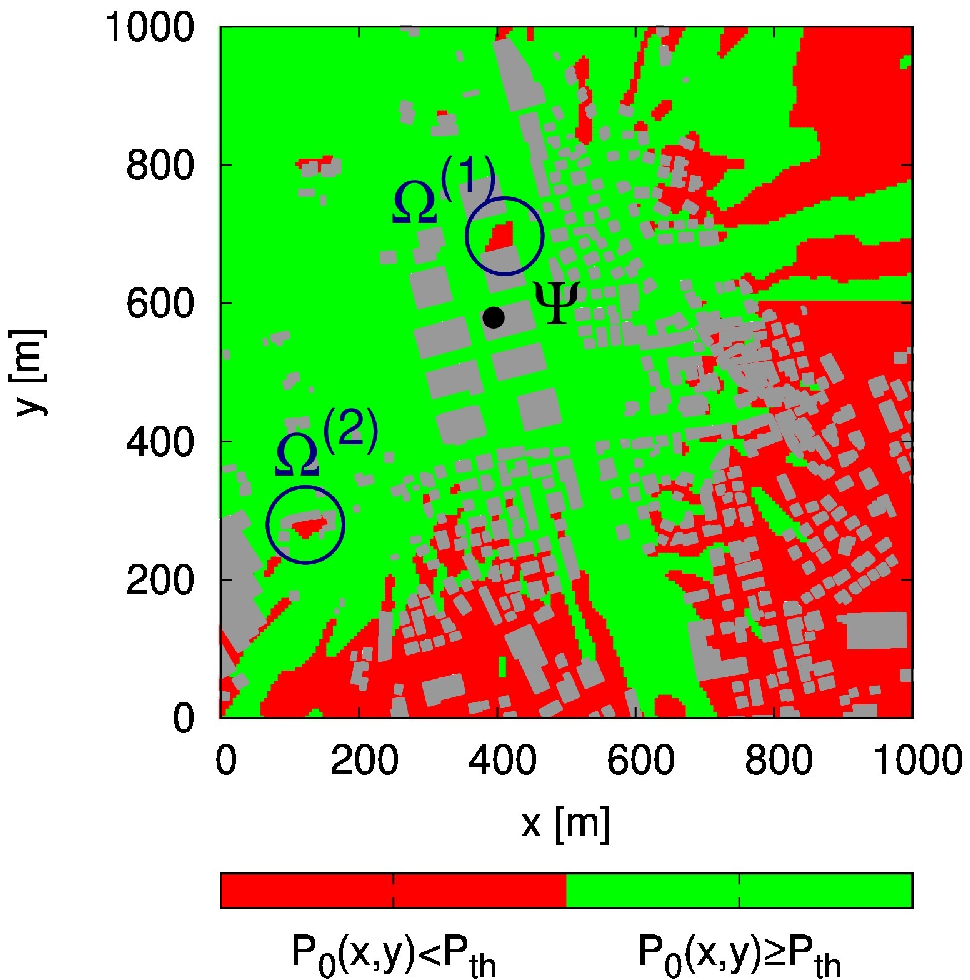}~~~~~~~~\tabularnewline
(\emph{b})\tabularnewline
\end{tabular}\end{center}

\begin{center}~\vfill\end{center}

\begin{center}\textbf{Fig. 7 - A. Benoni} \textbf{\emph{et al.}}\textbf{,}
\textbf{\emph{{}``}}Planning of \emph{EM} Skins for ...''\end{center}

\newpage
\begin{center}~\vfill\end{center}

\begin{center}\begin{tabular}{cc}
\includegraphics[%
  width=0.45\columnwidth,
  keepaspectratio]{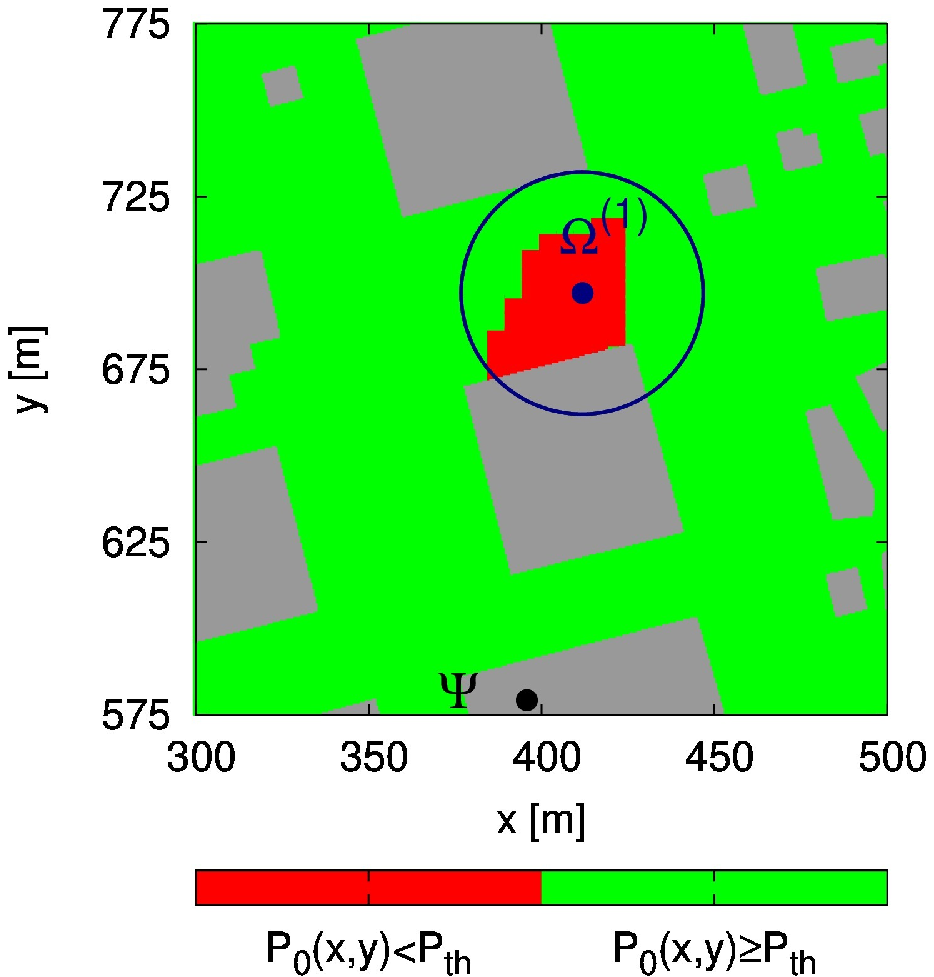}&
\includegraphics[%
  width=0.45\columnwidth,
  keepaspectratio]{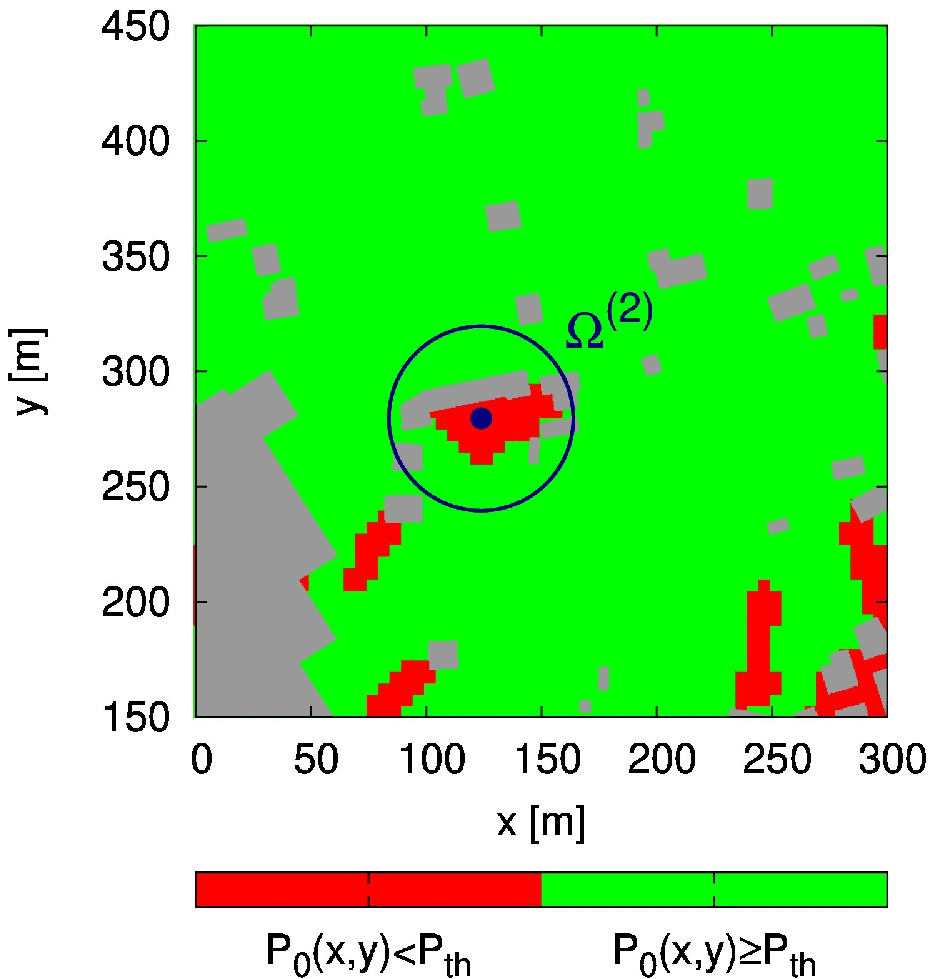}\tabularnewline
(\emph{a})&
(\emph{b})\tabularnewline
&
\tabularnewline
\includegraphics[%
  width=0.45\columnwidth]{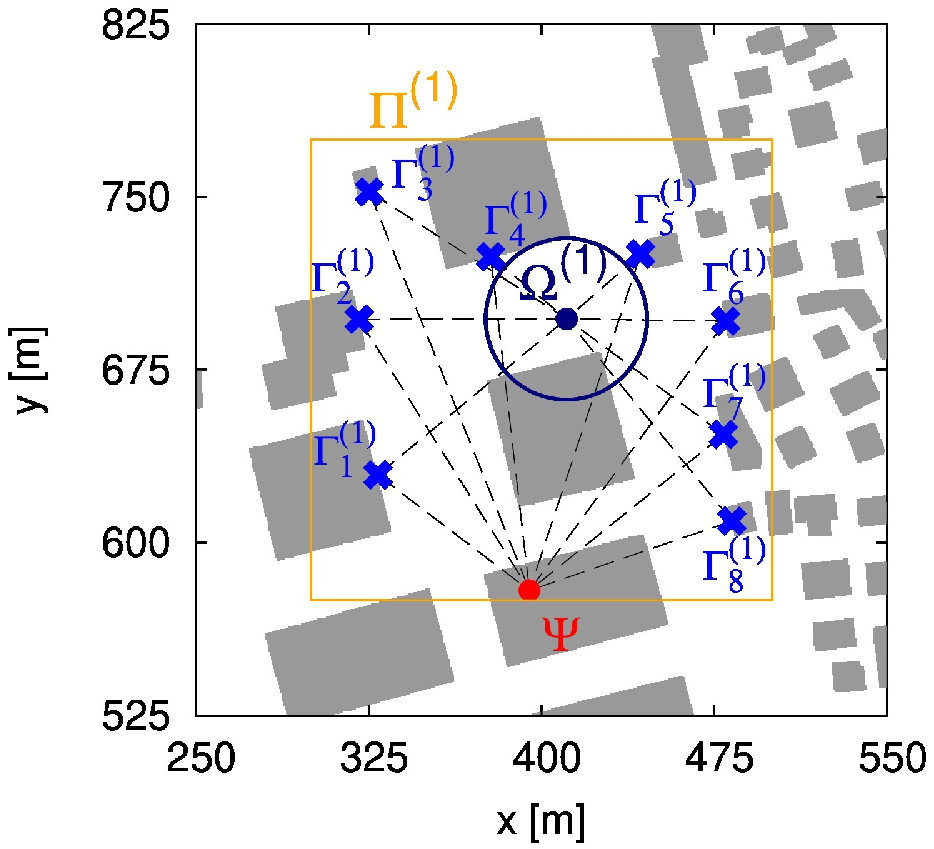}&
\includegraphics[%
  width=0.45\columnwidth]{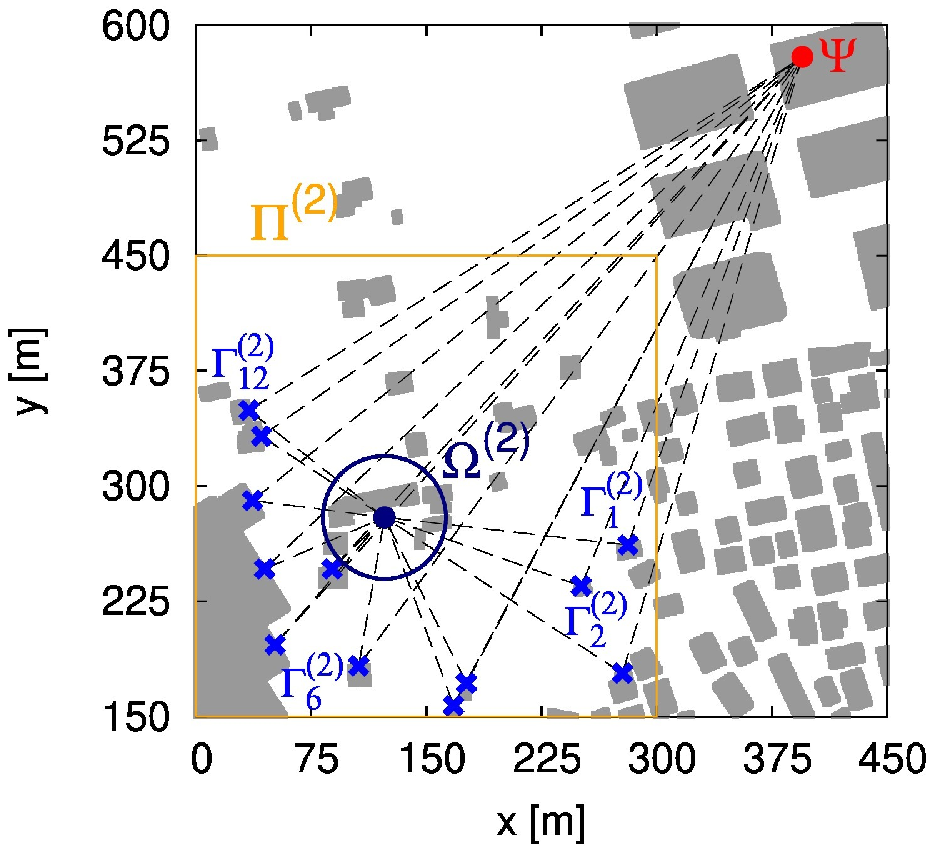}\tabularnewline
(\emph{c})&
(\emph{d})\tabularnewline
\end{tabular}\end{center}

\begin{center}~\vfill\end{center}

\begin{center}\textbf{Fig. 8 - A. Benoni} \textbf{\emph{et al.}}\textbf{,}
\textbf{\emph{{}``}}Planning of \emph{EM} Skins for ...''\end{center}

\newpage
\begin{center}~\vfill\end{center}

\begin{center}\begin{tabular}{cc}
\includegraphics[%
  width=0.40\columnwidth]{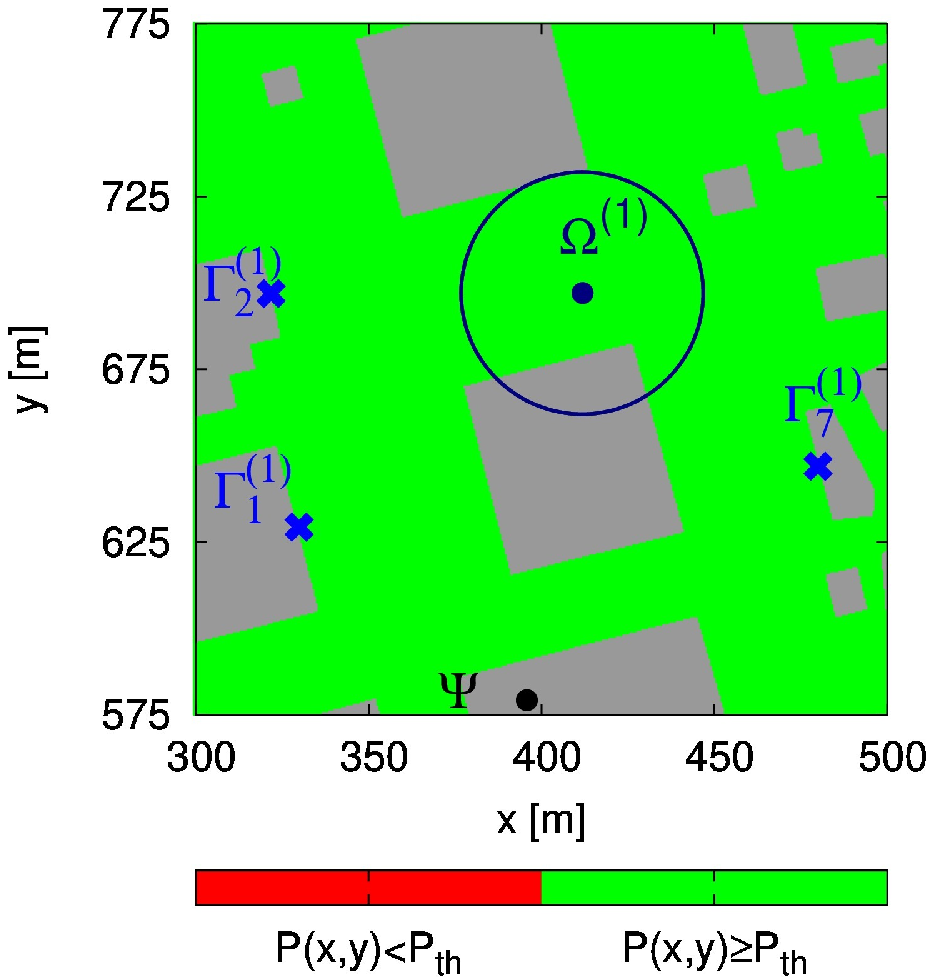}&
\includegraphics[%
  width=0.40\columnwidth]{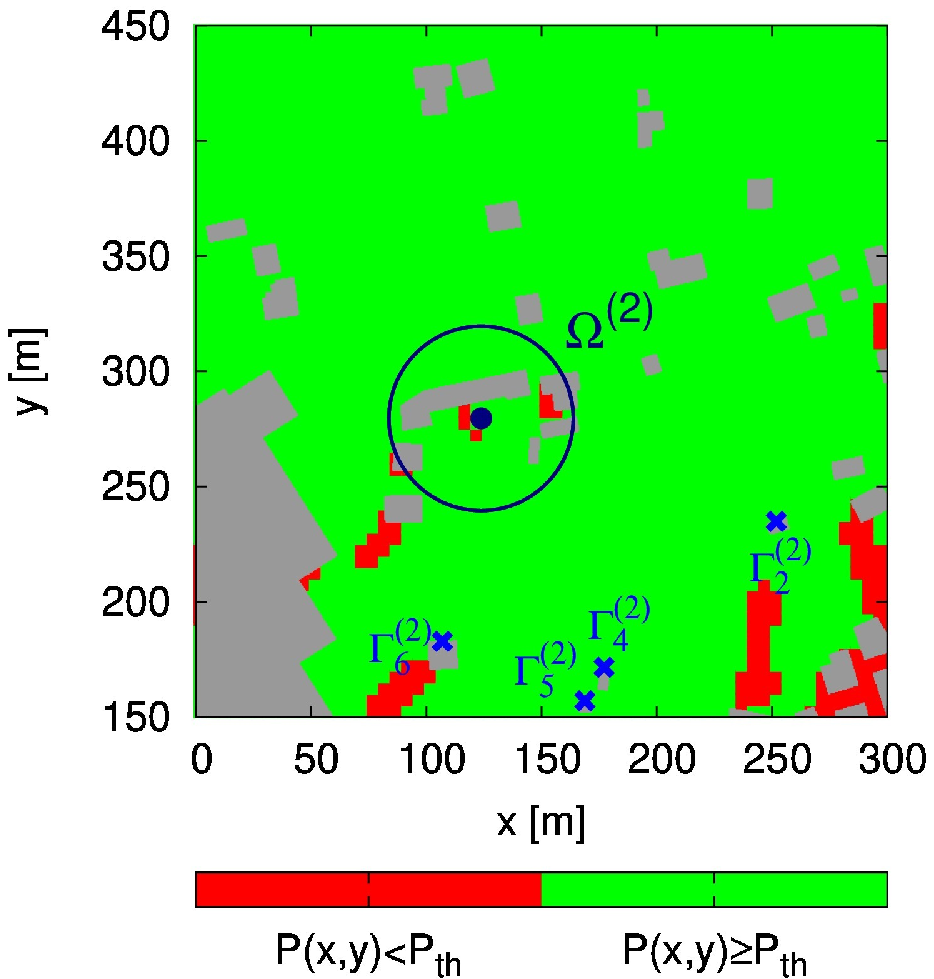}\tabularnewline
(\emph{a})&
(\emph{b})\tabularnewline
\includegraphics[%
  width=0.40\columnwidth]{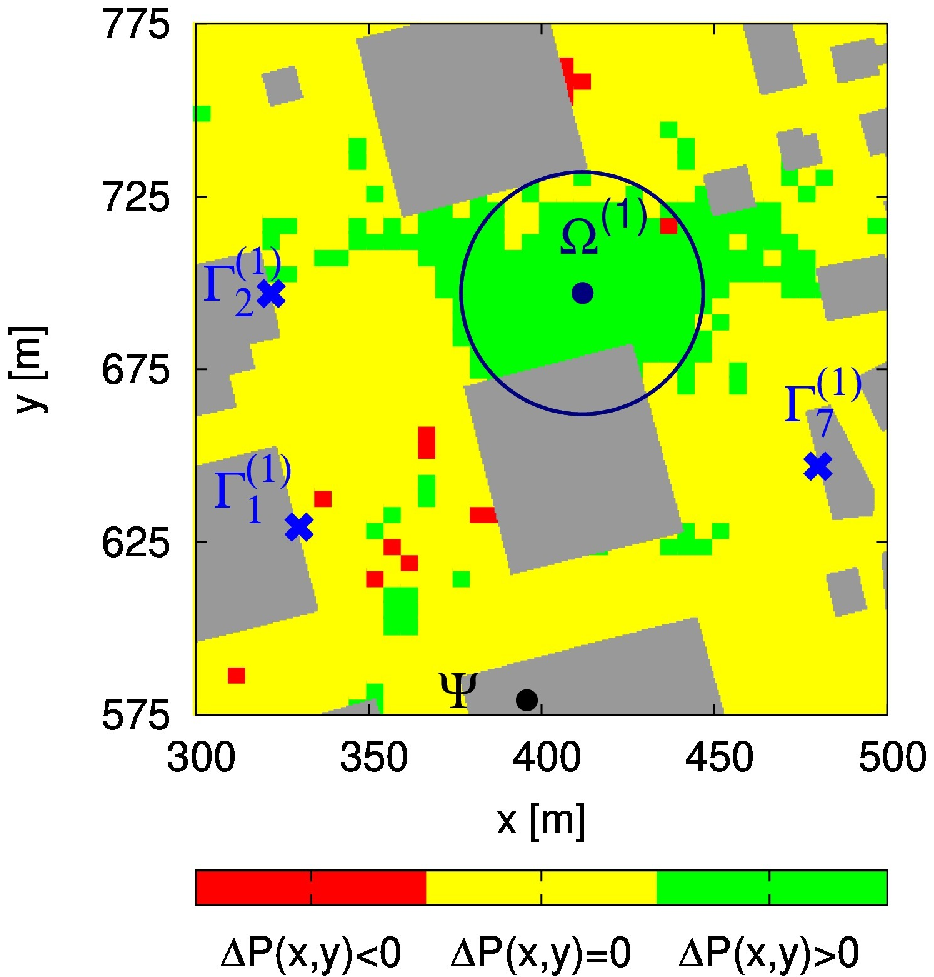}&
\includegraphics[%
  width=0.40\columnwidth]{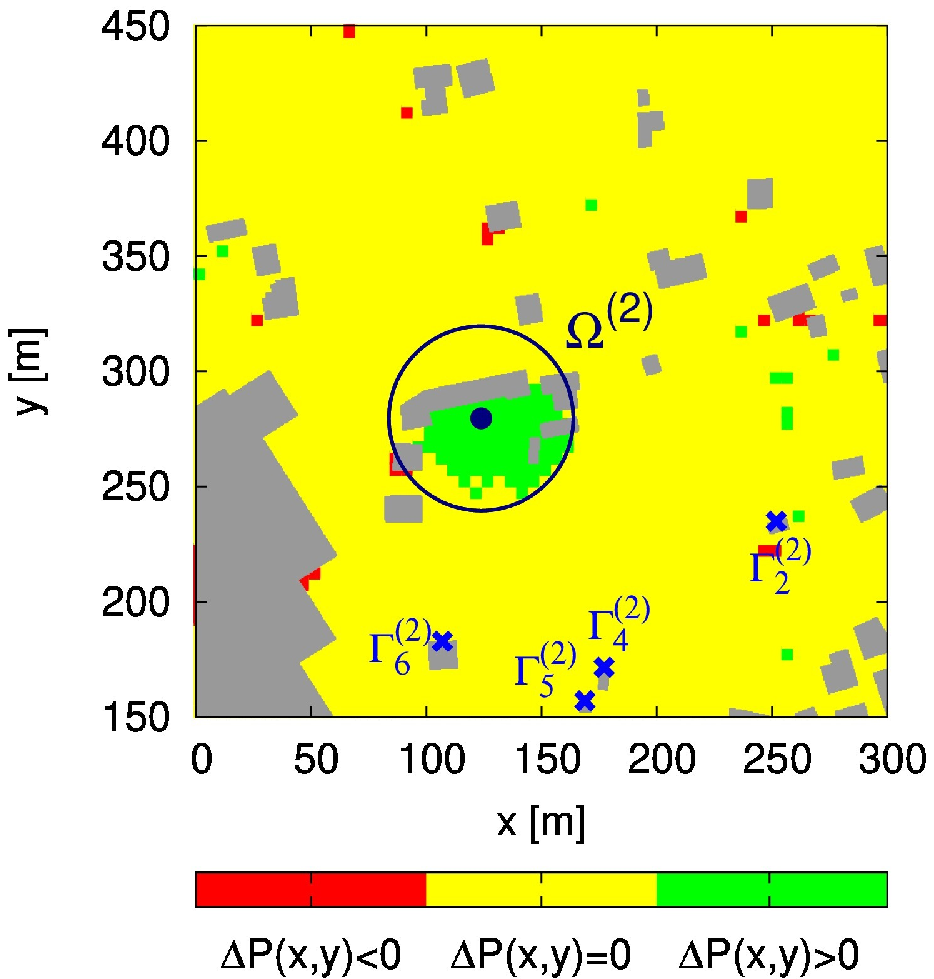}\tabularnewline
(\emph{c})&
(\emph{d})\tabularnewline
\end{tabular}\end{center}

\begin{center}~\vfill\end{center}

\begin{center}\textbf{Fig. 9 - A. Benoni} \textbf{\emph{et al.}}\textbf{,}
\textbf{\emph{{}``}}Planning of \emph{EM} Skins for ...''\end{center}

\newpage
\begin{center}~\vfill\end{center}

\begin{center}\includegraphics[%
  width=0.90\columnwidth]{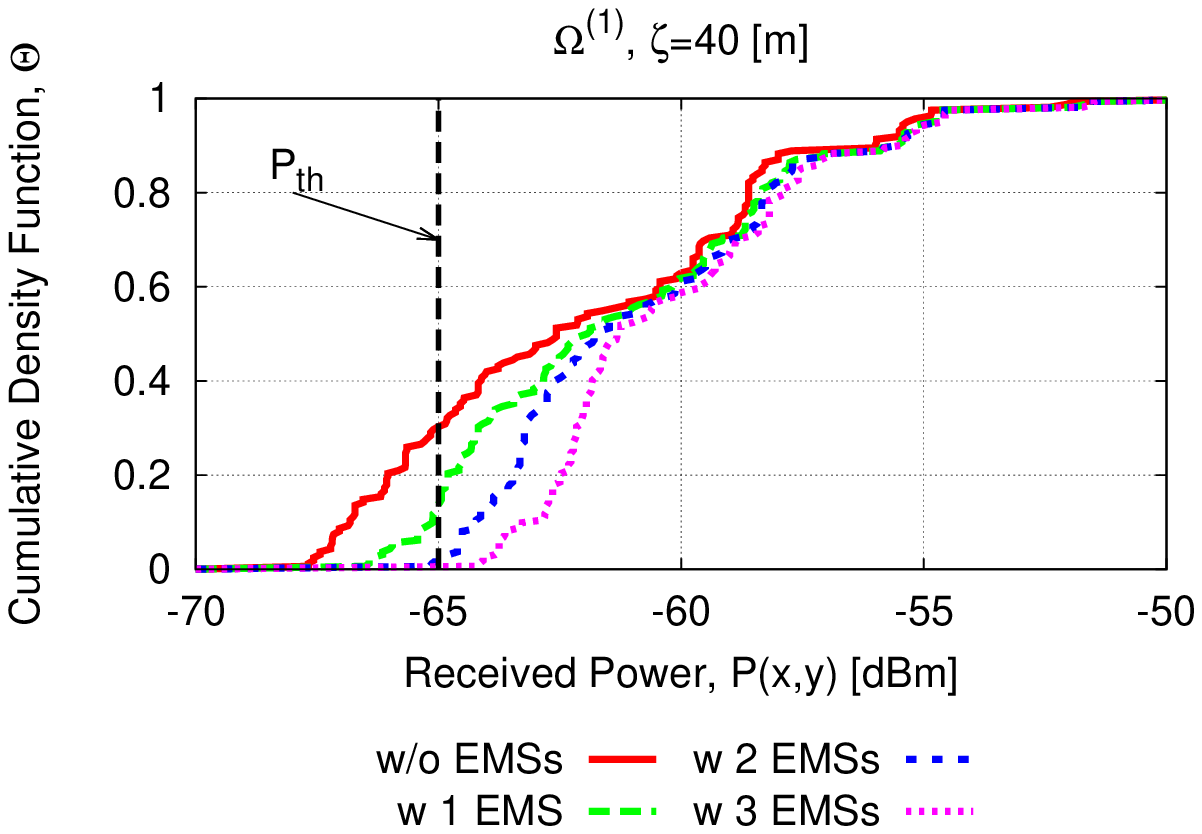}\end{center}

\begin{center}~\vfill\end{center}

\begin{center}\textbf{Fig. 10 - A. Benoni} \textbf{\emph{et al.}}\textbf{,}
\textbf{\emph{{}``}}Planning of \emph{EM} Skins for ...''\end{center}

\newpage
\begin{center}~\vfill\end{center}

\begin{center}\begin{tabular}{c}
\includegraphics[%
  width=0.55\columnwidth]{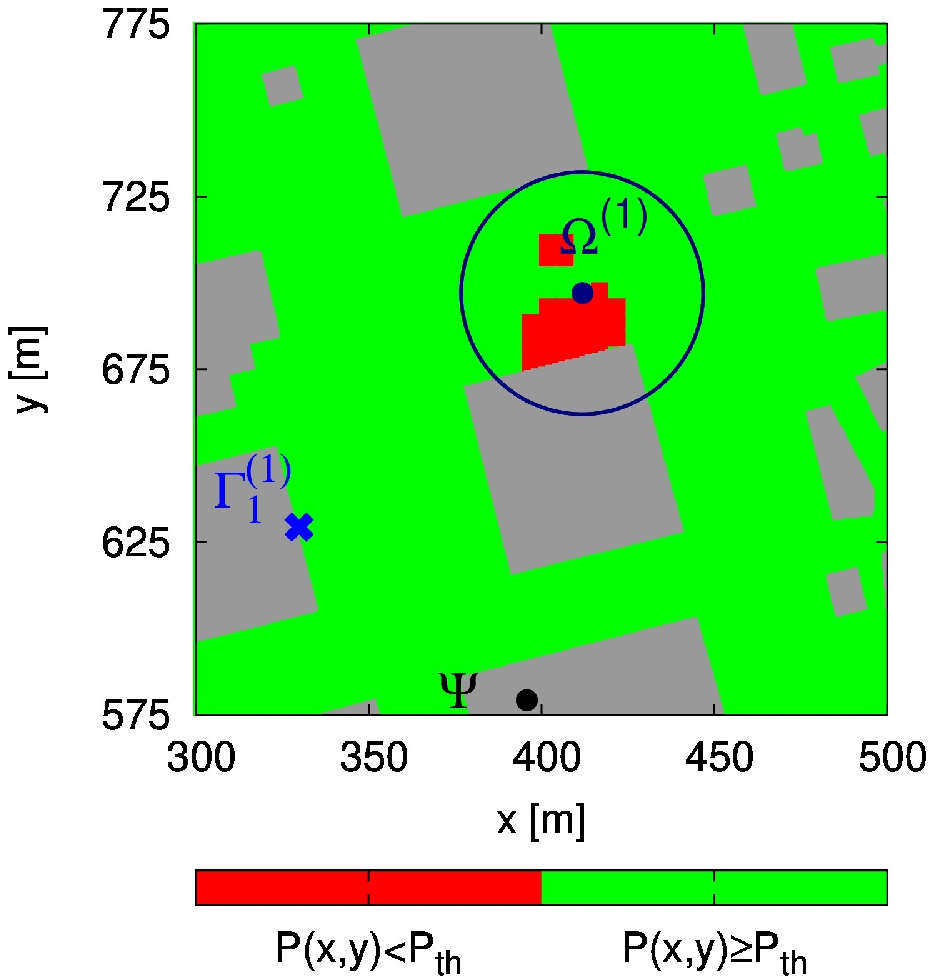}\tabularnewline
(\emph{a})\tabularnewline
\tabularnewline
\includegraphics[%
  width=0.55\columnwidth]{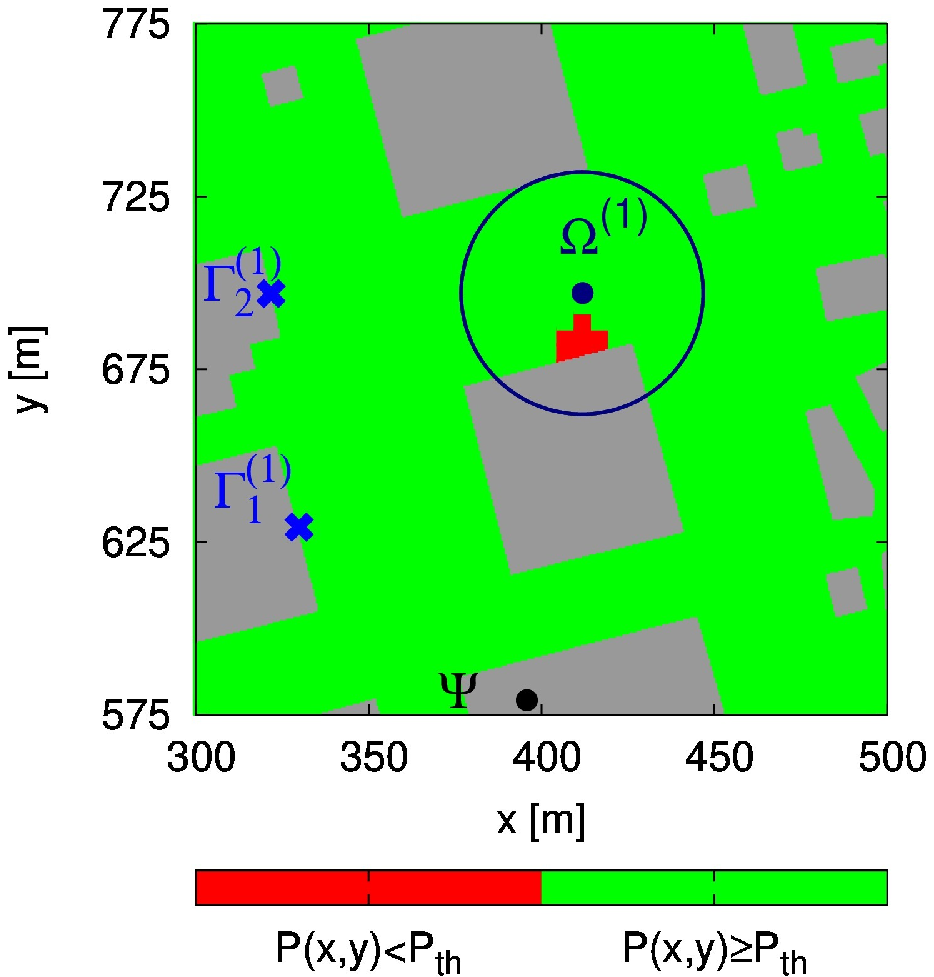}\tabularnewline
(\emph{b})\tabularnewline
\end{tabular}\end{center}

\begin{center}~\vfill\end{center}

\begin{center}\textbf{Fig. 11 - A. Benoni} \textbf{\emph{et al.}}\textbf{,}
\textbf{\emph{{}``}}Planning of \emph{EM} Skins for ...''\end{center}

\newpage
\begin{center}~\vfill\end{center}

\begin{center}\includegraphics[%
  width=0.90\columnwidth]{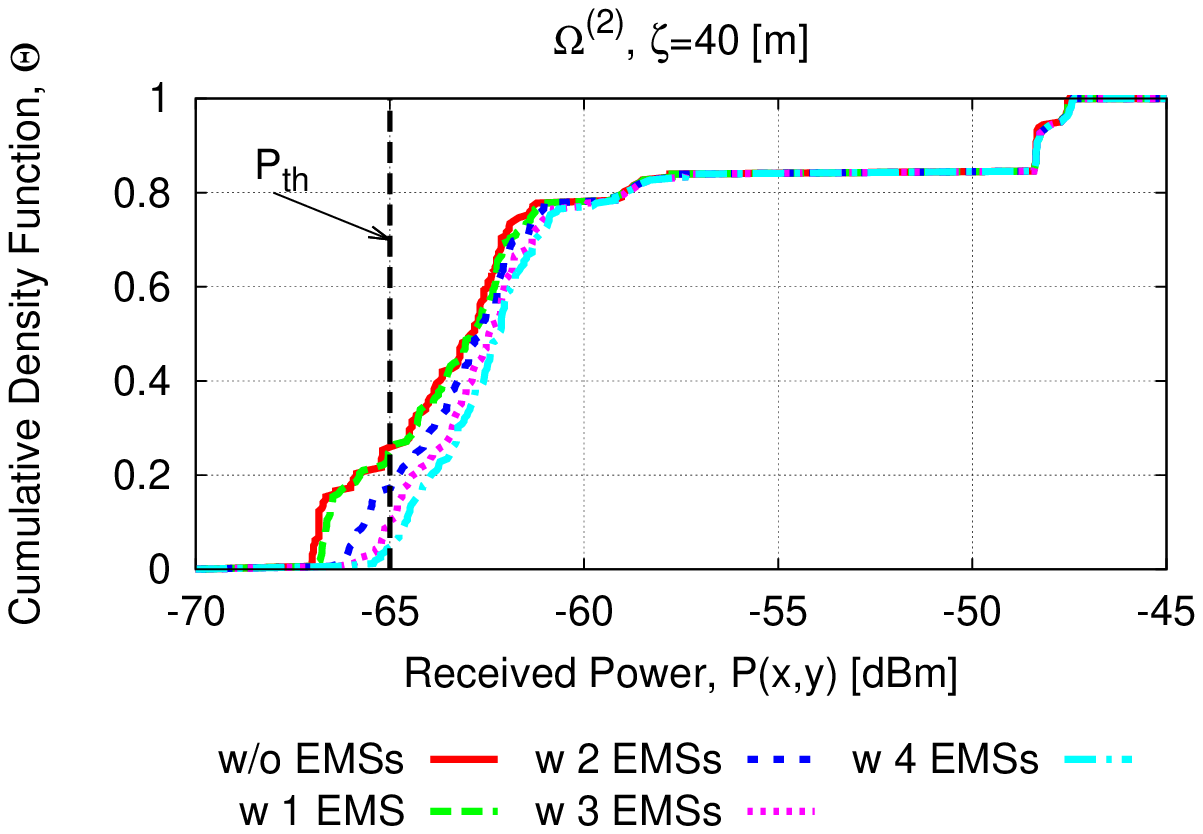}\end{center}

\begin{center}~\vfill\end{center}

\begin{center}\textbf{Fig. 12 - A. Benoni} \textbf{\emph{et al.}}\textbf{,}
\textbf{\emph{{}``}}Planning of \emph{EM} Skins for ...''\end{center}

\newpage
\begin{center}~\vfill\end{center}

\begin{center}\begin{tabular}{cc}
\includegraphics[%
  width=0.45\columnwidth]{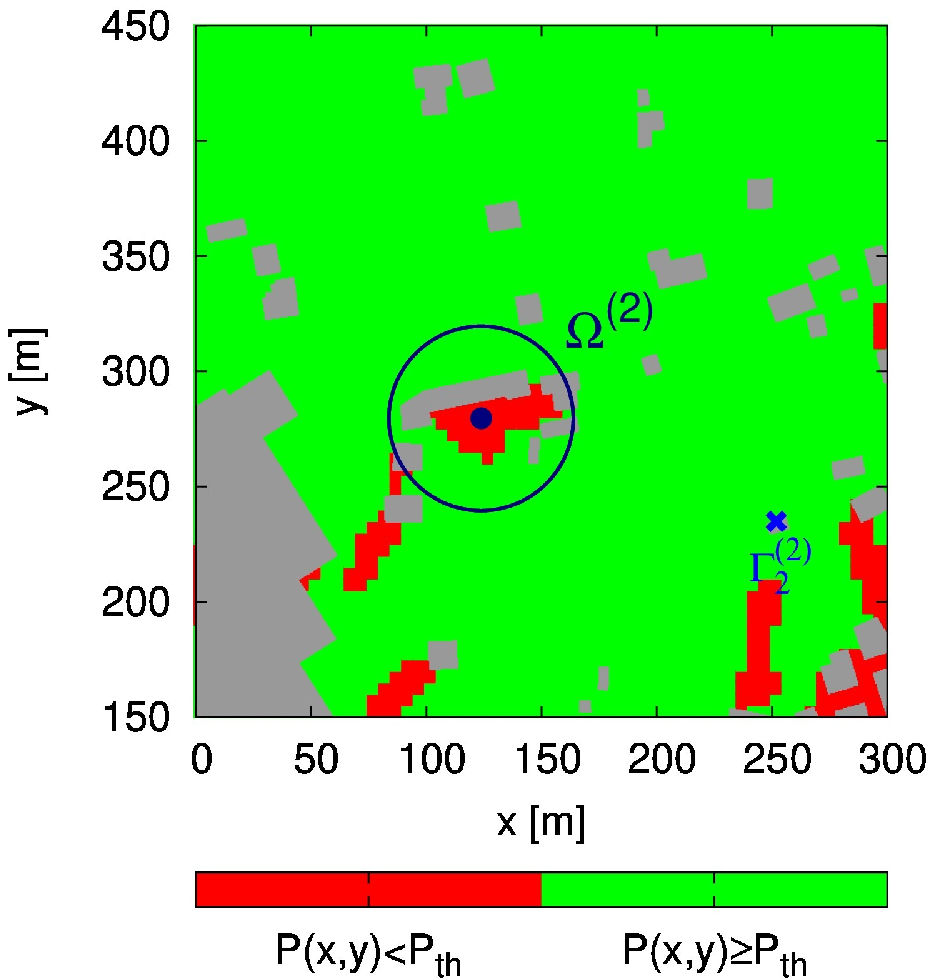}&
\includegraphics[%
  width=0.45\columnwidth]{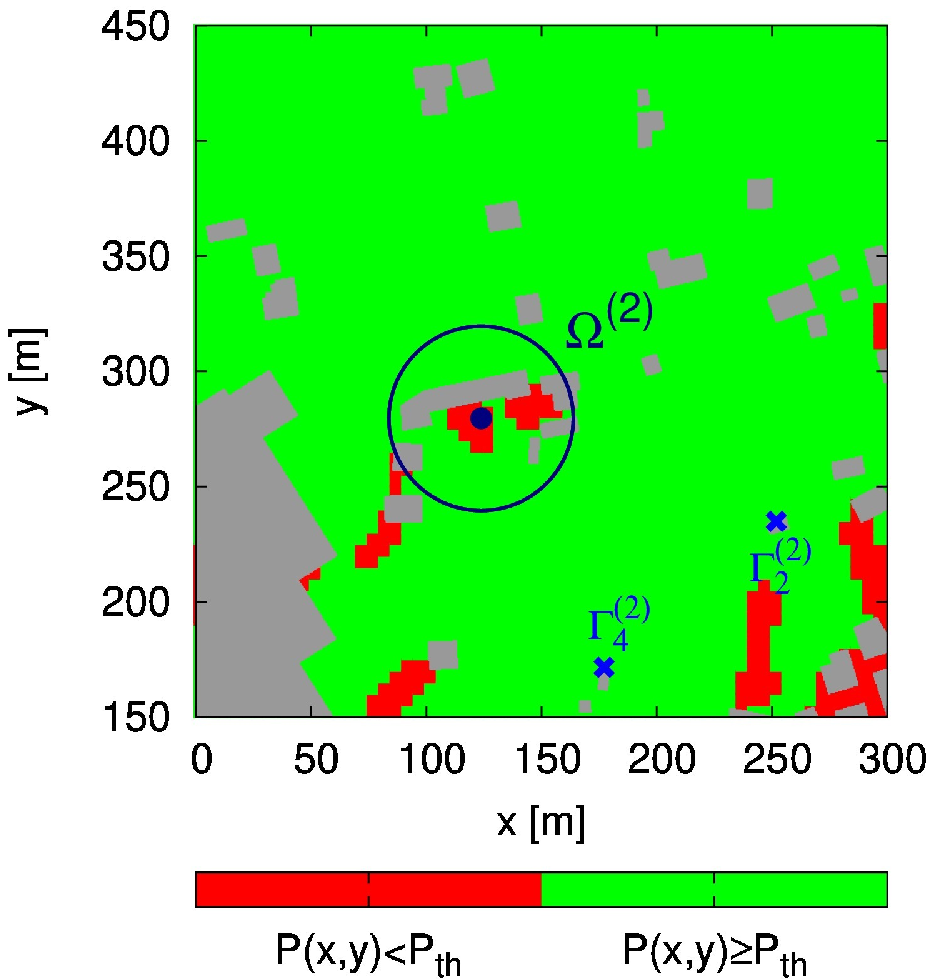}\tabularnewline
(\emph{a})&
(\emph{b})\tabularnewline
\multicolumn{2}{c}{}\tabularnewline
\multicolumn{2}{c}{\includegraphics[%
  width=0.45\columnwidth]{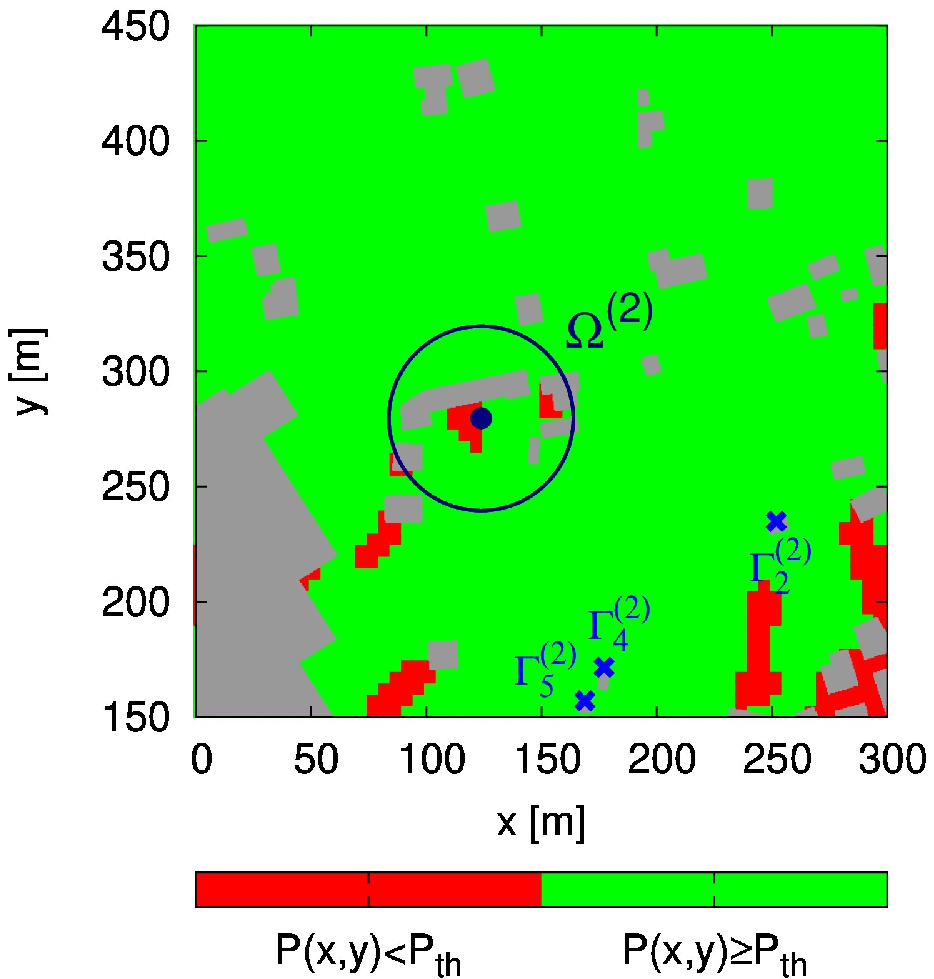}}\tabularnewline
\multicolumn{2}{c}{(\emph{c})}\tabularnewline
\end{tabular}\end{center}

\begin{center}~\vfill\end{center}

\begin{center}\textbf{Fig. 13 - A. Benoni} \textbf{\emph{et al.}}\textbf{,}
\textbf{\emph{{}``}}Planning of \emph{EM} Skins for ...''\end{center}

\newpage
\begin{center}~\vfill\end{center}

\begin{center}\includegraphics[%
  width=0.80\columnwidth]{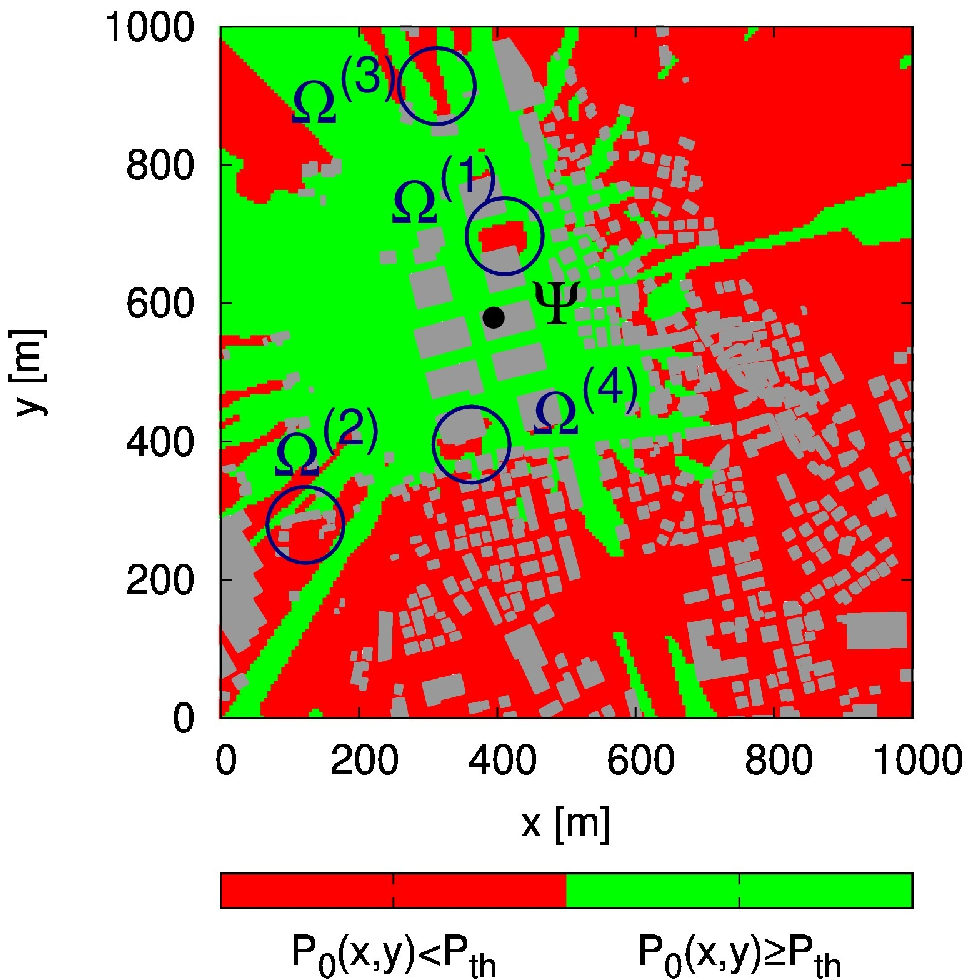}\end{center}

\begin{center}~\vfill\end{center}

\begin{center}\textbf{Fig. 14 - A. Benoni} \textbf{\emph{et al.}}\textbf{,}
\textbf{\emph{{}``}}Planning of \emph{EM} Skins for ...''\end{center}

\newpage
\begin{center}~\vfill\end{center}

\begin{center}\begin{tabular}{c}
\includegraphics[%
  width=0.65\columnwidth]{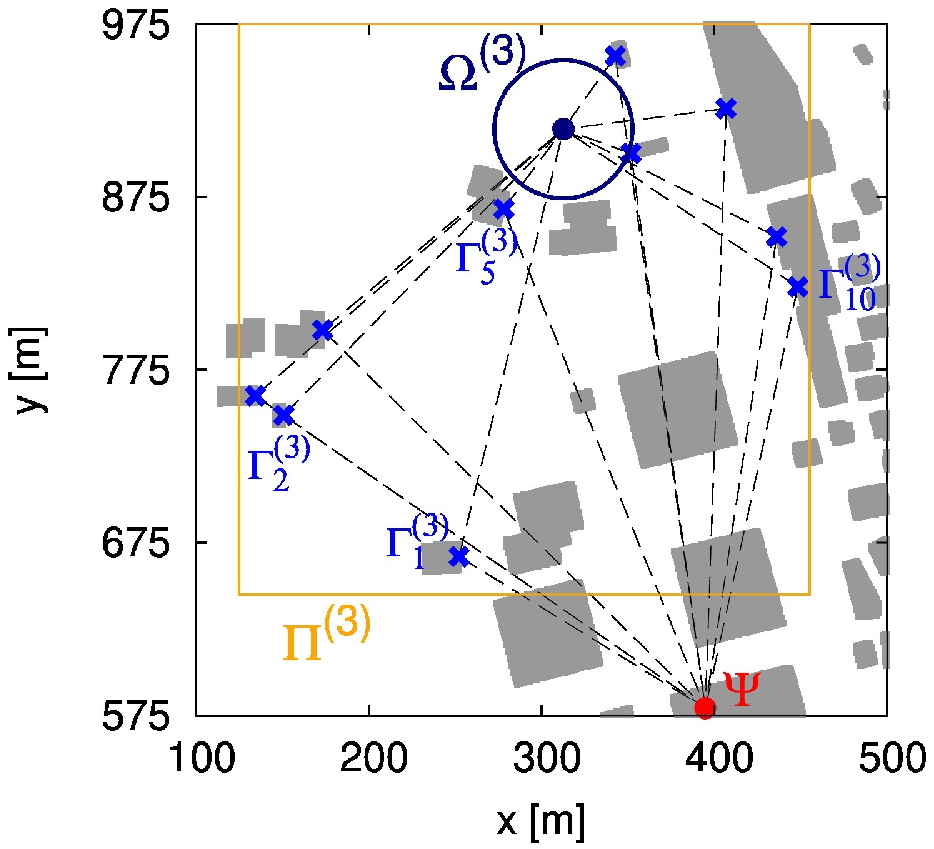}\tabularnewline
(\emph{a})\tabularnewline
\tabularnewline
\includegraphics[%
  width=0.65\columnwidth]{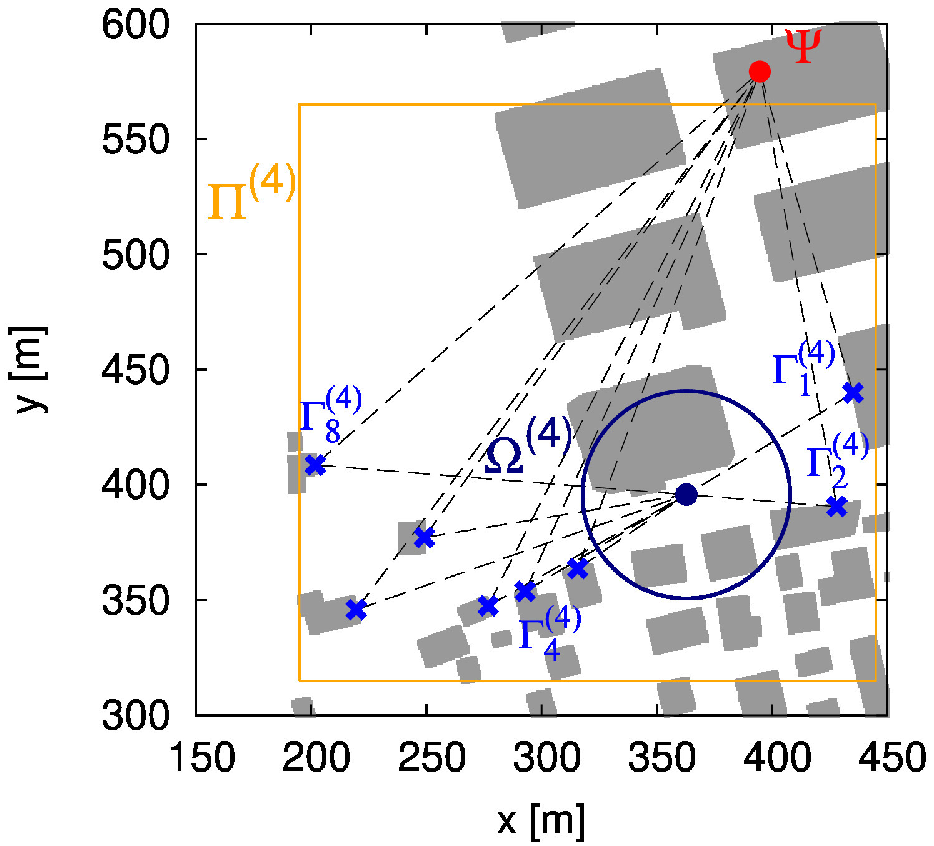}\tabularnewline
(\emph{b})\tabularnewline
\end{tabular}\end{center}

\begin{center}~\vfill\end{center}

\begin{center}\textbf{Fig. 15 - A. Benoni} \textbf{\emph{et al.}}\textbf{,}
\textbf{\emph{{}``}}Planning of \emph{EM} Skins for ...''\end{center}

\newpage
\begin{center}~\vfill\end{center}

\begin{center}\begin{tabular}{ccc}
\emph{w/o EMS}s&
\multicolumn{2}{c}{\emph{w EMS}s}\tabularnewline
\includegraphics[%
  width=0.24\columnwidth,
  keepaspectratio]{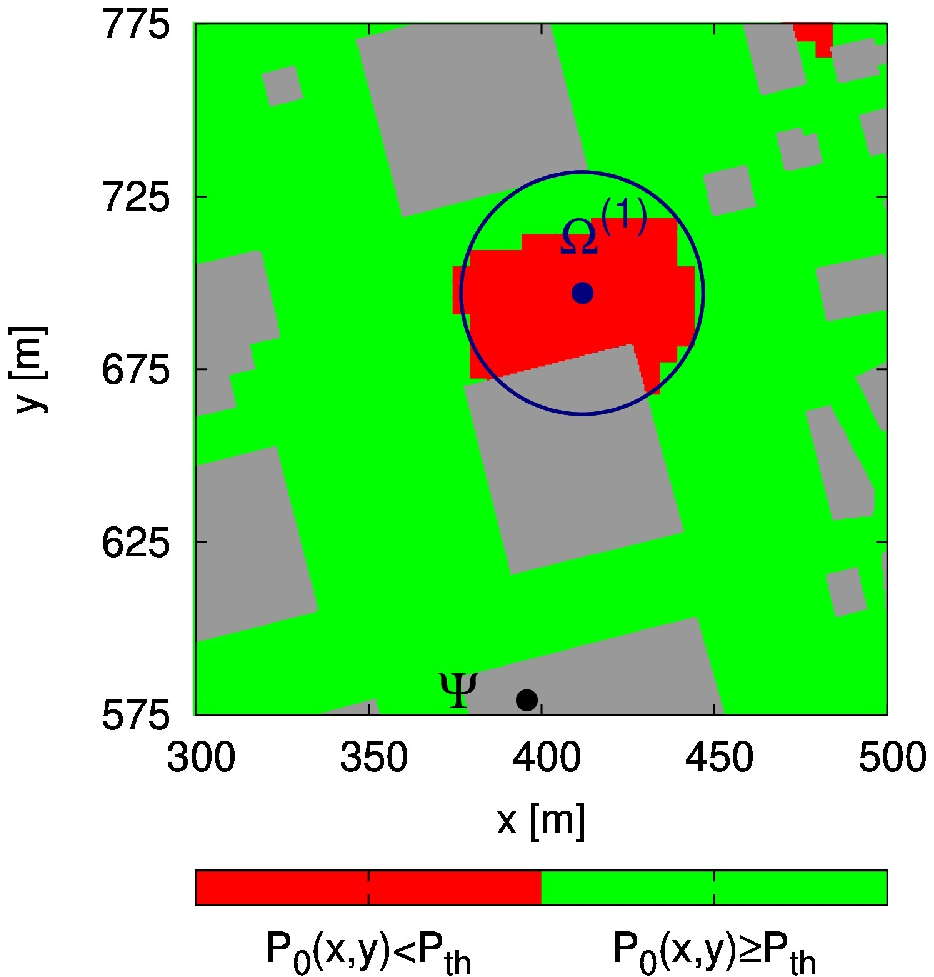}&
\includegraphics[%
  width=0.24\columnwidth]{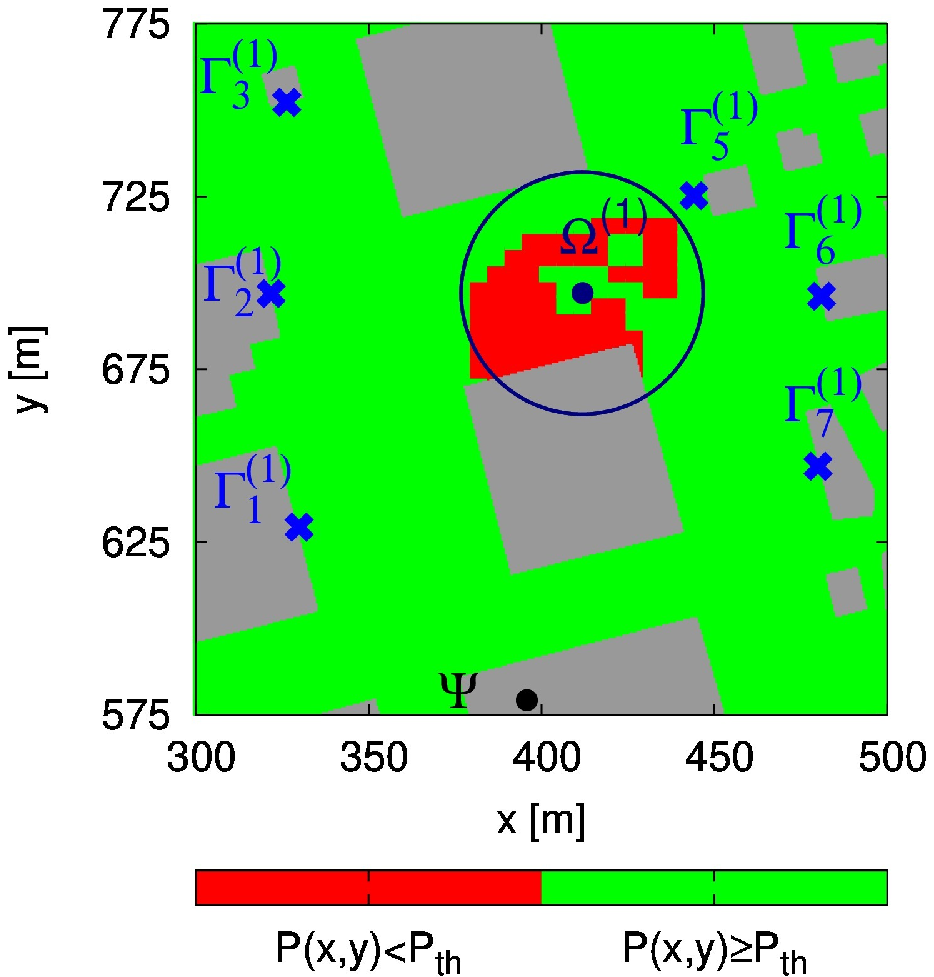}&
\includegraphics[%
  width=0.24\columnwidth]{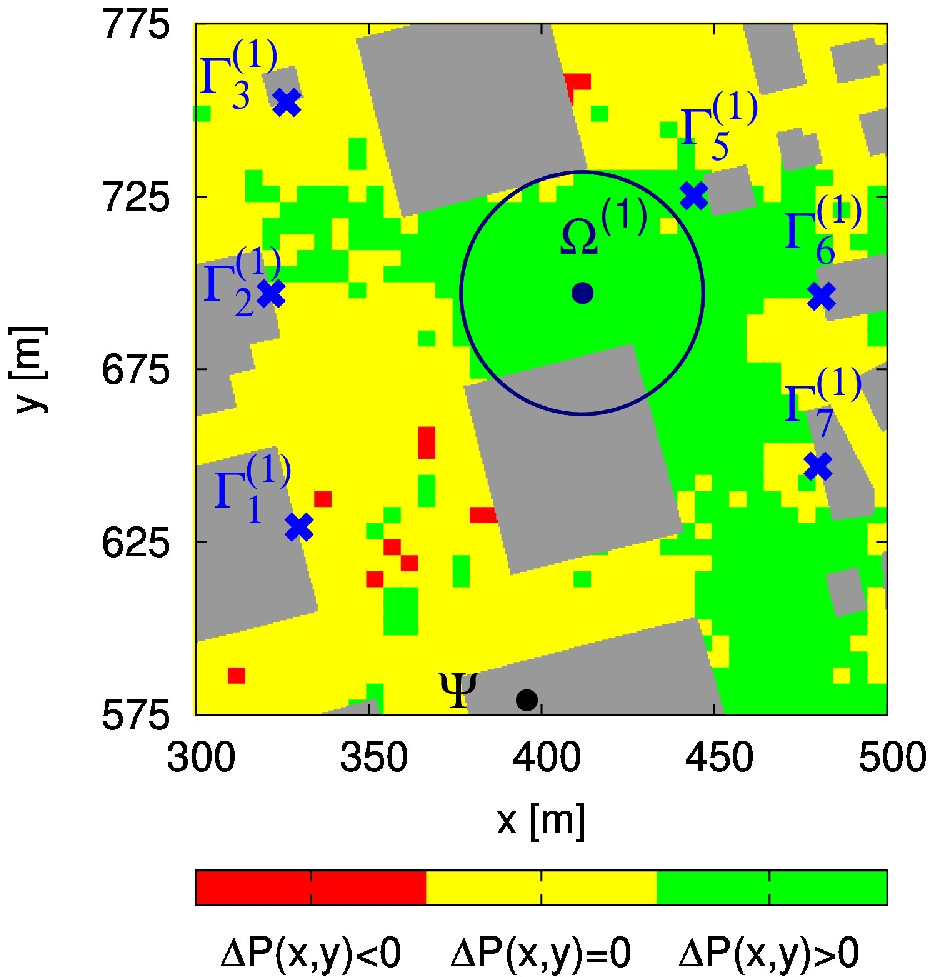}\tabularnewline
(\emph{a})&
(\emph{b})&
(\emph{c})\tabularnewline
\includegraphics[%
  width=0.24\columnwidth,
  keepaspectratio]{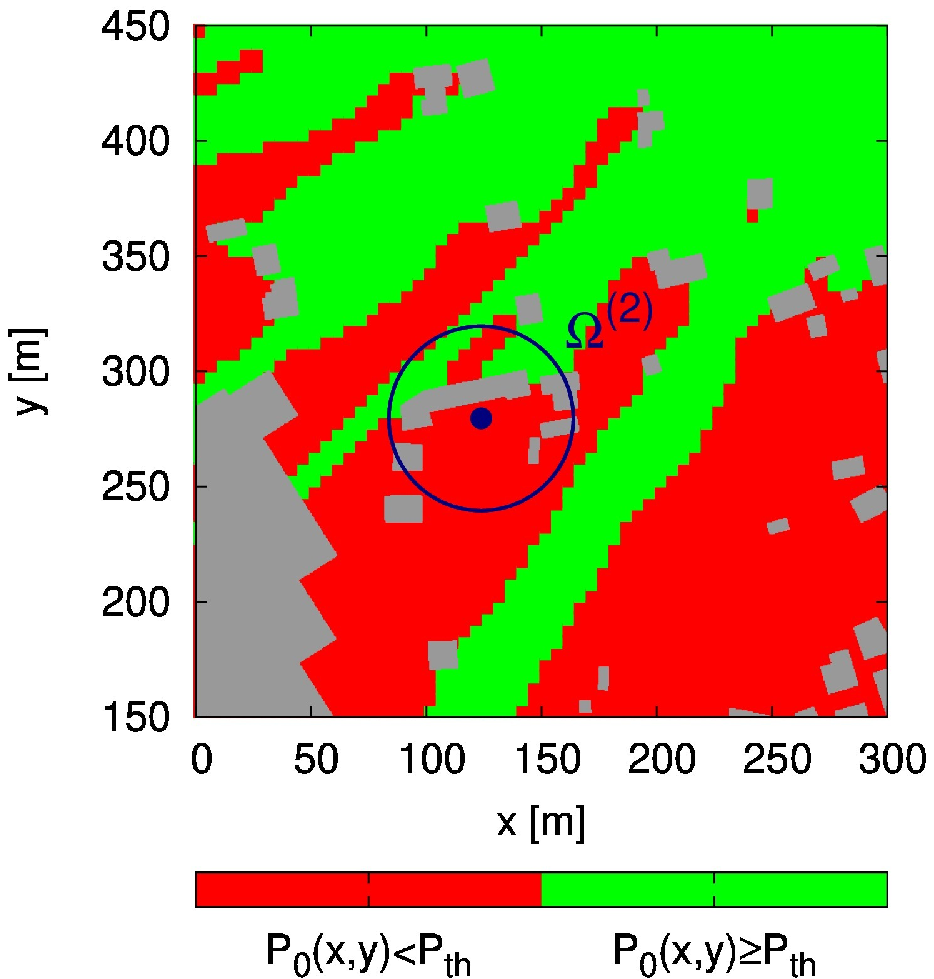}&
\includegraphics[%
  width=0.24\columnwidth]{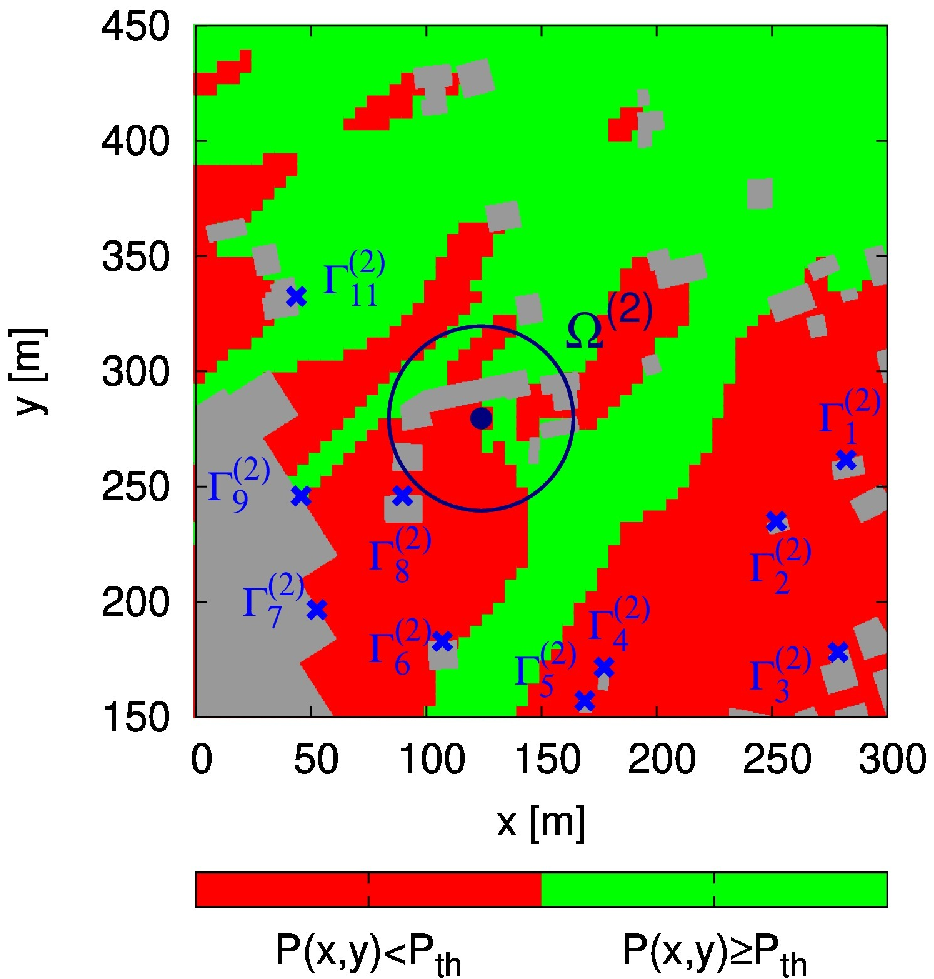}&
\includegraphics[%
  width=0.24\columnwidth]{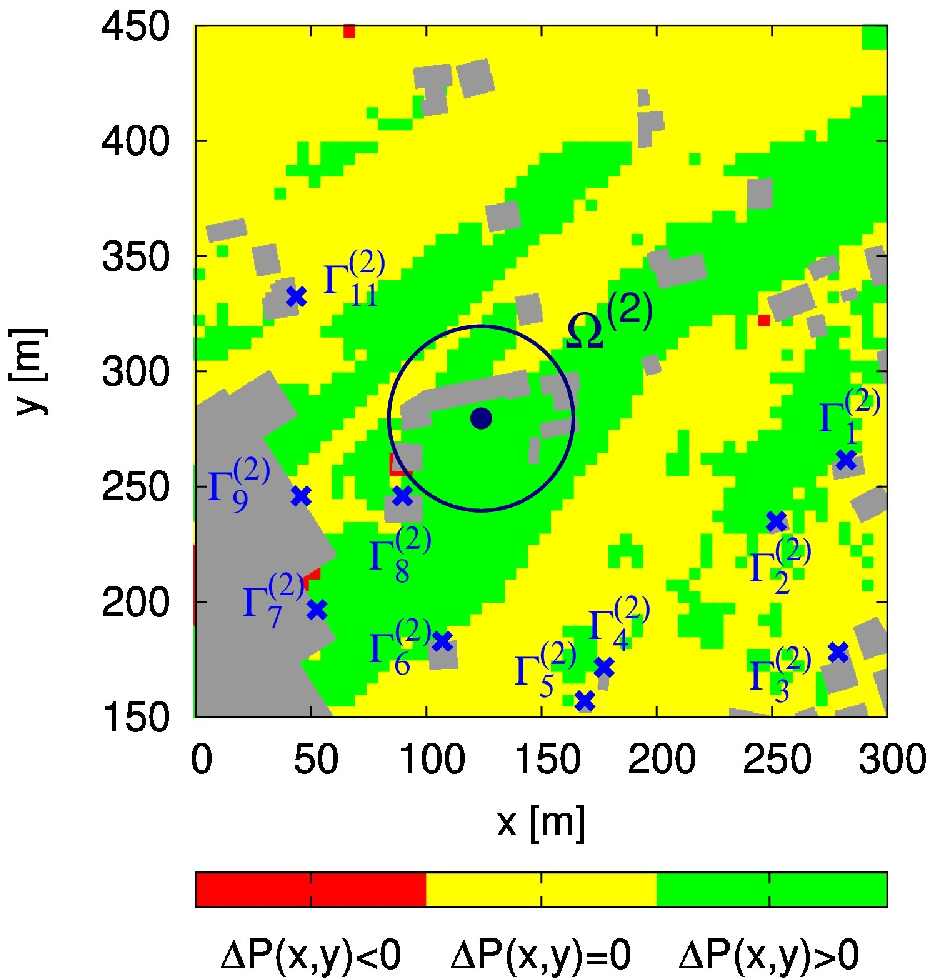}\tabularnewline
(\emph{d})&
(\emph{e})&
(\emph{f})\tabularnewline
\includegraphics[%
  width=0.24\columnwidth,
  keepaspectratio]{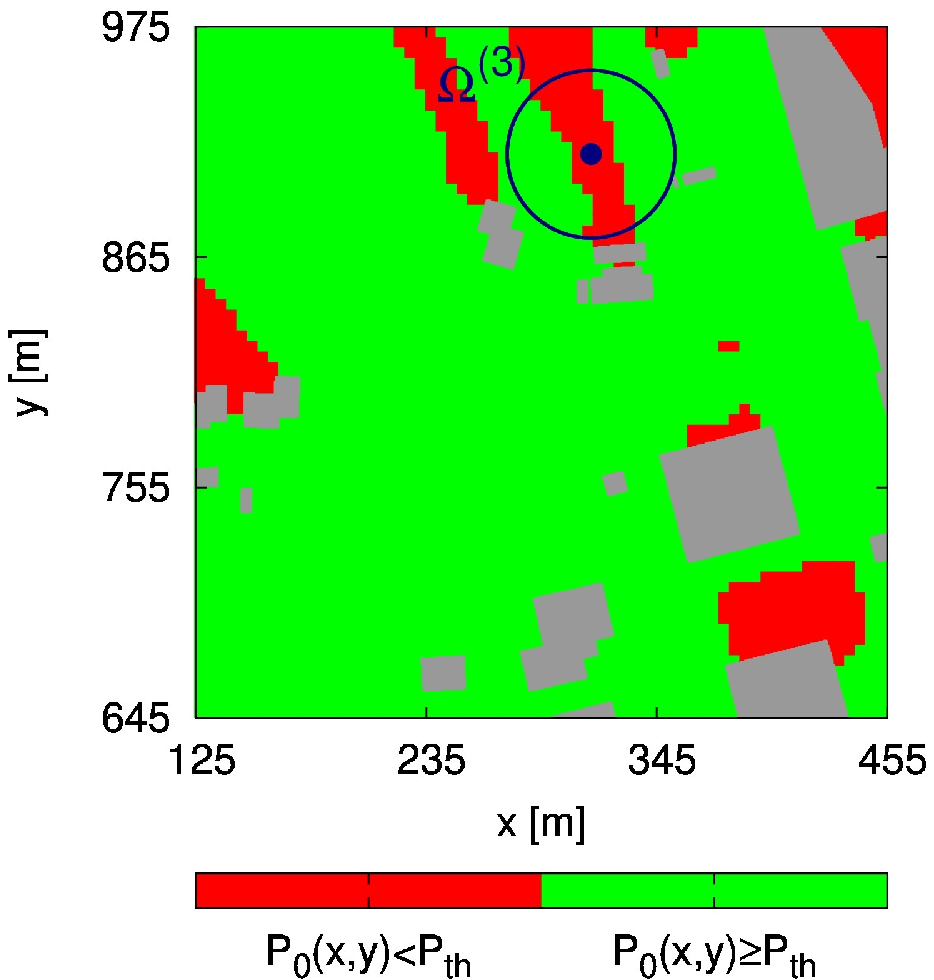}&
\includegraphics[%
  width=0.24\columnwidth]{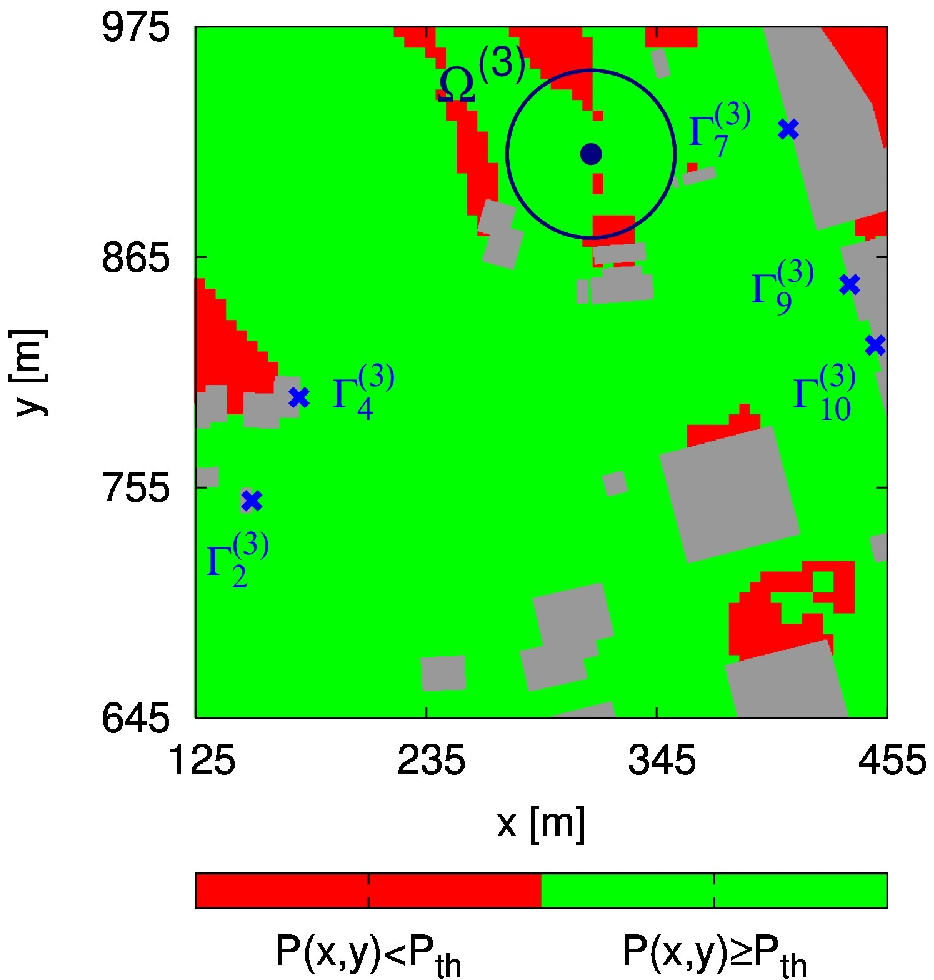}&
\includegraphics[%
  width=0.25\columnwidth]{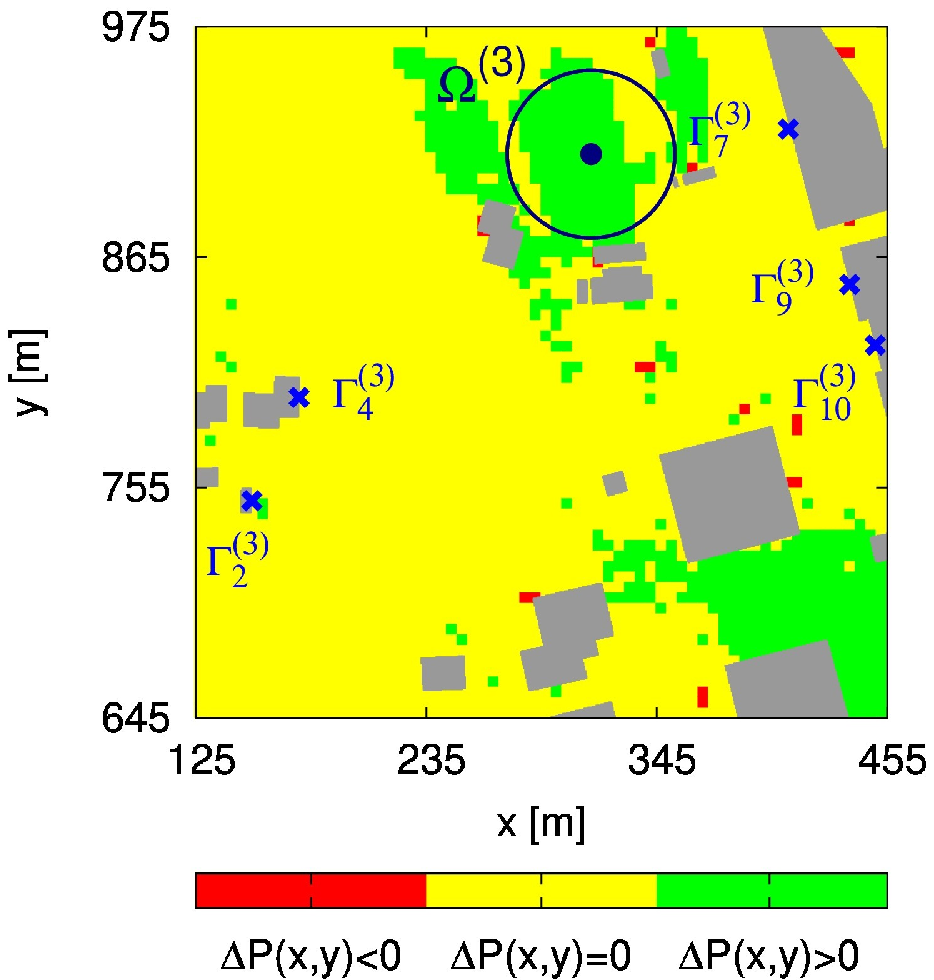}\tabularnewline
(\emph{g})&
(\emph{h})&
(\emph{i})\tabularnewline
\includegraphics[%
  width=0.24\columnwidth,
  keepaspectratio]{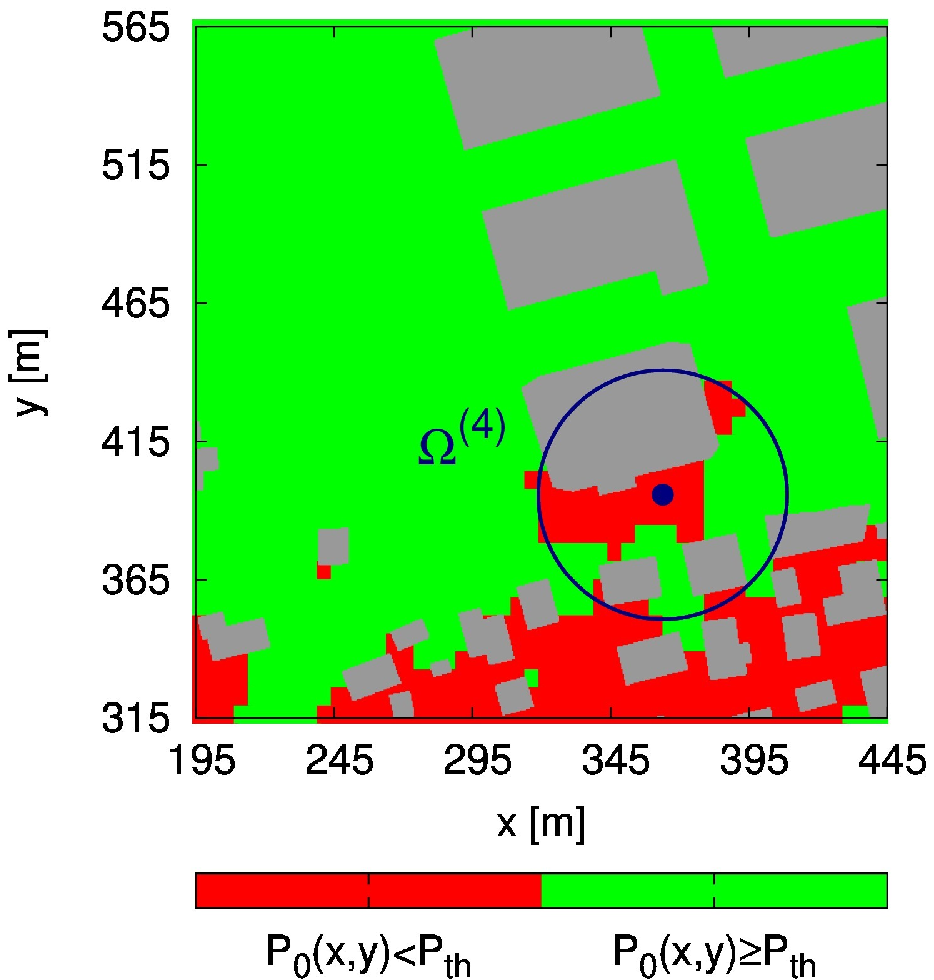}&
\includegraphics[%
  width=0.24\columnwidth]{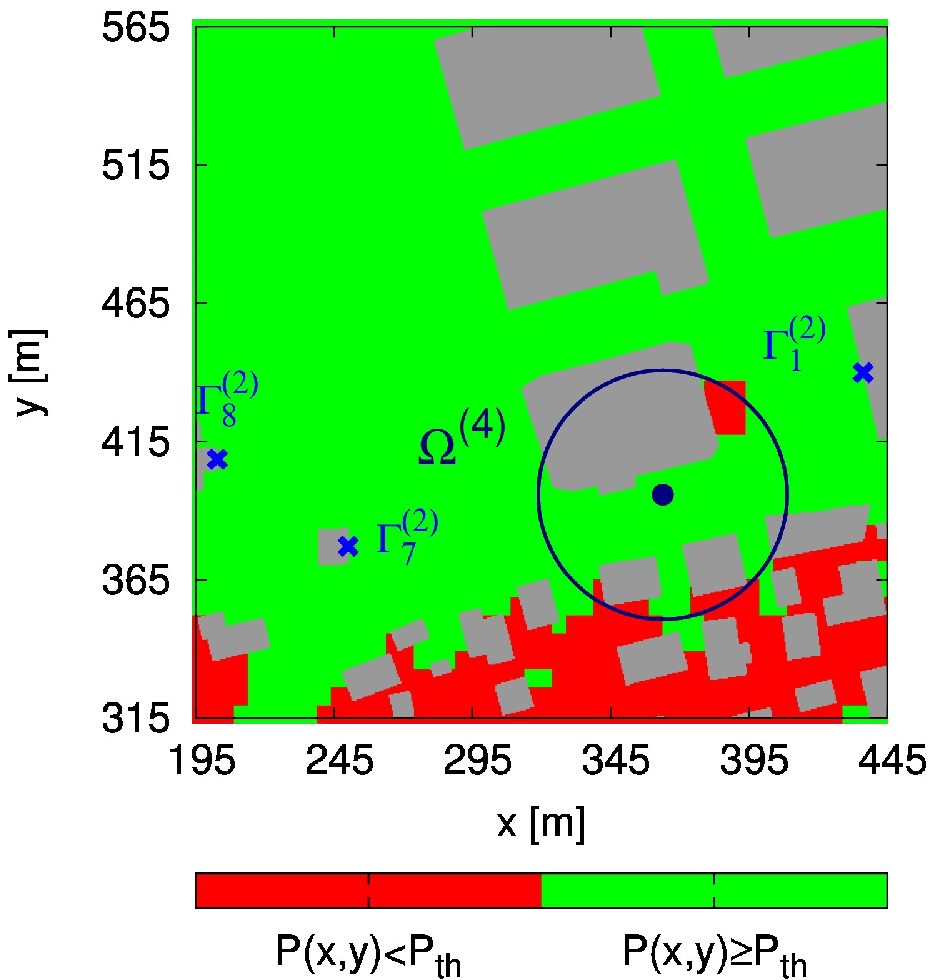}&
\includegraphics[%
  width=0.24\columnwidth]{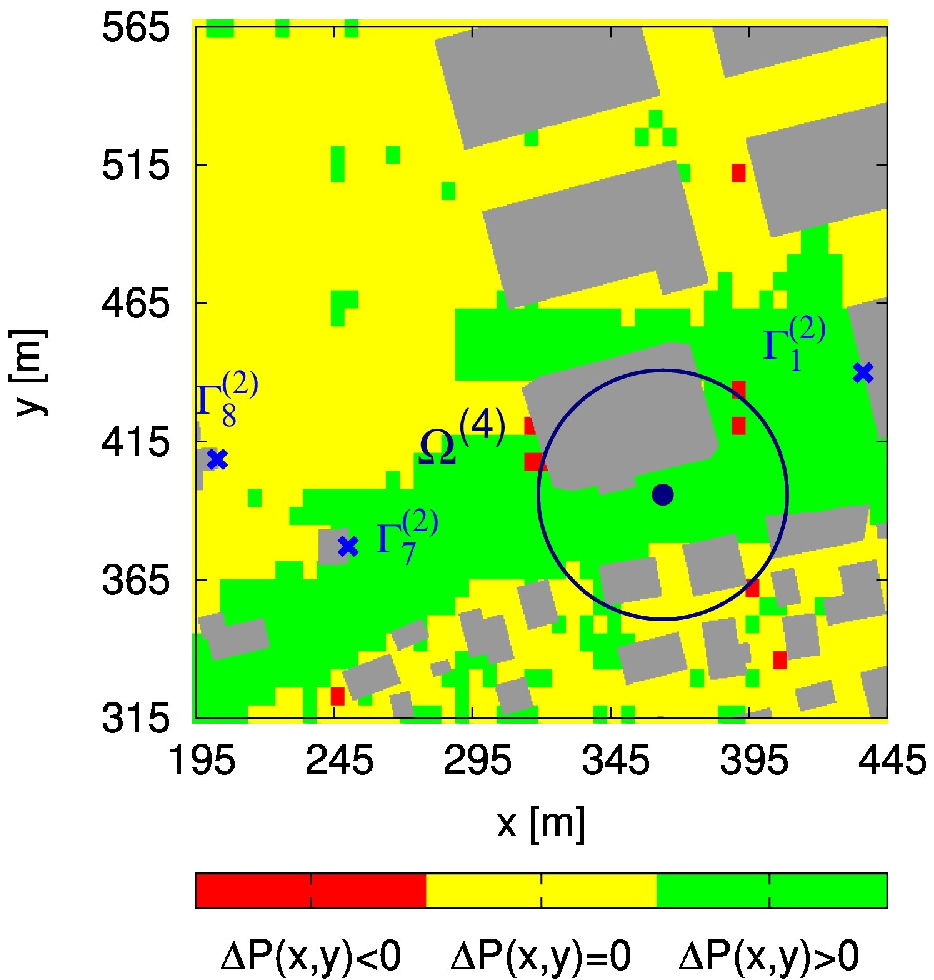}\tabularnewline
(\emph{l})&
(\emph{m})&
(\emph{n})\tabularnewline
\end{tabular}\end{center}

\begin{center}~\vfill\end{center}

\begin{center}\textbf{Fig. 16 - A. Benoni} \textbf{\emph{et al.}}\textbf{,}
\textbf{\emph{{}``}}Planning of \emph{EM} Skins for ...''\end{center}

\newpage
\begin{center}~\vfill\end{center}

\begin{center}\begin{tabular}{cc}
\includegraphics[%
  width=0.48\columnwidth]{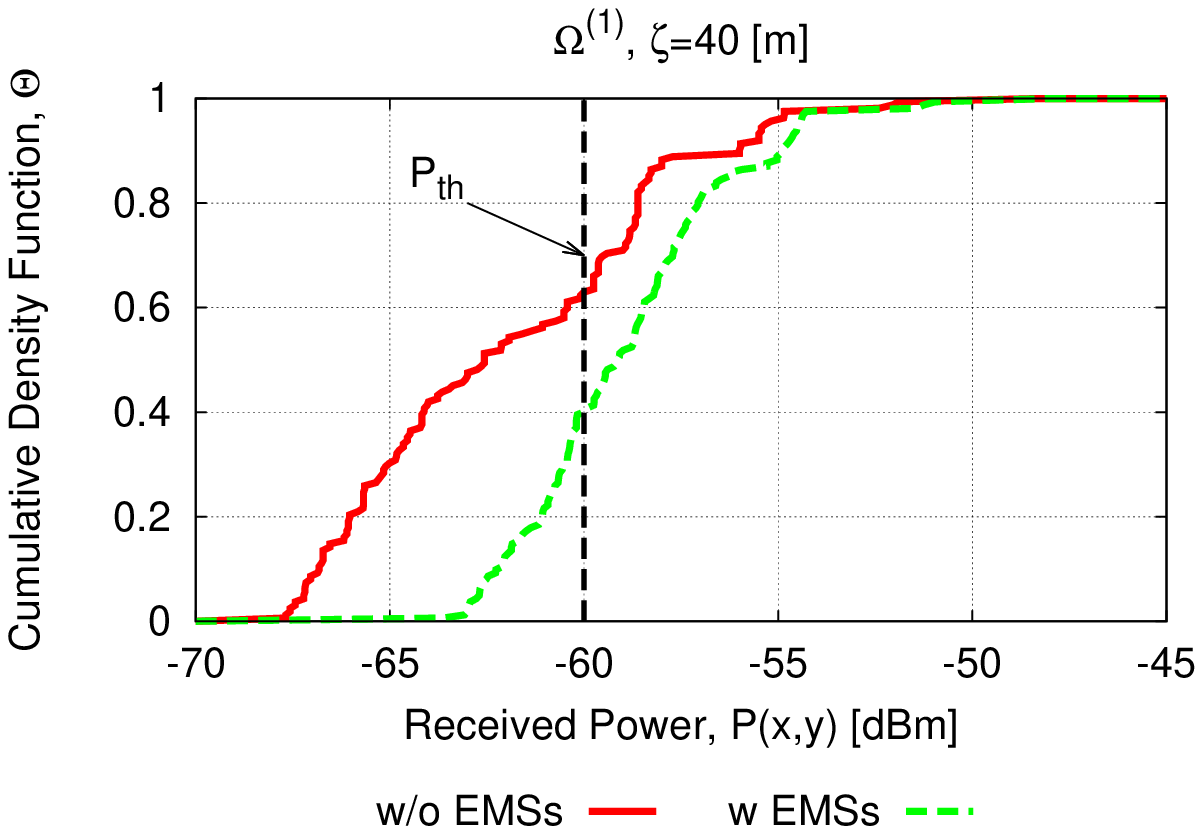}&
\includegraphics[%
  width=0.48\columnwidth]{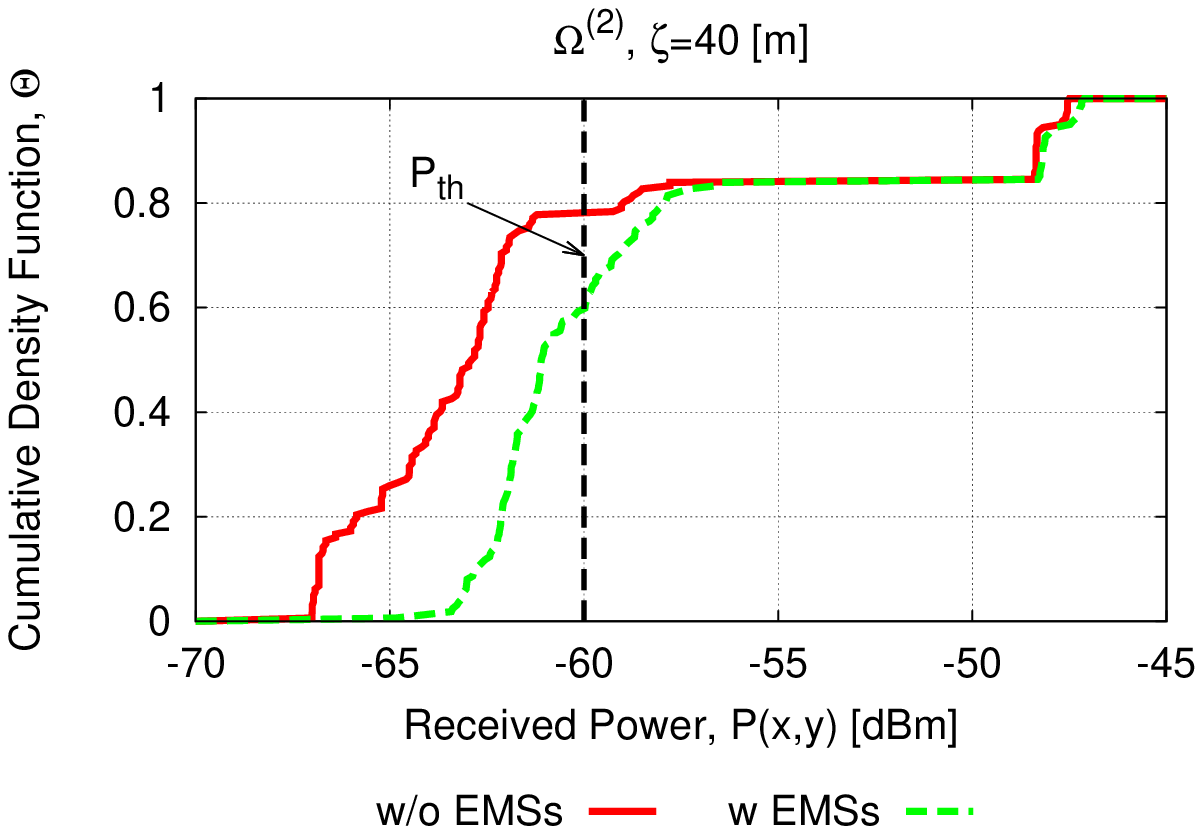}\tabularnewline
(\emph{a})&
(\emph{b})\tabularnewline
&
\tabularnewline
\includegraphics[%
  width=0.48\columnwidth]{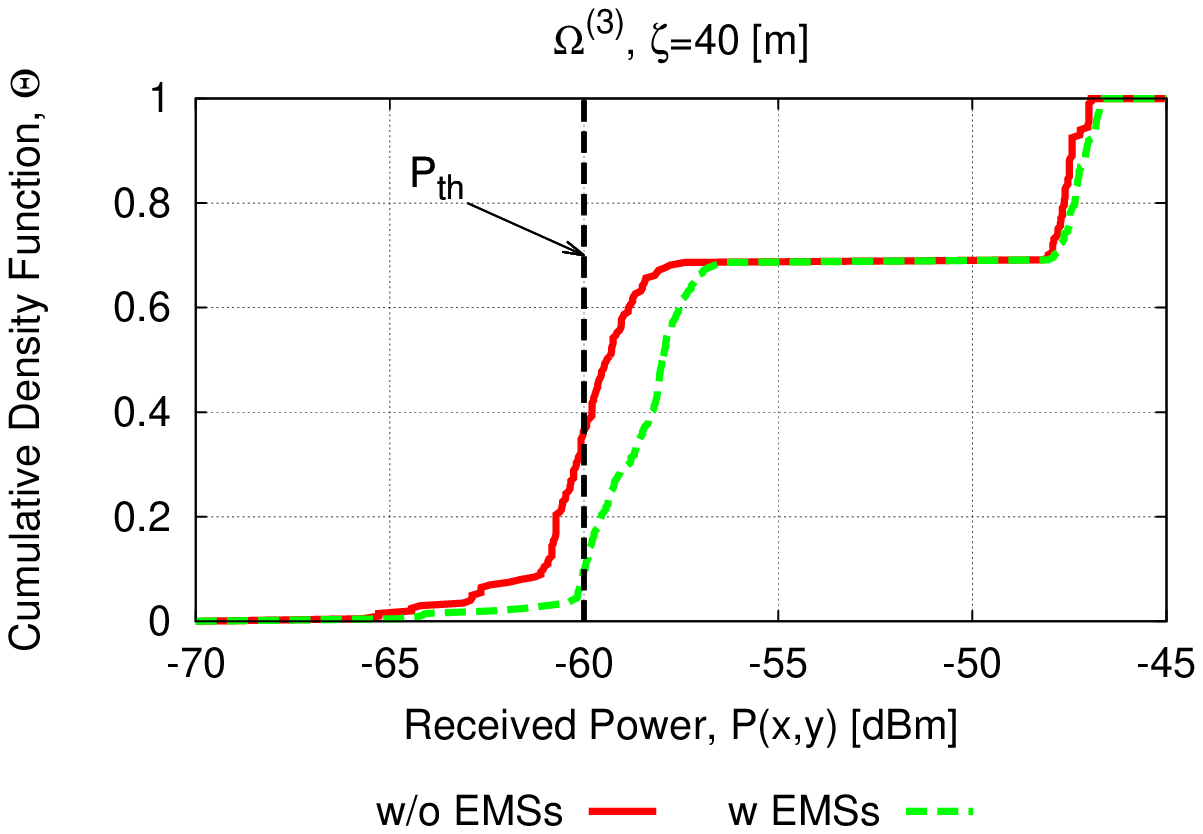}&
\includegraphics[%
  width=0.48\columnwidth]{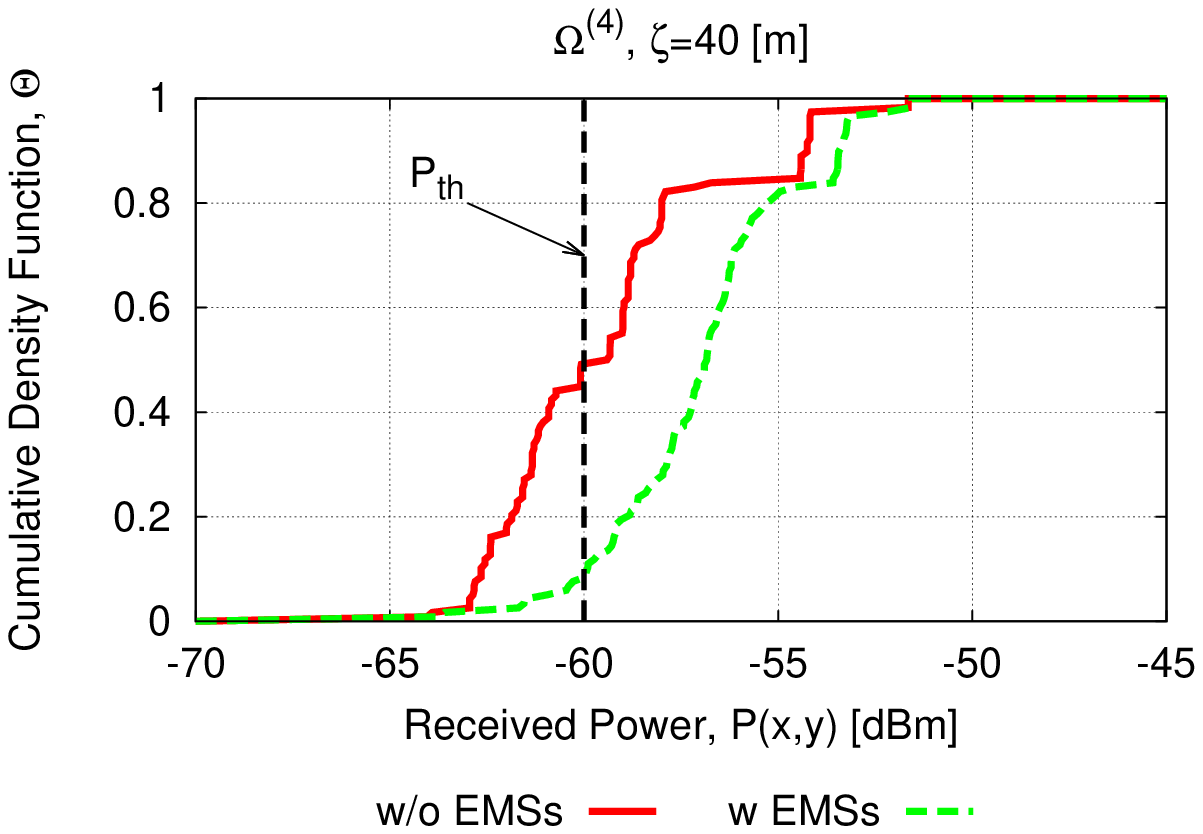}\tabularnewline
(\emph{c})&
(\emph{d})\tabularnewline
\end{tabular}\end{center}

\begin{center}~\vfill\end{center}

\begin{center}\textbf{Fig. 17 - A. Benoni} \textbf{\emph{et al.}}\textbf{,}
\textbf{\emph{{}``}}Planning of \emph{EM} Skins for ...''\end{center}

\newpage
\begin{center}~\vfill\end{center}

\begin{center}\begin{tabular}{|c||c|c|c|c|c|c|c|c|}
\hline 
{\footnotesize $\mathcal{P}_{th}$ {[}dBm{]}}&
{\footnotesize $S$}&
{\footnotesize $K$}&
{\footnotesize $B$}&
{\footnotesize $\Phi_{cov}\left\{ \underline{\chi}=\underline{0}\right\} $}&
{\footnotesize $Q^{\left(opt\right)}$}&
{\footnotesize $\Phi\left\{ \underline{\chi}^{\left(opt\right)}\right\} $}&
{\footnotesize $\Phi_{cov}\left\{ \underline{\chi}^{\left(opt\right)}\right\} $}&
{\footnotesize $\Phi_{cost}\left\{ \underline{\chi}^{\left(opt\right)}\right\} $}\tabularnewline
\hline
\hline 
{\footnotesize $-65$ }&
{\footnotesize $2$}&
{\footnotesize $20$}&
{\footnotesize $1.05\times10^{6}$}&
{\footnotesize $2.10\times10^{-2}$}&
{\footnotesize $7$}&
{\footnotesize $3.50\times10^{-1}$}&
{\footnotesize $2.50\times10^{-4}$}&
{\footnotesize $3.50\times10^{-1}$}\tabularnewline
\hline
{\footnotesize $-60$ }&
{\footnotesize $4$}&
{\footnotesize $38$}&
{\footnotesize $2.75\times10^{11}$}&
{\footnotesize $5.41\times10^{-2}$}&
{\footnotesize $24$}&
{\footnotesize $6.46\times10^{-1}$}&
{\footnotesize $1.40\times10^{-2}$}&
{\footnotesize $6.32\times10^{-1}$}\tabularnewline
\hline
\end{tabular}\end{center}

\begin{center}~\vfill\end{center}

\begin{center}\textbf{Tab. I - A. Benoni} \textbf{\emph{et al.}}\textbf{,}
\textbf{\emph{{}``}}Planning of \emph{EM} Skins for ...''\end{center}
\end{document}